\documentclass[11pt]{amsart}
\usepackage{amsfonts,amssymb}
\usepackage{amsthm}
\usepackage{amscd}
\usepackage{array}
\usepackage{color}
\usepackage{pstricks,pst-node}
\usepackage{pst-plot}
\catcode`\@=11
\def\marginnote#1{}

\newcount\hour
\newcount\minute
\newtoks\amorpm
\hour=\time\divide\hour by60 \minute=\time{\multiply\hour by60
\global\advance\minute by-\hour}
\edef\standardtime{{\ifnum\hour<12 \global\amorpm={am}%
        \else\global\amorpm={pm}\advance\hour by-12 \fi
        \ifnum\hour=0 \hour=12 \fi
        \number\hour:\ifnum\minute<10 0\fi\number\minute\the\amorpm}}
\edef\militarytime{\number\hour:\ifnum\minute<10 0\fi\number\minute}
\renewcommand{\theequation}{\thesection.\arabic{equation}}

\def\draftlabel#1{{\@bsphack\if@filesw {\let\thepage\relax
      \xdef\@gtempa{\write\@auxout{\string
          \newlabel{#1}{{\@currentlabel}{\thepage}}}}}\@gtempa \if@nobreak
    \ifvmode\nobreak\fi\fi\fi\@esphack} \gdef\@eqnlabel{#1}}
    \def\@eqnlabel{}
\def\@vacuum{}
\def\draftmarginnote#1{\marginpar{\raggedright\scriptsize\tt#1}}

\def\draft{
  \oddsidemargin -.5truein
  \def\@oddfoot{\footnotesize \sl preliminary draft \hfil
    \rm\thepage\hfil\sl\today\quad\militarytime}
  \let\@evenfoot\@oddfoot \overfullrule 3pt
    \let\label=\draftlabel
    \let\marginnote=\draftmarginnote
  \def\@eqnnum{(\theequation)\rlap{\kern\marginparsep\tt\@eqnlabel}%
    \global\let\@eqnlabel\@vacuum}

  }

\voffset= - 0.5in \hoffset= - 1.0in

\def\be{\begin{equation}}
\def\ee{\end{equation}}
\def\bea{\begin{eqnarray}}
\def\eea{\end{eqnarray}}
\def\<{\langle}
\def\>{\rangle}

\def\Im{{\rm Im}}

\def\tr{{\mathrm{tr\,}}}

\def\CC{{\mathbb C}}

\def\ZZ{{\mathbb Z}}

\def\e{{\,\rm e}\,}

\def\bea{\begin{eqnarray}}
\def\eea{\end{eqnarray}}

\def\nn{\nonumber}
\def\beq{\begin{equation}}
\def\eeq{\end{equation}}
\def\ba{\beq\begin{array}{c}}
\def\ea{\end{array}\eeq}

\theoremstyle{plain}
\newtheorem{theorem}{Theorem}[section]
\newtheorem{lemma}[theorem]{Lemma}
\newtheorem{corollary}[theorem]{Corollary}
\newtheorem{proposition}[theorem]{Proposition}

\theoremstyle{definition}
\newtheorem{definition}{Definition}[section]
\newtheorem{remark}[definition]{Remark}
\newtheorem{example}[definition]{Example}

\let\operatorname=\mathrm
\let\text=\mathrm

\let\wtd=\widetilde

\def\mod{\operatorname{mod}}

\def\beq{\begin{equation}}
\def\eeq{\end{equation}}
\def\bea{\begin{eqnarray}}
\def\eea{\end{eqnarray}}

\def\Tr{{\tr}}

\newcommand{\cpict}[3]{
\dimen1=#1\advance\dimen1 by-\hsize\divide\dimen1 by-2 \vtop to #2{
\noindent\hskip\dimen1{\special{em:graph #3.bmp}} \vfil}\hskip-2cm }

\newcommand{\tcr}{\textcolor{red}}

\long\def\rem#1{}
\let\@@savethanks\thanks
\def\thanks#1{\gdef\thefootnote{\alph{footnote}}\@@savethanks{#1}}

\begin{document}

\title[Colliding holes in Riemann surfaces and quantum cluster algebras]
{Colliding holes in Riemann surfaces and quantum cluster algebras}

\author{Leonid Chekhov$^\ast$ and Marta Mazzocco$^\dagger$}\thanks{$^\ast$Steklov Mathematical Institute, Moscow, Russia. Email: chekhov@mi.ras.ru.$^\dagger$Department of Mathematical Sciences, Loughborough University, UK. Email: m.mazzocco@lboro.ac.uk, Phone: +44 (0)1509 223187, 
Fax: +44 (0)1509 223969.}
\maketitle

\begin{abstract}
In this paper, we describe a new type of surgery for non-compact Riemann surfaces that naturally appear when colliding two holes or two sides of the same hole in an orientable Riemann surface with boundary (and possibly orbifold points). As a result of this surgery, bordered cusps appear on the boundary components of the Riemann surface. In Poincar\'e uniformization, these bordered cusps correspond to ideal triangles in the fundamental domain.
We introduce the notion of \emph{bordered cusped Teichm\"uller space} and endow it with a Poisson structure, quantization of which is achieved with a canonical quantum ordering.
We give a complete combinatorial description of the bordered cusped Teichm\"uller space by introducing the notion of {\it maximal cusped lamination,}\/ a lamination consisting of geodesics arcs between bordered cusps and closed geodesics homotopic to the boundaries such that it triangulates the Riemann surface. We show that each bordered cusp carries a natural decoration, i.e. a choice of a horocycle, so that the lengths of the arcs in the maximal cusped lamination are defined as $\lambda$-lengths in Thurston--Penner terminology. We compute the Goldman bracket explicitly in terms of these $\lambda$-lengths and show that the Mapping Class Group acts as a generalized cluster algebra mutation. From the physical point of view, our construction provides an explicit coordinatization of moduli spaces of open/closed string worldsheets {and their quantization}.
\end{abstract}

\section{Introduction}
In this paper, we describe a new type of surgery for non-compact Riemann surfaces that naturally appear when colliding two holes or two sides of the same hole in an orientable Riemann surface with boundary (and possibly orbifold points). We define this process in such a way that only the portion of the Riemann surface between the two holes, or between the two sides of the same hole, is affected. We call this portion of surface {\it chewing-gum.}\/ We regularise the chewing-gum by introducing two lines, called {\it collars,} which separate the chewing-gum from the rest of the Riemann surface.
As the collision process starts, we impose that the  chewing-gum hyperbolic area is preserved so that  the chewing-gum becomes longer and thinner (in hyperbolic metric sense). We prove that upon taking the limit of the chewing-gum length to infinity, the chewing gum breaks into two {\it bordered cusps}\/ whereas the collars
become horocycles decorating these cusps on the newly obtained Riemann surface (or surfaces if the result is disconnected).
In Poincar\'e uniformisation, these bordered cusps correspond to ideal triangles in the fundamental domain. We call  \emph{bordered cusped Riemann surface} a Riemann surface $\Sigma_{g,s,n}$ of genus $g$, at least one hole and a total of $s\ge 1$ holes and orbifold points, and with additional  $n\ge 1$ \emph{bordered cusps} situated on holes to which these cusps are assigned.

In the limit of the chewing-gum length to infinity, closed geodesics that were passing along the chewing gum become geodesic \emph{arcs} -- infinitely long geodesics that start and terminate at the bordered cusps. The main objects describing Riemann surfaces in a Mapping-Class-Group (MCG) invariant way are \emph{geodesic  functions}, i.e.  $2\cosh\frac{l_\gamma}{2}$ where $l_\gamma$ is the length of a closed geodesic $\gamma$. In the limiting process, the geodesic functions  of closed geodesics passing through the chewing-gum become exponentiated signed half-lengths of the parts of arcs confined between the horocycles associated to the cusps, or in other words, genuine $\lambda$ lengths in the Penner--Thurston description.

We introduce the notion of \emph{bordered cusped Teichm\"uller space} $\widehat{\mathfrak T}_{g,s,n}$ and give a complete combinatorial description of it by introducing the notion of {\it extended shear coordinates.} We compute the Goldman bracket on the extended shear coordinates and construct a set of functionally independent
$\lambda$-lengths that completely coordinatize the bordered cusped
Teichm\"uller space in such a way that the Goldamn bracket is closed and combinatorially explicit (see Theorem \ref{th-WP-cusp} and Corollary \ref{cor:arcs}) and the MCG action corresponds to the generalized cluster algebra structure introduced in \cite{ChSh}.

The problem of producing a closed Poisson algebra of geodesic  functions on a Riemann surface $\Sigma_{g,s}$ for any genus $g$  and any number $s_h>1$ of holes and any number of $s_o$ of orbifold points remained open (here $s=s_o+s_h$). In this paper we fully characterise the Poisson algebra of geodesic functions on $\Sigma_{g,s}$  as a specific Poisson sub-algebra of the set of $\lambda$-lengths on the bordered cusped
Teichm\"uller space  $\widehat{\mathfrak T}_{g,s,1}$ of Riemann surfaces of genus $g$ with the same number of holes $s_h$ and of orbifold points $s_o$ and one {\it  bordered cusp} on the boundary (see Subsection \ref{se:no-cusp}).

By complexification, we find Darboux coordinates (quantum tori) for the moduli spaces of
non-compact Riemann surfaces, so that our results will find applications in the theory of open intersection numbers (see \cite{PST, BH, B1} ).

The case of $g=0$ is treated in great detail in \cite{CMR} due to
its links with the theory of the Painlev\'e differential equations. It
is interesting to observe that in these cases the chewing gum moves produce the
confluence scheme of the Painlev\'e differential equations, and at quantum level it
produces the confluence of the spherical sub-algebras of the confluent Cherednik
algebras defined in \cite{MM}. The role of cluster algebras in the Cherednik algebra
setting will be investigated further in subsequent publications.

In physical terms, we provide an explicit coordinatization of open/closed string world-sheets described as \emph{windowed surfaces}
by R. Kaufmann and Penner in \cite{KP} where they considered laminations of Riemann surfaces $\Sigma^{w}_{g,s,n}$ of genus $g$ with $s>0$ holes (boundary components) and with $n\ge 0$ \emph{windows} -- i.e. domains stretched between marked points located on the boundaries of the holes. These laminations comprised both closed curves and curves starting and terminating at windows thus describing foliations of $\Sigma_{g,s,n}^{w}$.
In our construction, we decorate $\Sigma_{g,s,n}$ by horocycles based at the endpoints of the bordered cusps; laminations on the
windowed surfaces $\Sigma_{g,s,n}^{w}$ then correspond to sets of arcs stretched between bordered cusps, so, literally, the windows of Kaufmann and Penner are segments of horocycles confined between two bordering geodesic curves separating windows.
In the present paper, we describe the Teichm\"uller spaces $\widehat{\mathfrak T}_{g,s,n}$ of $\Sigma_{g,s,n}$; the Kaufmann--Penner coordinates on the space of laminations are then the projective (tropical) limit of the extended shear coordinates on  $\widehat{\mathfrak T}_{g,s,n}$ introduced in this paper. We thus provide a convenient parameterization of the open/closed string world-sheets and their quantization (see subsection \ref{ss:KP}).

\vskip 2mm
We also attack the problem of quantum ordering of the product of non--commuting operators obtained by quantising this picture.

To solve the problem of quantum ordering we introduce the notion of {\it cusped geodesic lamination}\/  (CGL) comprising both
closed geodesics and geodesic arcs such that  they have no intersections nor self-intersections in the interior of a Riemann surface, but can be incident to the same bordered cusp (note that this condition establishes a linear ordering on the set of ends of arcs belonging to the same CGL and incident to the same bordered cusp).  We prove the following theorem:

\vskip10pt

\noindent {\bf Theorem.}\ For any Riemann surface $\Sigma_{g,s,n}$ of genus $g$ with $s_h\ge 1$ holes, $s_o$ orbifold points, $s=s_o+s_h$, and with $n\ge 1$ \emph{bordered cusps}, there always exists a maximal CGLs denoted by CGL$_{\mathfrak a}^{\text{max}}$ that comprises exactly $6g-6+3s+2n$ elements that are arcs and $\omega$-cycles (closed loops around orbifold points or holes not containing bordered cusps) with the following properties:
\begin{enumerate}
\item Arcs from CGL$_{\mathfrak a}^{\text{max}}$ are edges of an ideal triangle partition of $\Sigma_{g,s,n}$ in which every hole that does not contain bordered cusps and every orbifold point is enclosed in a monogon.
\item The $\lambda$-lengths of the arcs in CGL$_{\mathfrak a}^{\text{max}}$  satisfy homogeneous Poisson brackets or homogeneous commutation relations (see formula (\ref{eq:comb-p})).
\item  The $\omega$ coefficients corresponding to $\omega$-cycles are $2\cosh(P/2)$ for holes with the perimeter $P$ and $2\cos(\pi/r)$ for $\mathbb Z_r$-orbifold points; these coefficients are Casimirs and are invariant under the MCG action.
\item Given any closed geodesic $\gamma$ (geodesic arc $\mathfrak a$) in the Riemann surface $\Sigma_{g,s,n}$, its geodesic length function ($\lambda$-length) is a Laurent polynomial with positive coefficients of the $\omega$-cycles  and of the $\lambda$--lengths of the arcs in  CGL$_{\mathfrak a}^{\text{max}}$.
\end{enumerate}

For any given CGL$_{\mathfrak a}^{\text{max}}$, the fat graph $\widehat{\mathcal G}$ dual to the triangle partition defined by it  is a spine of $\Sigma_{g,s,n}$ ($n\ge1$) in which all holes without bordered cusps and all orbifold points are
contained in loops, at every bordered cusp we have exactly one one-valent vertex, and all other vertices are three-valent. On this fat-graph, we introduce
the {\it extended shear coordinates $\{Z_\alpha,\pi_j\}$}, where $Z_\alpha$ denote the standard shear coordinates of the inner edges of the fat-graph and $\pi_j$ are new shear coordinates of the open edges. We describe explicitly the $1:1$ correspondence between these extended shear coordinates and
$\lambda$-lengths
of arcs in the CGL$_{\mathfrak a}^{\text{max}}$ corresponding to the ideal triangle partition dual to this $\widehat{\mathcal G}$. In this correspondence, every loop in
$\widehat{\mathcal G}$ corresponds to an $\omega$-cycle in CGL$_{\mathfrak a}^{\text{max}}$ containing the an un-cusped hole or a orbifold point.
The edge of $\widehat{\mathcal G}$ incident to a loop corresponds to the arc
bordering a monogon from CGL$_{\mathfrak a}^{\text{max}}$, every edge of $\widehat{\mathcal G}$ joining two different three-valent vertices intersects with exactly one arc of CGL$_{\mathfrak a}^{\text{max}}$, and every edge of $\widehat{\mathcal G}$ terminating at a bordered cusp corresponds to the
bordering arc immediately to the left of this cusp. We show (see Theorem~\ref{prop:monoidal}) that $\lambda$-lengths of arcs from $CGL_{\mathfrak a}^{\text{max}}$ are monomials in $e^{Z_\alpha/2}$, $e^{\pi_j/2}$ and, vice versa, all $e^{Z_\alpha/2}$, $e^{\pi_j/2}$ are monomials in $\lambda_{\mathfrak a}^{\pm 1/2}$.

This monomiality property is crucial for quantisation and dictates the quantum ordering, allowing us to prove that the quantized $\lambda$-lengths of the arcs in CGL$_{\mathfrak a}^{\text{max}}$  satisfy homogeneous commutation relations (see formula (\ref{homog})).

The two Poisson and quantum algebras
of $\{Z_\alpha,\pi_j\}$ and of arc functions in CGL$_{\mathfrak a}^{\text{max}}$ therefore imply one another and we prove that quantum algebras of arc functions
from the same CGL$_{\mathfrak a}^{\text{max}}$ satisfy the same quantum commutation relations as in the quantum cluster algebras by  Berenstein--Zelevinsky  \cite{BerZel}. We can therefore identify a CGL$_{\mathfrak a}^{\text{max}}$ with a seed of a quantum cluster algebra in such a way that the
quantum cluster algebras we obtain---let us call them {\it quantum cluster algebras of geometric type}---satisfy the main axioms of the Berenstein--Zelevinsky
construction. However,  the mutation transformations in our quantum cluster algebras of geometric type
include also generalized cluster
transformations from \cite{ChSh}  besides the standard Ptolemy-type mutations. Moreover, the Laurent
and positivity properties for geodesic functions expressed in the extended shear coordinates directly imply the Laurent and positivity
properties for our  quantum cluster algebras of geometric type. It is interesting to mention that in the case of
the bordered cusped Teichm\"uller spaces ${\mathcal T}_{g,s,n}$ with $n\ge1$ both the extended shear coordinates and $\lambda$-lengths of arcs from a CGL$_{\mathfrak a}^{\text{max}}$ of $\Sigma_{g,s,n}$ satisfy homogeneous $q$-commutation relations being therefore quantum tori. We can obtain Poisson and quantum algebras of $\lambda$-lengths only for arcs starting and terminating at bordered cusps; presumably no such algebras can be defined for arcs starting and/or terminating at punctures (holes) so we eliminate such arcs from CGLs by imposing the monogon condition. This allows us simultaneously avoid the issue of tagging the ends of arcs terminating at punctures~\cite{FST},~\cite{FT}.

Before explaining the structure of the paper, let us recall some important results on which this paper is based.

Darboux coordinates  for moduli spaces of Riemann surfaces
with holes (and no bordered cusps) were identified in \cite{ChF1} with the shear
coordinates for an ideal triangle decomposition obtained in  \cite{Fock1} by
generalising the results for punctured
Riemann surfaces proved in \cite{Penn1}.

An explicit combinatorial construction of the corresponding classical geodesic  functions in terms of
shear coordinates of decorated Teichm\"uller spaces for Riemann surfaces with holes (and no bordered cusps)
was proposed in \cite{ChF2}: it was shown there that all
geodesic  functions are Laurent polynomials of exponentiated coordinates with positive integer coefficients;
this remains true for Riemann surfaces with $\mathbb Z_2$ and $\mathbb Z_3$ orbifold points~\cite{Ch1a},~\cite{Ch2}; the integrality
condition breaks in general in the case of orbifold points of arbitrary order \cite{ChSh} but positivity remains in this case
as well. In  \cite{ChSh}, a general combinatorial construction of geodesic  functions in terms of shear coordinates for
orbifold Riemann surfaces was constructed; as a byproduct of this constructions,  new generalised cluster transformations (cluster algebras with coefficients)
were introduced.

Upon quantisation, the observables of a quantum Riemann surface are given by an algebra of quantum
geodesic  functions. The  shear coordinates were quantized in \cite{ChF1} and in the Liouville-type parameterisation in \cite{Kashaev}.
Universally, the quantum mapping-class group transformations (or, the quantum flip
morphsms) that satisfy the quantum pentagon identity were based on the quantum dilogarithm function \cite{Faddeev}.
Shear coordinates can be identified with the $Y$-type cluster variables \cite{FZ},~\cite{FZ2}.

As regarding quantum geodesic functions, the problem of quantum ordering
was first mentioned in \cite{ChF2} where the determining conditions of mapping-class-group (MCG) invariance and
satisfaction of the quantum skein relations were formulated. The compatibility of these two conditions was
implicitly proved by Kashaev~\cite{Kashaev-Dehn} who constructed unitary operators of quantum
Dehn twists whose action on operators of quantum geodesic functions obviously preserves their quantum
algebra. It remained however the problem of formulating a recipe for obtaining a quantum operator in
an explicit form, likewise the Kulish, Sklyanin, and Nazarov recipe (see \cite{KulSk}, \cite{Naz}) for constructing Yangian central elements extended
to the case of twisted Yangians by Molev, Ragoucy, and Sorba (the quantum ordering for
twisted Yangians was constructed in \cite{MR} for the $O(n)$ case and in \cite{MRS} for the
$Sp(2n)$ case).

We remark that a quantitative description of surfaces {with marked points on the boundary} could be deduced  from works by Fock and Goncharov \cite{FG1}, Musiker, Schiffler and Williams \cite{MSW1}, \cite{MW}, and S. Fomin, M. Shapiro, and D. Thurston \cite{FST}, \cite{FT}. In particular the authors of \cite{FST} considered systems of (tagged) arcs starting and terminating either at bordered cusps or at punctures (holes) of $\Sigma_{g,s,n}$. Due to the satisfaction of the Ptolemy relations for $\lambda$-lengths of arcs \cite{Penn1}, the correspondence to cluster algebras was immediate; in \cite{MSW1} the positivity property for $\lambda$-lengths of arcs connecting marked points (bordered cusps in our terminology) was proved in a technically rather elaborated way with the use of Ptolemy relations only. Then, in \cite{MW},  a nice quantitative description of simple arcs was attained: their $\lambda$-lengths were identified with upper-right elements (denoted $K$-traces in the present text) of products of $2\times 2$-matrices from $PSL(2,\mathbb R)$ and a part of skein relations between these elements were constructed (Lemma~6.11 of \cite{MW}).\footnote{Curiously, the missing relation in \cite{MW} was just the Ptolemy relation: following the authors of \cite{MW}, let $ur(M)$ denote the upper-right element of the matrix $M$, then, for any four $(2\times 2)$-matrices $M_i$ with unit determinants, $ur(M_1M_2)ur(M_3M_4)=ur(M_1M_4)ur(M_3M_2)
+ur(M_1M_3^{-1})ur(M_2^{-1}M_4)$.} {Our approach differs from these papers in the fact that by considering bordered cusps rather than marked points on the boundary, we have extra structure that allows a completely combinatorial approach without the need of elaborate machinery.}

This paper is organised as follows. In Sec.~\ref{s:preliminaries}, we present some known facts about quantum geodesics, quantum MCG transformations, and quantum ordering for shear coordinates and $\lambda$-lengths or Riemann surfaces with holes, orbifold points and no cusps.

The new material starts in Sec.~\ref{s:limit} with the geometrical picture in Poincar\'e geometry in which we define the new surgery derived from colliding holes, the ``chewing gum'' construction. We prove that in the limit of broken chewing gum we obtain the Ptolemy relations for arc functions of the newly obtained arcs out of skein relations satisfied by the geodesic functions before taking the limit.

In Sec.~\ref{s:graph-bordered} we define the {\it bordered cusped Teichm\"uller space} of bordered cusped Riemann surfaces and
 provide the explicit fat-graph (combinatorial) description of
arcs (and, therefore, for $\lambda$-lengths and for the corresponding $X$-cluster variables) in terms of the extended set of shear coordinates of the new Riemann surface with decorated bordered cusps. We consider cusped geodesic laminations CGL that are collections of closed geodesics $\gamma$ and geodesic arcs $\mathfrak a$
such that they have no (self)intersections inside the Riemann surface, but different arcs can be incident to the same bordered cusp, and introduce the corresponding algebraic objects. We introduce the concept of CGL$_{\mathfrak a}^{\text{max}}$ and we explicitly write the 1-1 correspondence between arc functions of arcs from a
CGL$_{\mathfrak a}^{\text{max}}$ and extended shear coordinates of the fat graph dual to this lamination.
We also describe how the Kaufmann--Penner coordinatized lamination space of windowed Riemann surfaces appears as a projective limit of our $\lambda$-length description and compare our approach with that of Fomin, M. Shapiro, and D. Thurston (see \cite{FST,FT}). We conclude
this section with the description of Poisson algebras of arc functions proving that arc functions from the same CGL have homogeneous Poisson brackets.

In Sec.~\ref{s:q}, we formulate the quantum MCG transformations for shear coordinates and the quantum mutations for arcs for Riemann surfaces with bordered cusps and find quantum commutation relations between the new shear coordinates that are invariant w.r.t. these MCG transformations. For arcs, we explicitly construct the quantum ordering that is invariant under the action of the
quantum MCG and show that this quantum ordering coincides with the natural ordering.  The quantum commutation relations between arc functions from the same CGL
become homogeneous thus defining a quantum torus. We then write quantum mutation relations induced by quantum MCG transformations exclusively in terms of quantum $\lambda$-lengths of arcs from the same CGL$_{\mathfrak a}^{\text{max}}$. We can therefore identify any CGL$_{\mathfrak a}^{\text{max}}$ with a \emph{seed} of a Berenstein--Zelevinsky quantum cluster algebra \cite{BerZel}; these seeds are related by quantum mutations, which include besides the standard binomial terms also terms corresponding to generalized cluster transformations. We thus provide a geometric setting for the quantum cluster algebras.

\section{Important facts on quantum geodesics, quantum Teichm\"uller spaces, and related $\lambda$-lengths (cluster variables)}\label{s:preliminaries}
\setcounter{equation}{0}

Following \cite{Penn1,Fock1,Kashaev,ChF1,ChF2,ChSh,ChM2}, in this section we recall the combinatorial description of the Teichm\"uller space ${\mathfrak T}_{g,s}$ of Riemann surfaces of genus $g$ with $s_h$ holes and
$s_o$ orbifold points ($s=s_h+s_o$), the coordinate description of the Poisson structure on ${\mathfrak T}_{g,s}$, the action of the mapping class group and the quantisation procedure. We adapt the standard notations and theory to treat holes and orbifold points on the same footing and to prepare the ground for our generalisation to the case of bordered cusps.

\subsection{Combinatorial description of ${\mathfrak T}_{g,s}$}\label{s:graph}

We first describe the relation between fat graphs endowed with elements of $PSL(2,{\mathbb R})$ and Fuchsian groups.

\subsubsection{Fat graph description for Riemann surfaces with holes and ${\mathbb Z}_p$ orbifold points}

\begin{definition}\label{def-pend}
We call a fat graph (a graph with the prescribed cyclic ordering of edges
entering each vertex) ${\mathcal G}_{g,s_h+s_o}$ a {\em spine of the Riemann surface} $\Sigma_{g,s_h+s_o}$
with $g$ handles, $s_h>0$ holes, and $s_o$ orbifold points of the corresponding orders $p_i$, $i=1,\dots,s_o$, if
\begin{itemize}
\item[(a)] this graph can be embedded  without self-intersections in $\Sigma_{g,s_h+s_o}$;
\item[(b)] all vertices of ${\mathcal G}_{g,s_h+s_o}$ are three-valent;
\item[(c)] upon cutting along all edges of ${\mathcal G}_{g,s_h+s_o}$ the Riemann surface
$\Sigma_{g,s_h+s_o}$ splits into $s=s_h+s_o$ polygons each containing exactly one hole or an orbifold point
and being simply connected upon contracting  this hole or removing the orbifold point. All polygons containing orbifold
points (and some, but not all polygons containing holes) are monogons, that is, every such monogon is bounded by an edge that
starts and terminates at the same three-valent vertex of the spine.
\end{itemize}
\end{definition}

\begin{remark}\label{rm:pending}
In our previous papers \cite{ChM,ChM2}, we {associated ``pending'' edges to orbifold points. In this paper instead,} we attach
additional loops to ends of these edges. This enables us treating holes and orbifold points on equal footing, so we often write
${\mathcal G}_{g,s}$ and $\Sigma_{g,s}$ with $s=s_h+s_o$ without distinguishing between holes and orbifold points.
\end{remark}

The edges in the above graph are labeled by distinct integers $\alpha=1,2,\dots,6g-6+3s$, and we set
a real number $Z_\alpha$ into correspondence to the $\alpha$th edge if it is not a loop. To each edge that
is a loop we set into correspondence the number $\omega_i$ such that
\be
\omega_i=\left\{
\begin{array}{ll}
 2\cosh(P_i/2) & \hbox{if the monogon contains a hole with the perimeter $P_i\ge 0$},    \\
 2\cos(\pi/p_i) & \hbox{if the monogon contains an orbifold point of order  $p_i\in {\mathbb Z}_+$, $p_i\ge 2$}.
\end{array}
\right.
\label{omegai}
\ee

The first homotopy groups $\pi_1(\Sigma_{g,s_h+s_o})$ and $\pi_1({\mathcal G}_{g,s_h+s_o})$ coincide because
each closed path in $\Sigma_{g,s_h+s_o}$ can be homotopically transformed to a closed path in ${\mathcal G}_{g,s_h+s_o}$
(taking into account paths that go around orbifold points)
in a unique way. The standard statement in hyperbolic geometry is that conjugacy classes of hyperbolic elements of
a Fuchsian group $\Delta_{g,s_h+s_o}\subset PSL(2,\mathbb R)$ are in the 1-1 correspondence with homotopy
classes of closed paths in the Riemann surface $\Sigma_{g,s_h+s_o}={\mathbb H}/\Delta_{g,s_h+s_o}$ so that we can refer to the
``length $\ell_\gamma$
of a hyperbolic element $\gamma\in\Delta_{g,s_h+s_o}$'' to mean the minimum length of
curves from the corresponding homotopy class; it is then the length of a unique closed
geodesic line belonging to this class.

The real numbers $Z_\alpha$ in Definition~\ref{def-pend} are the $h$-lengths (logarithms of cross-ratios) in
\cite{Penn1}: they are called the {\em (Thurston) shear
coordinates} \cite{ThSh},\cite{Bon2} in the case of punctured Riemann surface (when all $P_i=0$). We identify these
shear coordinates with coordinates of the decorated Teichm\"uller space ${\mathfrak T}_{g,s_h+s_o}$.
It was proved in \cite{ChSh} that any metrizable Riemann surface of genus $g$ with exactly $s_o$ orbifold points of the prescribed
orders $p_i$ and $s_h$ holes with the prescribed perimeters $P_i$ corresponds, up to the action of a discretely acting MCG group, to
a fat graph ${\mathcal G}_{g,s_h+s_o}$ whose edges are endowed with the real numbers $Z_\alpha$ and $\omega_i$ and, vice versa, for any
choice of the above real numbers we have a metrizable Riemann surface corresponding to such fat graph. The correspondence is
understood as the coincidence of \emph{spectra} of the above objects: the sets of lengths of closed geodesics (geodesic functions) on the
Riemann surface and on the graph.

In the case of  surfaces with punctures, the dual to the above
fat graph description is an \emph{ideal triangle decomposition} constructed in \cite{Penn1}. This description
was generalised to surfaces with holes in \cite{Fock1} and to surfaces with holes and orbifold points in \cite{Ch1a,Ch2,ChSh}. Each edge of a dual graph (a side of an ideal triangle decorated by horocycles based at its vertices) carries a $\lambda$-length, which is by definition
\be
\lambda_\alpha =e^{l_\alpha/2}
\label{lambda}
\ee
where $l_\alpha$ is the signed length of the part of the ideal triangle edge confined between two horocycles based at the ends of this edge (the sign is negative if these horocycles intersect)\footnote{We remind the reader that in this approach the choice of a horocycle at each vertex of the ideal  triangulation is fixed once for ever and that the Euclidean diameters of such horocycles constitute the ``decoration"  in \cite{Penn1}.}. The mapping class group acts on the set of lambda lengths by morphisms, that are dual to those for
$h$-lengths and correspond to mutations of cluster variables, thus implying a natural identification of lambda lengths with a subclass of cluster varieties of a geometrical origin in \cite{FZ2}.

The $h$-lengths are related to the $\lambda$-lengths through the cross-ratio relation (see the left-hand side of Fig.~\ref{fi:cross})
\be
e^{Z_e}=\frac{\lambda_b\lambda_d}{\lambda_a\lambda_c}.
\label{cross-l}
\ee
The geometrical meaning of $Z_e$ is the signed geodesic distance between perpendiculars to the common side $e$ of
two adjacent ideal triangles through the vertices of these triangles (see examples in Fig.~\ref{fi:cross}).

\subsubsection{The Fuchsian group $\Delta_{g,s_h+s_o}$ and geodesic functions}\label{ss:geodesic}

The combinatorial description of conjugacy classes of the Fuchsian group $\Delta_{g,s_h+s_o}$ is attained in terms of
(closed) paths on ${\mathcal G}_{g,s_h+s_o}$ to which we set into correspondence products of matrices from
$PSL(2,\mathbb R)$.
Every time the path homeomorphic to a (closed) geodesic $\gamma$ passes along the edge with the label $\alpha$ we
insert~\cite{Fock1} the so-called {\it edge matrix}:
\be
\label{XZ} X_{Z_\alpha}=\left(
\begin{array}{cc} 0 & -\e^{Z_\alpha/2}\\
                \e^{-Z_\alpha/2} & 0\end{array}\right)
\ee
into the corresponding string of matrices. We also have the ``right'' and ``left'' turn matrices
to be set in proper places when a path makes corresponding turns at three-valent vertices (except those incident to loops),
\be
\label{R}
R=\left(\begin{array}{cc} 1 & 1\\ -1 & 0\end{array}\right), \qquad
L= R^2=\left(\begin{array}{cc} 0 & 1\\ -1 &
-1\end{array}\right).
\ee

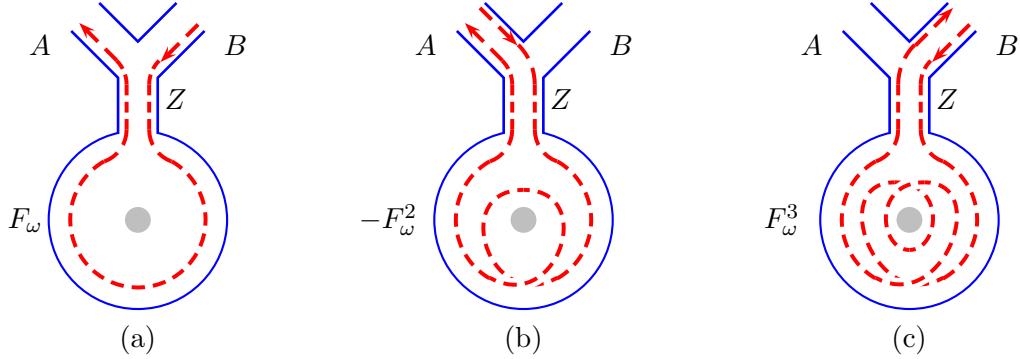
\begin{figure}[tb]
{\psset{unit=1}
\begin{pspicture}(-2.5,-3)(2.5,2)
\pcline[linecolor=blue, linewidth=16pt](0,-0.3)(0,1)
\pcline[linecolor=blue, linewidth=16pt](0.7,1.7)(0,1)
\pcline[linecolor=blue, linewidth=16pt](-0.7,1.7)(0,1)
\pscircle[linecolor=blue,fillstyle=solid,fillcolor=white](0,-1){1.2}
\pcline[linecolor=white, linewidth=14pt](0,-0.35)(0,1)
\pcline[linecolor=white, linewidth=14pt](0.8,1.8)(0,1)
\pcline[linecolor=white, linewidth=14pt](-0.8,1.8)(0,1)
\pscircle[linecolor=white,fillstyle=solid,fillcolor=lightgray](0,-1){0.2}
\pcline[linecolor=red, linestyle=dashed, linewidth=1.5pt]{<-}(0.3,1.1)(0.8,1.6)
\psarc[linecolor=red, linestyle=dashed, linewidth=1.5pt](0.6,0.8){.45}{135}{180}
\pcline[linecolor=red, linestyle=dashed, linewidth=1.5pt](0.15,0.8)(.15,0.2)
\psarc[linecolor=red, linestyle=dashed, linewidth=1.5pt](0.6,0.2){.45}{180}{250}
\psarc[linecolor=red, linestyle=dashed, linewidth=1.5pt](0,-1){0.9}{-240}{60}
\pcline[linecolor=red, linestyle=dashed, linewidth=1.5pt]{->}(-0.28,1.12)(-0.78,1.62)
\psarc[linecolor=red, linestyle=dashed, linewidth=1.5pt](-0.6,0.8){.45}{0}{45}
\pcline[linecolor=red, linestyle=dashed, linewidth=1.5pt](-0.15,0.8)(-.15,0.2)
\psarc[linecolor=red, linestyle=dashed, linewidth=1.5pt](-0.6,0.2){.45}{290}{360}
\rput(-1.3,1.2){\makebox(0,0)[cb]{$A$}}
\rput(1.3,1.2){\makebox(0,0)[cb]{$B$}}
\rput(0.5,.6){\makebox(0,0){$Z$}}
\rput(-1.5,-1){\makebox(0,0){$F_\omega$}}
\rput(0,-2.6){\makebox(0,0){(a)}}
\end{pspicture}
\begin{pspicture}(-2.5,-3)(2.5,2)
\pcline[linecolor=blue, linewidth=16pt](0,-0.3)(0,1)
\pcline[linecolor=blue, linewidth=16pt](0.7,1.7)(0,1)
\pcline[linecolor=blue, linewidth=16pt](-0.7,1.7)(0,1)
\pscircle[linecolor=blue,fillstyle=solid,fillcolor=white](0,-1){1.2}
\pcline[linecolor=white, linewidth=14pt](0,-0.35)(0,1)
\pcline[linecolor=white, linewidth=14pt](0.8,1.8)(0,1)
\pcline[linecolor=white, linewidth=14pt](-0.8,1.8)(0,1)
\pscircle[linecolor=white,fillstyle=solid,fillcolor=lightgray](0,-1){0.2}
\pcline[linecolor=red, linestyle=dashed, linewidth=1.5pt]{<-}(-0.08,1.32)(-0.58,1.82)
\psarc[linecolor=red, linestyle=dashed, linewidth=1.5pt](-0.6,0.8){.75}{0}{45}
\pcline[linecolor=red, linestyle=dashed, linewidth=1.5pt](0.15,0.8)(.15,0.2)
\psarc[linecolor=red, linestyle=dashed, linewidth=1.5pt](0.6,0.2){.45}{180}{250}
\psarc[linecolor=red, linestyle=dashed, linewidth=1.5pt](0,-1){0.9}{-240}{-120}
\psarc[linecolor=red, linestyle=dashed, linewidth=1.5pt](0,-1){0.9}{-60}{60}
\psbezier[linecolor=red, linestyle=dashed, linewidth=1.5pt](0.45,-1.77)(-0.45,-2.27)(-1,-0.6)(0,-0.6)
\psbezier[linecolor=white, linewidth=5pt](-0.45,-1.77)(0.45,-2.27)(1,-0.6)(0,-0.6)
\psbezier[linecolor=red, linestyle=dashed, linewidth=1.5pt](-0.45,-1.77)(0.45,-2.27)(1,-0.6)(0,-0.6)
\pcline[linecolor=red, linestyle=dashed, linewidth=1.5pt]{->}(-0.28,1.12)(-0.78,1.62)
\psarc[linecolor=red, linestyle=dashed, linewidth=1.5pt](-0.6,0.8){.45}{0}{45}
\pcline[linecolor=red, linestyle=dashed, linewidth=1.5pt](-0.15,0.8)(-.15,0.2)
\psarc[linecolor=red, linestyle=dashed, linewidth=1.5pt](-0.6,0.2){.45}{290}{360}
\rput(-1.3,1.2){\makebox(0,0)[cb]{$A$}}
\rput(1.3,1.2){\makebox(0,0)[cb]{$B$}}
\rput(0.5,.6){\makebox(0,0){$Z$}}
\rput(-1.8,-1){\makebox(0,0){$-F^2_\omega$}}
\rput(0,-2.6){\makebox(0,0){(b)}}
\end{pspicture}
\begin{pspicture}(-2.5,-3)(2.5,2)
\pcline[linecolor=blue, linewidth=16pt](0,-0.3)(0,1)
\pcline[linecolor=blue, linewidth=16pt](0.7,1.7)(0,1)
\pcline[linecolor=blue, linewidth=16pt](-0.7,1.7)(0,1)
\pscircle[linecolor=blue,fillstyle=solid,fillcolor=white](0,-1){1.2}
\pcline[linecolor=white, linewidth=14pt](0,-0.35)(0,1)
\pcline[linecolor=white, linewidth=14pt](0.8,1.8)(0,1)
\pcline[linecolor=white, linewidth=14pt](-0.8,1.8)(0,1)
\pscircle[linecolor=white,fillstyle=solid,fillcolor=lightgray](0,-1){0.2}
\pcline[linecolor=red, linestyle=dashed, linewidth=1.5pt]{<-}(0.3,1.1)(0.8,1.6)
\psarc[linecolor=red, linestyle=dashed, linewidth=1.5pt](0.6,0.8){.45}{135}{180}
\pcline[linecolor=red, linestyle=dashed, linewidth=1.5pt](0.15,0.8)(.15,0.2)
\psarc[linecolor=red, linestyle=dashed, linewidth=1.5pt](0.6,0.2){.45}{180}{250}
\psarc[linecolor=red, linestyle=dashed, linewidth=1.5pt](0,-1){0.9}{-240}{-120}
\psarc[linecolor=red, linestyle=dashed, linewidth=1.5pt](0,-1){0.9}{-60}{60}
\psbezier[linecolor=red, linestyle=dashed, linewidth=1.5pt](0.45,-1.77)(-0.45,-2.27)(-1.1,-0.5)(-0.2,-0.5)
\psbezier[linecolor=red, linestyle=dashed, linewidth=1.5pt](0,-1.4)(0.4,-1.4)(0.5,-0.5)(-0.2,-0.5)
\psbezier[linecolor=white, linewidth=5pt](-0.45,-1.77)(0.45,-2.27)(1.1,-0.5)(0.2,-0.5)
\psbezier[linecolor=white, linewidth=5pt](0,-1.4)(-0.4,-1.4)(-0.5,-0.5)(0.2,-0.5)
\psbezier[linecolor=red, linestyle=dashed, linewidth=1.5pt](0,-1.4)(-0.4,-1.4)(-0.5,-0.5)(0.2,-0.5)
\psbezier[linecolor=red, linestyle=dashed, linewidth=1.5pt](-0.45,-1.77)(0.45,-2.27)(1.1,-0.5)(0.2,-0.5)
\pcline[linecolor=red, linestyle=dashed, linewidth=1.5pt]{->}(0.08,1.32)(0.58,1.82)
\psarc[linecolor=red, linestyle=dashed, linewidth=1.5pt](0.6,0.8){.75}{135}{180}
\pcline[linecolor=red, linestyle=dashed, linewidth=1.5pt](-0.15,0.8)(-.15,0.2)
\psarc[linecolor=red, linestyle=dashed, linewidth=1.5pt](-0.6,0.2){.45}{290}{360}
\rput(-1.3,1.2){\makebox(0,0)[cb]{$A$}}
\rput(1.3,1.2){\makebox(0,0)[cb]{$B$}}
\rput(0.5,.6){\makebox(0,0){$Z$}}
\rput(-1.7,-1){\makebox(0,0){$F^3_\omega$}}
\rput(0,-2.6){\makebox(0,0){(c)}}
\end{pspicture}
}
\caption{\small Part of a graph with a loop.
The variable $Z$ corresponds to a unique edge incident to the loop.
We present
three typical examples of geodesics undergoing single (a), double (b),
and triple (c) clockwise rotations.}
\label{fi:corner}
\end{figure}

When orbifold points are present, the Fuchsian group contains besides hyperbolic elements also elliptic
elements corresponding to rotations about these orbifold points.
The corresponding generators ${\wtd F}_i$, $i=1,\dots,s_o$, of the rotations through $2\pi/p_i$
are conjugates of the matrices $F_{\omega_i}$,
\be
\label{F-p}
{\wtd F}_i=U_iF_{\omega_i}U_i^{-1},\qquad F_{\omega_i}:=
\left(\begin{array}{cc} 0 & 1\\ -1 & -w_i\end{array}\right),\quad \omega_i=2\cos(\pi/p_i).
\ee
Following \cite{ChSh}, we introduce special matrices corresponding to going along a loop labeled by $\omega_i$ (\ref{omegai})
(without differing between orbifold points and holes contained inside the loop):
every time a path goes clockwise around the loop (see Fig.~\ref{fi:corner}(a)),
we insert the matrix (\ref{F-p}) in the corresponding string of matrices (\emph{without}
adding the matrices of left/right turns, i.e., the corresponding string has the form
$\cdots X_{Z_\alpha} {F}_{\omega_i}  X_{Z_\alpha}\cdots $,
where $\alpha$ is the label of the unique edge attached to the loop). When going along a loop
$k$ times clockwise we insert the matrix $(-1)^{k+1}F_\omega^k$ into the product of $2\times2$-matrices.
For example, parts of geodesic functions
in the three cases in Fig.~\ref{fi:corner} where we denote the shear coordinates by $A,B,Z$, read:
\be
\label{XZFXZ}
\begin{array}{ll}
\hbox{(a)}\quad & \dots  X_ALX_ZF_\omega X_ZLX_B\dots, \\
\hbox{(b)}\quad & \dots  X_ALX_Z(-F^2_\omega)X_ZRX_A\dots, \\
\hbox{(c)}\quad & \dots  X_BRX_Z(F^3_\omega)X_ZLX_B\dots. \\
\end{array}
\ee
Note that  $F_{\omega_p}^p=(-1)^{p-1}{\mathbb E}$ when $\omega_p=2\cos{\pi/p}$,
so going around the ${\mathbb Z}_p$ orbifold point $p$ times merely corresponds to avoiding this loop.
(For the ${\mathbb Z}_2$ orbifold points
this pattern was first proposed by Fock and Goncharov \cite{FG}; the graph morphisms were described in \cite{Ch2}.)

If a loop circumnavigates a hole, not an orbifold point, then, when going around it counterclockwise, we must insert the matrix
$-F_\omega^{-1}=\left(\begin{array}{cc} w & 1\\ -1 & 0\end{array}\right)$, etc.

As explained in Remark \ref{rm:pending}, this convention saves us from distinguishing between holes and orbifold points in all our computations. To revert to the usual setting in which the portion of the geodesic going clockwise around the loop like  Fig.~\ref{fi:corner}(a) is described by $X_Z L X_P L X_Z$ rather than by $X_Z  F_\omega  X_Z$, we just need to shift the shear coordinate $Z$ by ${P}/{2}$. In other words,
$$
X_{Z} L X_P L X_{Z}= X_{Z+P/2} F_\omega X_{Z+P/2}.
$$

Resuming, an element $P_\gamma$ in the Fuchsian group has then the typical structure:
\be
\label{Pgamma}
P_{\gamma}=LX_{Z_n}RX_{Z_{n-1}}\cdots RX_{Z_{j+1}}LX_{Z_j}(-1)^{k+1}F^k_{\omega_i} X_{Z_{j-1}}R\dots X_{Z_1}.
\ee
In the corresponding {\em geodesic function}
\be
\label{G}
G_{\gamma}\equiv \tr P_\gamma=2\cosh(\ell_\gamma/2),
\ee
$\ell_\gamma$ is the actual length of the closed
geodesic on the Riemann surface.

\begin{remark}\label{rm-positivity}
Note that the combinations
\bea
&{}&RX_Z=\left(\begin{array}{cc} e^{-Z/2} & -e^{Z/2}\\ 0 & e^{Z/2}\end{array}\right),
\quad
LX_Z=\left(\begin{array}{cc} e^{-Z/2} & 0 \\ -e^{-Z/2} & e^{Z/2}\end{array}\right),\nonumber\\
&{}&RX_ZF_\omega X_Z=\left(\begin{array}{cc} e^{-Z}+\omega & -e^{Z}\\ -\omega & e^{Z}\end{array}\right),\quad
LX_ZF_\omega X_Z=\left(\begin{array}{cc} e^{-Z} & 0\\ -e^{-Z}-\omega & e^{Z}\end{array}\right),\nonumber\\
&{}&
RX_Z(-F_\omega^{-1})X_Z=\left(\begin{array}{cc} e^{-Z} & -e^{Z}-\omega\\ 0 & e^{Z}\end{array}\right),\quad
LX_Z(-F_\omega^{-1})X_Z=\left(\begin{array}{cc} e^{-Z} & -\omega\\ -e^{-Z} & e^{Z}+\omega\end{array}\right),
\nonumber
\eea
as well as products of any number of these matrices have the sign structure
$\left(\begin{array}{cc} + & -\\ - & +\end{array}\right)$, so the trace of any element $P_\gamma$ with first powers of
$F_\omega$ and/or $-F_\omega^{-1}$ is a
sum of exponentials with positive integer coefficients. This observation will be important
when proving positivity of cluster transformations in Sec.~\ref{s:graph-bordered}.
\end{remark}

The group generated by the elliptic elements (\ref{F-p}) together with the hyperbolic elements corresponding to
translations along $A$- and $B$-cycles of the Riemann surface
and around holes is not necessarily Fuchsian because its action
may not be discrete. The necessary and
sufficient conditions under which we obtain a {\em regular} (that is, locally smooth everywhere expect exactly $s_o$ orbifold points)
Riemann surface were formulated in terms of graphs in \cite{ChSh} where it was proven
that we obtain a regular Riemann surface for any set of real
numbers $Z_\alpha$ from Definition~\ref{def-pend} and vice versa. For a given Riemann surface, this
set is not unique and equivalent sets are related by discrete modular group action, so we
identify the $(6g-6+3s_h+2s_o)$-tuple of real coordinates $\{Z_\alpha\}$
with the coordinates of the decorated
Teichm\"uller space ${\mathfrak T}_{g,s_h+s_o}$ (the decoration assigns positive or negative signs to
every hole with nonzero perimeter). The
lengths of geodesics on $\Sigma_{g,s_h+s_o}$ are given by traces of products (\ref{Pgamma})
corresponding to paths in the corresponding spine.

Transitions between different parameterizations are formulated in terms of flip morphisms
(mutations) of edges: any two spines from the given topological class are related by a finite sequence of flips.
We therefore identify flips on edges with the MCG action.


\subsection{Poisson structure}\label{ss:Poisson}

One of the most attractive properties of the graph description is a very simple Poisson algebra on the set
of coordinates $Z_\alpha$, $\alpha=1,\dots, 6g-6+3s_h+2s_o$.

\begin{theorem}\label{th-WP} In the coordinates $Z_\alpha $ on any fixed spine
corresponding to a surface with or without orbifold points,
the Weil--Petersson bracket $B_{{\mbox{\tiny WP}}}$ reads
\be
\label{WP-PB}
\bigl\{f({\mathbf Z}),g({\mathbf Z})\bigr\}=\sum_{{\hbox{\small 3-valent} \atop \hbox{\small vertices $\alpha=1$} }}^{4g+2s+n-4}
\,\sum_{i=1}^{3 \mod 3}
\left(\frac{\partial f}{\partial Z_{\alpha_i}} \frac{\partial g}{\partial Z_{\alpha_{i+1}}}
- \frac{\partial g}{\partial Z_{\alpha_i}} \frac{\partial f}{\partial Z_{\alpha_{i+1}}}\right),
\ee
where the sum ranges all three-valent vertices of a graph and
$\alpha_i$ are the labels of the cyclically (clockwise)
ordered ($\alpha_4\equiv \alpha_1 $) edges incident to the vertex
with the label $\alpha$. This bracket gives rise to the {\em Goldman
bracket} on the space of geodesic length functions \cite{Gold}.
\end{theorem}

Note that formula (\ref{WP-PB}) is insensitive to whether we include or remove vertices incident to loops into this sum because the term
in brackets is identically zero for such a vertex. The quantities $\omega_i$ are therefore central and we interpret them as parameters.

The center of the Poisson algebra {\rm(\ref{WP-PB})} is generated by
elements of the form $\sum Z_\alpha$, where the sum ranges all edges
of ${\mathcal G}_{g,s_h+s_o} $ (taken with multiplicities) belonging to the same boundary component
(which can also be a monogon containing a hole). The dimension of this center is obviously $s_h$.

\subsection{Flip morphisms of fat graphs}\label{ss:flip}

There are two sorts of flip morphisms (see Theorem \ref{thm:flip-m}): morphisms induced by flips of inner edges (see Fig. \ref{fi:flip}) and morphisms induced by flips of edges that are adjacent to a loop (see Fig. \ref{fi:interchange-p-dual}); we describe these two cases in the following two sub-sections. For convenience here below we drop the indices $g,s_o,s_h$ as these numbers are preserved by the flip morphisms.

\subsubsection{Flipping inner edges}\label{sss:mcg}

Given a spine ${\mathcal G}$ of $\Sigma$, if an internal
edge $\alpha$ is neither a loop nor is adjacent to a loop, we may produce
another spine ${\mathcal G} _\alpha$ of $\Sigma$ by contracting and expanding edge $\alpha$ of
${\mathcal G} $, the edge labeled $Z$ in Figure~\ref{fi:flip}.
We say that ${\mathcal G} _\alpha$ arises from ${\mathcal G}$ by a
{\it Whitehead move} (or flip) along the edge $\alpha$.
A labeling of edges of the spine ${\mathcal G}$ implies a natural labeling of edges of the
spine ${\mathcal G}_\alpha$; we then obtain a morphism between the spines ${\mathcal G}$ and ${\mathcal G}_\alpha$.

\begin{figure}[tb]
\begin{pspicture}(-3,-3)(4,3){
\newcommand{\FLIP}{%
{\psset{unit=1}
\psline[linewidth=18pt,linecolor=blue](0,-1)(0,1)
\psline[linewidth=18pt,linecolor=blue](0,1)(1.5,2)
\psline[linewidth=18pt,linecolor=blue](0,1)(-1.5,2)
\psline[linewidth=18pt,linecolor=blue](0,-1)(1.5,-2)
\psline[linewidth=18pt,linecolor=blue](0,-1)(-1.5,-2)
\psline[linewidth=14pt,linecolor=white](0,-1)(0,1)
\psline[linewidth=14pt,linecolor=white](0,1)(1.5,2)
\psline[linewidth=14pt,linecolor=white](0,1)(-1.5,2)
\psline[linewidth=14pt,linecolor=white](0,-1)(1.5,-2)
\psline[linewidth=14pt,linecolor=white](0,-1)(-1.5,-2)
}
}
\rput(-2.5,0){\FLIP}
\psline[linewidth=2pt]{<->}(-0.5,0)(0.5,0)
\rput{90}(3.5,0){\FLIP}
\rput(-2.5,0){
\rput(-1.1,2.2){\makebox(0,0)[lb]{$A$}}
\rput(1.1,2.2){\makebox(0,0)[rb]{$B$}}
\rput(-0.8,0){\makebox(0,0)[lc]{$Z$}}
\rput(1.1,-2.2){\makebox(0,0)[rt]{$C$}}
\rput(-1.1,-2.2){\makebox(0,0)[lt]{$D$}}
\psline[linecolor=red](-1.5,-2)(-.15,-1.1)
\psline[linecolor=red](1.5,2)(.15,1.1)
\psbezier[linecolor=red](-.15,-1.1)(0.15,-0.9)(-0.15,0.9)(.15,1.1)
\rput(-.1,.15){\psline[linecolor=red](-1.5,-2)(-.15,-1.1)}
\rput(-.1,-.15){\psline[linecolor=red](-1.5,2)(-.15,1.1)}
\psbezier[linecolor=red](-.25,-.95)(-0.1,-0.85)(-0.1,0.85)(-.25,.95)
\rput(.08,-.12){\psline[linecolor=red](-1.5,-2)(-.15,-1.1)}
\rput(-.08,-.12){\psline[linecolor=red](1.5,-2)(.15,-1.1)}
\psbezier[linecolor=red](-.07,-1.22)(0,-1.176)(0,-1.176)(.07,-1.22)
}
\rput(3.5,0){
\rput(-2.2,-2){\makebox(0,0)[lt]{$D - \phi(-Z)$}}
\rput(2.2,-2){\makebox(0,0)[rt]{$C+\phi(Z)$}}
\rput(2.2,2){\makebox(0,0)[rb]{$B-\phi(-Z)$}}
\rput(-2.2,2){\makebox(0,0)[lb]{$A+\phi(Z)$}}
\rput(0,0.5){\makebox(0,0)[cb]{$-Z$}}
}
\rput{90}(3.5,0){
\psline[linecolor=red](1.5,-2)(.15,-1.1)
\psline[linecolor=red](-1.5,2)(-.15,1.1)
\psbezier[linecolor=red](.15,-1.1)(-0.15,-0.9)(0.15,0.9)(-.15,1.1)
\rput(-.1,.15){\psline[linecolor=red](-1.5,-2)(-.15,-1.1)}
\rput(-.1,-.15){\psline[linecolor=red](-1.5,2)(-.15,1.1)}
\psbezier[linecolor=red](-.25,-.95)(-0.1,-0.85)(-0.1,0.85)(-.25,.95)
}
\rput{270}(3.5,0){
\rput(.08,-.12){\psline[linecolor=red](-1.5,-2)(-.15,-1.1)}
\rput(-.08,-.12){\psline[linecolor=red](1.5,-2)(.15,-1.1)}
\psbezier[linecolor=red](-.07,-1.22)(0,-1.176)(0,-1.176)(.07,-1.22)
}
\rput(-2.5,0){
\put(-1.8,1.9){\makebox(0,0)[cc]{\hbox{\tcr{\small$1$}}}}
\put(-1.8,1.9){\pscircle[linecolor=red]{.2}}
\put(1.7,2.1){\makebox(0,0)[cc]{\hbox{\tcr{\small$2$}}}}
\put(1.7,2.1){\pscircle[linecolor=red]{.2}}
\put(1.6,-2.3){\makebox(0,0)[cc]{\hbox{\tcr{\small$3$}}}}
\put(1.6,-2.3){\pscircle[linecolor=red]{.2}}
}
\rput(3.5,0){
\put(-2.2,1.7){\makebox(0,0)[cc]{\hbox{\tcr{\small$1$}}}}
\put(-2.2,1.7){\pscircle[linecolor=red]{.2}}
\put(2.2,1.7){\makebox(0,0)[cc]{\hbox{\tcr{\small$2$}}}}
\put(2.2,1.7){\pscircle[linecolor=red]{.2}}
\put(2.1,-1.6){\makebox(0,0)[cc]{\hbox{\tcr{\small$3$}}}}
\put(2.1,-1.6){\pscircle[linecolor=red]{.2}}
}
}
\end{pspicture}
\caption{\small Flip on the shear coordinate $Z$. The edge undergoing the flip is assumed to be an internal
edge that is neither a loop nor adjacent to a loop. We indicate the correspondences between geodesic paths
undergoing the flip.}
\label{fi:flip}
\end{figure}
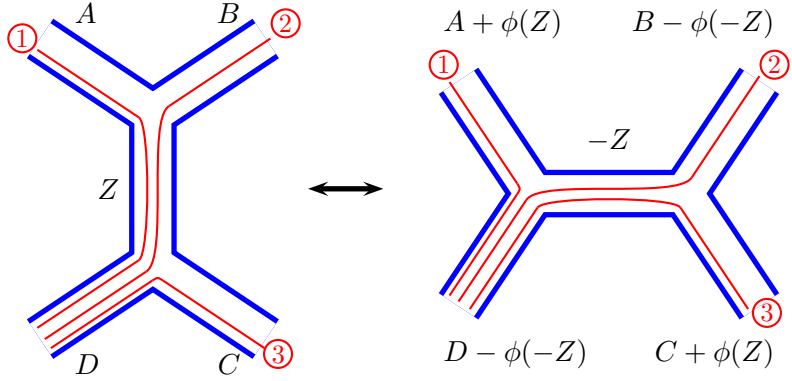

\begin{figure}[tb]
\begin{pspicture}(-3,-3)(4,3){
\newcommand{\DISC}{%
{\psset{unit=1}
\rput{30}(0,0){\psarc[linewidth=1.5pt,linecolor=green,linestyle=dashed](6,0){5.2}{150}{210}}
\rput{210}(0,0){\psarc[linewidth=1.5pt,linecolor=green,linestyle=dashed](6,0){5.2}{150}{210}}
\rput{120}(0,0){\psarc[linewidth=1.5pt,linecolor=green,linestyle=dashed](3.46,0){1.73}{120}{240}}
\rput{300}(0,0){\psarc[linewidth=1.5pt,linecolor=green,linestyle=dashed](3.46,0){1.73}{120}{240}}
\rput{90}(0,0){\pscircle[linecolor=white,fillstyle=solid,fillcolor=white](2.25,0){0.75}}
\rput{-30}(0,0){\pscircle[linecolor=white,fillstyle=solid,fillcolor=white](2.5,0){0.5}}
\rput{-90}(0,0){\pscircle[linecolor=white,fillstyle=solid,fillcolor=white](2.5,0){0.5}}
\rput{150}(0,0){\pscircle[linecolor=white,fillstyle=solid,fillcolor=white](2.6,0){0.4}}
\rput{90}(0,0){
\psclip{\pscircle[linecolor=white](2.25,0){0.75}}
\rput{-60}(0,0){\psarc[linewidth=.5pt,linecolor=green](6,0){5.2}{150}{180}}
\rput{30}(0,0){\psarc[linewidth=.5pt,linecolor=green](3.46,0){1.73}{180}{240}}
\endpsclip
}
\rput{-30}(0,0){
\psclip{\pscircle[linecolor=white](2.5,0){0.5}}
\rput{60}(0,0){\psarc[linewidth=.5pt,linecolor=green](6,0){5.2}{180}{210}}
\rput{-30}(0,0){\psarc[linewidth=.5pt,linecolor=green](3.46,0){1.73}{120}{180}}
\endpsclip
}
\rput{-90}(0,0){
\psclip{\pscircle[linecolor=white](2.5,0){0.5}}
\rput{-60}(0,0){\psarc[linewidth=.5pt,linecolor=green](6,0){5.2}{150}{180}}
\rput{30}(0,0){\psarc[linewidth=.5pt,linecolor=green](3.46,0){1.73}{180}{240}}
\endpsclip
}
\rput{150}(0,0){
\psclip{\pscircle[linecolor=white](2.6,0){0.4}}
\rput{60}(0,0){\psarc[linewidth=.5pt,linecolor=green](6,0){5.2}{180}{210}}
\rput{-30}(0,0){\psarc[linewidth=.5pt,linecolor=green](3.46,0){1.73}{120}{180}}
\endpsclip
}
\rput{90}(0,0){\pscircle[linecolor=black,linewidth=1.5pt,linestyle=dashed](2.25,0){0.75}}
\rput{-30}(0,0){\pscircle[linecolor=black,linewidth=1.5pt,linestyle=dashed](2.5,0){0.5}}
\rput{-90}(0,0){\pscircle[linecolor=black,linewidth=1.5pt,linestyle=dashed](2.5,0){0.5}}
\rput{150}(0,0){\pscircle[linecolor=black,linewidth=1.5pt,linestyle=dashed](2.6,0){0.4}}
\rput(0,0){\pscircle[linecolor=black,linestyle=dashed]{3}}
}
}
\rput(-4,0){\DISC
\rput{-30}(0,0){\psline[linewidth=1.5pt,linecolor=blue,linestyle=dashed](-2.2,0)(2,0)}
\rput{-30}(0,0){\psline[linewidth=.5pt,linecolor=blue](-3,0)(-2.2,0)}
\rput{-30}(0,0){\psline[linewidth=.5pt,linecolor=blue](2,0)(3,0)}
}
\psline[linewidth=2pt]{->}(-0.5,0)(0.5,0)
\rput(4,0){\DISC
\rput{90}(0,0){\psline[linewidth=1.5pt,linecolor=red,linestyle=dashed](-2,0)(1.5,0)}
\rput{90}(0,0){\psline[linewidth=.5pt,linecolor=red](-3,0)(-2,0)}
\rput{90}(0,0){\psline[linewidth=.5pt,linecolor=red](1.5,0)(3,0)}
}
\rput(-4,0){
\rput{150}(0,0){\psarc[linewidth=.5pt](6,0){5.2}{180}{210}}
\rput{-30}(0,0){\psarc[linewidth=.5pt](6,0){5.2}{180}{210}}
\rput{-30}(0,0){\psline[linewidth=3pt,linecolor=blue](-0.8,0.1)(.8,0.1)}
\rput(-1.1,1.5){\makebox(0,0)[rb]{$\lambda_a$}}
\rput(1,0.8){\makebox(0,0)[rb]{$\lambda_b$}}
\rput(0.2,0.3){\makebox(0,0)[lc]{$Z$}}
\rput(0,-0.2){\makebox(0,0)[rc]{$\lambda_e$}}
\rput{-30}(0,0){\rput(0.78,-0.15){\makebox(0,0)[lc]{$\square$}}}
\rput{-30}(0,0){\rput(-0.78,0.15){\makebox(0,0)[rc]{$\square$}}}
\rput(1.1,-1.5){\makebox(0,0)[lt]{$\lambda_c$}}
\rput(-1,-0.8){\makebox(0,0)[lt]{$\lambda_d$}}
}
\rput(4,0){
\rput{90}(0,0){\psarc[linewidth=.5pt](6,0){5.2}{150}{180}}
\rput{-90}(0,0){\psarc[linewidth=.5pt](6,0){5.2}{150}{180}}
\rput{90}(0,0){\psline[linewidth=3pt,linecolor=red](-0.8,0.1)(.8,0.1)}
\rput(-1.1,1.5){\makebox(0,0)[rb]{$\lambda_a$}}
\rput(1,0.8){\makebox(0,0)[rb]{$\lambda_b$}}
\rput(1.1,-1.5){\makebox(0,0)[lt]{$\lambda_c$}}
\rput{90}(0,0){\rput(0.78,0.15){\makebox(0,0)[lc]{$\square$}}}
\rput{90}(0,0){\rput(-0.78,-0.15){\makebox(0,0)[rc]{$\square$}}}
\rput(-1,-0.8){\makebox(0,0)[lt]{$\lambda_d$}}
\rput(-0.3,0.2){\makebox(0,0)[rc]{$-Z$}}
\rput(0.2,-0.2){\makebox(0,0)[lc]{$\lambda_f$}}
}
}
\end{pspicture}
\caption{\small The transformation dual to the flip in Fig.~\ref{fi:flip}: the flip,
or mutation transformation, for the $\lambda$-lengths subject to the Ptolemy relation
$\lambda_e\lambda_f=\lambda_a\lambda_c+\lambda_b\lambda_d$. Here the shear coordinates
$Z$ and $-Z$ of Fig.~\ref{fi:flip} (indicated by bold lines) are logarithms of
the corresponding cross-ratios of $\lambda$-lengths (indicated by dashed lines)
and we use the Poincar\'e disc model to represent the hyperbolic plane.}
\label{fi:cross}
\end{figure}
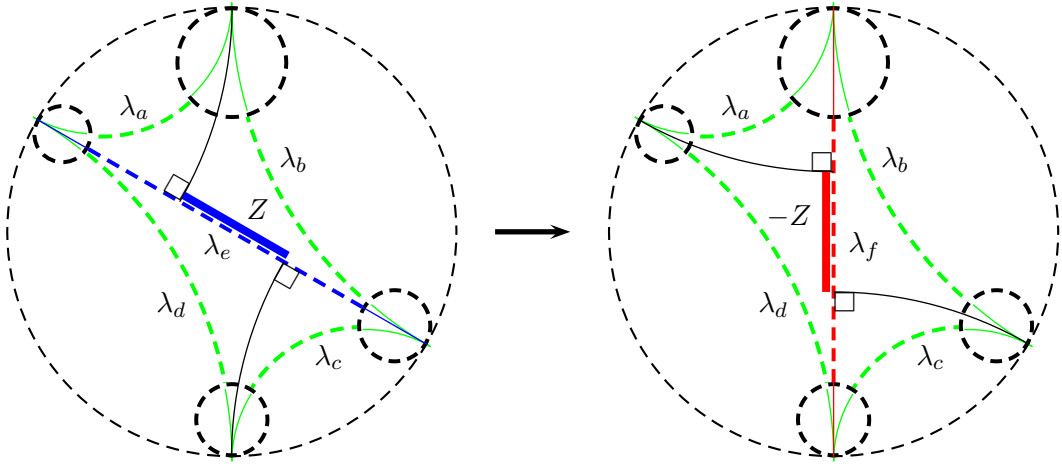

It was shown in \cite{ChF1} that setting $\phi (Z)={\rm log}(1+e^Z)$ and adopting the notation of Fig.~\ref{fi:flip}
for shear coordinates of nearby edges, the effect of a flip is
\bea
W_Z\,:\ (A,B,C,D,Z)&\to& (A+\phi(Z), B-\phi(-Z), C+\phi(Z), D-\phi(-Z), -Z)\nonumber\\
&:=&({\tilde A},{\tilde B},{\tilde C},{\tilde D},{\tilde Z}).
\label{abc}
\eea

The same flip morphism for the dual $\lambda$-lengths is depicted in Fig.~\ref{fi:cross}. The corresponding \emph{mutation}
is originated from the Ptolemy relation $\lambda_e\lambda_f=\lambda_a\lambda_c+\lambda_b\lambda_d$ \cite{Penn1}
valid for every decorated ideal quadrangle.

The following lemma establishes the properties of
invariance of geodesic functions w.r.t. the flip morphisms~\cite{ChF2}.

\begin{lemma} \label{lem-abc}
The transformation~{\rm(\ref{abc})} preserves
the traces of products over paths {\rm(\ref{G})} (the geodesic functions) and
transformation~{\rm(\ref{abc})} simultaneously preserves
Poisson structure {\rm(\ref{WP-PB})} on the shear coordinates.
\end{lemma}

\proof The proof of this lemma is an elementary consequence of the following matrix equalities
that can be established by simple calculations:
\bea
X_DRX_ZRX_{A}&=&X_{\tilde A}RX_{\tilde D},\label{mutation1}\\
X_DRX_ZLX_B&=&X_{\tilde D}LX_{\tilde Z}RX_{\tilde B},\label{mutation2}\\
X_CLX_D&=&X_{\tilde C}LX_{\tilde Z}LX_{\tilde D}.\label{mutation3}
\eea
Note that each of the above equalities corresponds to three geodesic cases in Fig.~\ref{fi:flip}). \endproof

\subsubsection{Flipping the edge incident to a loop}\label{sss:pending}

\begin{figure}[tb]
\begin{pspicture}(-3,-3)(4,3){
\newcommand{\FLIP}{%
{\psset{unit=1}
\psline[linewidth=18pt,linecolor=blue](-2,0)(0,0)
\psline[linewidth=18pt,linecolor=blue](0,0)(1,1.5)
\psline[linewidth=18pt,linecolor=blue](0,0)(1,-1.5)
\pscircle[linewidth=2pt,linecolor=blue,fillstyle=solid,fillcolor=white](-2.8,0){1}
\psline[linewidth=14pt,linecolor=white](-2,0)(0,0)
\psline[linewidth=14pt,linecolor=white](0,0)(1,1.5)
\psline[linewidth=14pt,linecolor=white](0,0)(1,-1.5)
\rput(-2.8,0){\pscircle[linecolor=white,fillstyle=solid,fillcolor=lightgray](0,0){0.3}}
\psarc[linecolor=red](-2,0.3){0.1}{200}{270}
\psarc[linecolor=red](-2,0.3){0.2}{200}{270}
\psarc[linecolor=red](-2,-0.3){0.1}{90}{160}
\psarc[linecolor=red](-2,-0.3){0.2}{90}{160}
\psline[linecolor=red](-2,0.1)(-0.01,0.1)
\psline[linecolor=red](-2,0.2)(-0,0.2)
\psline[linecolor=red](-2,-0.1)(-0.01,-0.1)
\psline[linecolor=red](-2,-0.2)(-0,-0.2)
\psarc[linecolor=red](-2.8,0){0.75}{20}{340}
\psarc[linecolor=red](-2.8,0){0.65}{20}{340}
}
}
\rput(-2.5,0){\FLIP}
\psline[linewidth=2pt]{<->}(-0.5,0)(0.5,0)
\rput{180}(2.5,0){\FLIP}
\rput(-2.5,0){
\rput(0.2,1.5){\makebox(0,0)[lb]{$A$}}
\rput(-1,0.5){\makebox(0,0)[cb]{$Z$}}
\rput(0.2,-1.5){\makebox(0,0)[lt]{$B$}}
\rput(-1.2,1.5){\makebox(0,0)[rb]{$\displaystyle\left\{ {w=2\cos(\pi/p) \atop w=2\cosh(P/2) }\right.$}}
\rput(-4,0){\makebox(0,0)[rc]{$\omega$}}
}
\rput(3.5,0){
\rput(-1.4,-1.5){\makebox(0,0)[lt]{$\displaystyle\left\{ {B - \phi(-Z+i\pi/p)-\phi(-Z-i\pi/p) \atop B - \phi(-Z+P/2)-\phi(-Z-P/2)}\right.$}}
\rput(-1.4,1.5){\makebox(0,0)[lb]{$\displaystyle\left\{ {A + \phi(Z+i\pi/p)+\phi(Z-i\pi/p) \atop A + \phi(Z+P/2)+\phi(Z-P/2)}\right.$}}
\rput(0,0.5){\makebox(0,0)[cb]{$-Z$}}
\rput(3,0){\makebox(0,0)[lc]{$\omega$}}
}
\rput{90}(3.5,0){
\rput(-.1,-.15){\psline[linecolor=red](-1.5,2)(-.09,1.06)}
\rput(.1,-.15){\psline[linecolor=red](1.5,2)(.09,1.06)}
}
\rput{270}(-3.5,0){
\rput(-.1,-.15){\psline[linecolor=red](-1.5,2)(-.09,1.06)}
\rput(.1,-.15){\psline[linecolor=red](1.5,2)(.09,1.06)}
}
\rput{270}(3.5,0){
\rput(.08,-.12){\psline[linecolor=red](-1.5,-2)(-.15,-1.1)}
\rput(-.08,-.12){\psline[linecolor=red](1.5,-2)(.15,-1.1)}
\psbezier[linecolor=red](-.07,-1.22)(0,-1.176)(0,-1.176)(.07,-1.22)
}
\rput{90}(-3.5,0){
\rput(.08,-.12){\psline[linecolor=red](-1.5,-2)(-.15,-1.1)}
\rput(-.08,-.12){\psline[linecolor=red](1.5,-2)(.15,-1.1)}
\psbezier[linecolor=red](-.07,-1.22)(0,-1.176)(0,-1.176)(.07,-1.22)
}
\rput(-2.5,0){
\rput(-.06,.04){\psline[linecolor=red](1,1.5)(0.1,.15)}
\rput(.06,-.04){\psline[linecolor=red](1,1.5)(0.1,.15)}
\psbezier[linecolor=red](.04,.19)(-0.02,0.1)(-0.02,0.1)(-0.1,0.1)
\psbezier[linecolor=red](.16,.11)(0.02,-0.1)(0.02,-.1)(-0.1,-.1)
}
\rput(2.5,0){
\rput(.06,.04){\psline[linecolor=red](-1,1.5)(-0.1,.15)}
\rput(-.06,-.04){\psline[linecolor=red](-1,1.5)(-0.1,.15)}
\psbezier[linecolor=red](-.04,.19)(0.02,0.1)(0.02,0.1)(0.1,0.1)
\psbezier[linecolor=red](-.16,.11)(-0.02,-0.1)(-0.02,-.1)(0.1,-.1)
}
\rput(-3.5,0){
\put(2.3,-1.6){\makebox(0,0)[cc]{\hbox{\tcr{\small$3$}}}}
\put(2.3,-1.6){\pscircle[linecolor=red]{.2}}
\put(2.2,1.7){\makebox(0,0)[cc]{\hbox{\tcr{\small$2$}}}}
\put(2.2,1.7){\pscircle[linecolor=red]{.2}}
\put(1.95,-1.8){\makebox(0,0)[cc]{\hbox{\tcr{\small$1$}}}}
\put(1.95,-1.8){\pscircle[linecolor=red]{.2}}
}
\rput(3.5,0){
\put(-2.3,-1.6){\makebox(0,0)[cc]{\hbox{\tcr{\small$1$}}}}
\put(-2.3,-1.6){\pscircle[linecolor=red]{.2}}
\put(-2.2,1.7){\makebox(0,0)[cc]{\hbox{\tcr{\small$2$}}}}
\put(-2.2,1.7){\pscircle[linecolor=red]{.2}}
\put(-1.95,-1.8){\makebox(0,0)[cc]{\hbox{\tcr{\small$3$}}}}
\put(-1.95,-1.8){\pscircle[linecolor=red]{.2}}
}
}
\end{pspicture}
\caption{\small
The transformation of {shear coordinates} when flipping an edge incident to a loop; either
$w=2\cos(\pi/p)$ or $w=2\cosh(P/2)$. We indicate how geodesic lines change upon flipping the edge.
}
\label{fi:interchange-p-dual}
\end{figure}

\begin{figure}[tb]
\begin{pspicture}(-3,-2)(2,3){
\newcommand{\DISCONE}{%
{\psset{unit=1}
\psbezier[linewidth=1.5pt,linecolor=green,linestyle=dashed](-3,0)(-2,0)(-2,2)(0,2)
\psbezier[linewidth=1.5pt,linecolor=green,linestyle=dashed](-3,0)(-2,0)(-2,-2)(0,-2)
\psbezier[linewidth=1.5pt,linecolor=green,linestyle=dashed](3,0)(2,0)(2,2)(0,2)
\psbezier[linewidth=1.5pt,linecolor=green,linestyle=dashed](3,0)(2,0)(2,-2)(0,-2)
\psbezier[linewidth=1.5pt,linecolor=blue,linestyle=dashed](-3,0)(-1.5,0)(-1.5,0.5)(0,0.5)
\psbezier[linewidth=1.5pt,linecolor=blue,linestyle=dashed](-3,0)(-1.5,0)(-1.5,-1.5)(0,-1.5)
\rput(0,0){\psarc[linewidth=1.5pt,linecolor=blue,linestyle=dashed](0,-0.5){1}{-90}{90}}
\rput(-0.02,-.95){\makebox(0,0)[lt]{\tiny$\square$}}
\rput(0.02,-1.5){\makebox(0,0)[rb]{\tiny$\square$}}
\rput(0,-0.7){\pscircle[linecolor=white,fillstyle=solid,fillcolor=lightgray](0,0){0.3}}
\rput(0,0){\psline[linewidth=.5pt](0,-1)(0,-1.5)}
\rput(0,0){\pscircle[linecolor=white,fillstyle=solid,fillcolor=white](2.25,0){0.75}}
\rput(0,0){\pscircle[linecolor=white,fillstyle=solid,fillcolor=white](-2.5,0){0.5}}
\psclip{\pscircle[linecolor=black,linewidth=1.5pt,linestyle=dashed](2.25,0){0.75}}
\psbezier[linewidth=.5pt,linecolor=green](3,0)(2,0)(2,2)(0,2)
\psbezier[linewidth=.5pt,linecolor=green](3,0)(2,0)(2,-2)(0,-2)
\endpsclip
\psclip{\pscircle[linecolor=black,linewidth=1.5pt,linestyle=dashed](-2.5,0){0.5}}
\psbezier[linewidth=.5pt,linecolor=green](-3,0)(-2,0)(-2,2)(0,2)
\psbezier[linewidth=.5pt,linecolor=green](-3,0)(-2,0)(-2,-2)(0,-2)
\psbezier[linewidth=.5pt,linecolor=blue](-3,0)(-1.5,0)(-1.5,0.5)(0,0.5)
\psbezier[linewidth=.5pt,linecolor=blue](-3,0)(-1.5,0)(-1.5,-1.5)(0,-1.5)
\endpsclip
\psbezier[linewidth=.5pt](0.9,0)(1.5,0.3)(2,0)(3,0)
\rput{25}(0.87,0){\rput(0,0){\makebox(0,0)[lb]{\tiny$\square$}}}
}
}
\newcommand{\DISCTWO}{%
{\psset{unit=1}
\psbezier[linewidth=1.5pt,linecolor=green,linestyle=dashed](-3,0)(-2,0)(-2,2)(0,2)
\psbezier[linewidth=1.5pt,linecolor=green,linestyle=dashed](-3,0)(-2,0)(-2,-2)(0,-2)
\psbezier[linewidth=1.5pt,linecolor=green,linestyle=dashed](3,0)(2,0)(2,2)(0,2)
\psbezier[linewidth=1.5pt,linecolor=green,linestyle=dashed](3,0)(2,0)(2,-2)(0,-2)
\psbezier[linewidth=1.5pt,linecolor=red,linestyle=dashed](3,0)(1.5,0)(1.5,0.5)(0,0.5)
\psbezier[linewidth=1.5pt,linecolor=red,linestyle=dashed](3,0)(1.5,0)(1.5,-1.5)(0,-1.5)
\rput(0,0){\psarc[linewidth=1.5pt,linecolor=red,linestyle=dashed](0,-0.5){1}{90}{270}}
\rput(-0.02,-.95){\makebox(0,0)[lt]{\tiny$\square$}}
\rput(0.02,-1.5){\makebox(0,0)[rb]{\tiny$\square$}}
\rput(0,-0.7){\pscircle[linecolor=white,fillstyle=solid,fillcolor=lightgray](0,0){0.3}}
\rput(0,0){\psline[linewidth=.5pt](0,-1)(0,-1.5)}
\rput(0,0){\pscircle[linecolor=white,fillstyle=solid,fillcolor=white](2.25,0){0.75}}
\rput(0,0){\pscircle[linecolor=white,fillstyle=solid,fillcolor=white](-2.5,0){0.5}}
\psclip{\pscircle[linecolor=black,linewidth=1.5pt,linestyle=dashed](2.25,0){0.75}}
\psbezier[linewidth=.5pt,linecolor=green](3,0)(2,0)(2,2)(0,2)
\psbezier[linewidth=.5pt,linecolor=green](3,0)(2,0)(2,-2)(0,-2)
\psbezier[linewidth=.5pt,linecolor=red](3,0)(1.5,0)(1.5,0.5)(0,0.5)
\psbezier[linewidth=.5pt,linecolor=red](3,0)(1.5,0)(1.5,-1.5)(0,-1.5)
\endpsclip
\psclip{\pscircle[linecolor=black,linewidth=1.5pt,linestyle=dashed](-2.5,0){0.5}}
\psbezier[linewidth=.5pt,linecolor=green](-3,0)(-2,0)(-2,2)(0,2)
\psbezier[linewidth=.5pt,linecolor=green](-3,0)(-2,0)(-2,-2)(0,-2)
\endpsclip
\psbezier[linewidth=.5pt](-0.9,0)(-1.5,0.3)(-2,0)(-3,0)
\rput{-25}(-0.87,0){\rput(0,0){\makebox(0,0)[rb]{\tiny$\square$}}}
}
}
\rput(-4,0){\DISCONE
}
\psline[linewidth=2pt]{->}(-0.5,0)(0.5,0)
\rput(4,0){\DISCTWO
}
\rput(-4,0){
\rput(0,0){\psarc[linewidth=3pt,linecolor=blue](0,-0.5){1.07}{-93}{30}}
\rput(-1.5,1.5){\makebox(0,0)[rb]{$\lambda_a$}}
\rput(-1.4,-1.5){\makebox(0,0)[rt]{$\lambda_b$}}
\rput(1.2,-0.8){\makebox(0,0)[lc]{$Z$}}
\rput(0,-0.2){\makebox(0,0)[rc]{$\omega$}}
\rput(-1.1,-1.1){\makebox(0,0)[lb]{$\lambda_c$}}
}
\rput(4,0){
\rput(0,0){\psarc[linewidth=3pt,linecolor=red](-0.02,-0.5){1.12}{152}{268}}
\rput(-1.5,1.5){\makebox(0,0)[rb]{$\lambda_a$}}
\rput(-1.4,-1.5){\makebox(0,0)[rt]{$\lambda_b$}}
\rput(-1.2,-0.8){\makebox(0,0)[rc]{$-Z$}}
\rput(0,-0.2){\makebox(0,0)[rc]{$\omega$}}
\rput(1.1,-1){\makebox(0,0)[rb]{$\lambda_d$}}
}
}
\end{pspicture}
\caption{\small The transformation dual to the flip in Fig.~\ref{fi:interchange-p-dual}: the flip,
or mutation transformation, for the $\lambda$-lengths in this case is described by the generalised cluster relation
$\lambda_c\lambda_d=\lambda_a^2+\lambda_b^2+\omega\lambda_a\lambda_b$. Here the shear coordinates
$Z$ and $-Z$ of Fig.~\ref{fi:interchange-p-dual} (indicated by bold lines) are logarithms of
the corresponding cross-ratios of $\lambda$-lengths (indicated by dashed lines).}
\label{fi:clustergeneral}
\end{figure}

\begin{lemma} \label{lem-pending1}{\rm (\cite{ChSh},\cite{ChM2})}
The transformation~in Fig.~\ref{fi:interchange-p-dual}
\bea
&{}&\{\tilde A,\tilde B,\tilde Z\}:= \{A+\phi(Z+i\pi/p)+\phi(Z-i\pi/p), B-\phi(-Z+i\pi/p)-\phi(-Z-i\pi/p),-Z\},\nonumber\\
&{}&\qquad\qquad w=2\cos(\pi/p);
\label{morphism-pending}\\
&{}&\{\tilde A,\tilde B,\tilde Z\}:= \{A+\phi(Z+P/2)+\phi(Z-P/2), B-\phi(-Z+P/2)-\phi(-Z-P/2),-Z\},\nonumber\\
&{}&\qquad\qquad w=2\cosh(P/2);
\label{morphism-pending-hole}
\eea
where $\phi(x)=\log(1+e^x)$,
is a morphism of the space
${\mathfrak T}_{g,s_h+s_o}^H$ that  preserves both Poisson structures {\rm(\ref{WP-PB})} and the geodesic
functions.
\end{lemma}

\proof Verifying the preservation of Poisson relations (\ref{WP-PB}) is simple, while to show
 that traces over paths are preserved we need to consider three different geodesic types like in Fig. \ref{fi:interchange-p-dual}, and
 in each of these cases we obtain the following
$2\times2$-{\em matrix} equalities that can be verified directly:
\bea
X_{A}LX_Z(F_\omega\Omega) X_ZLX_{B}&=&
X_{{\tilde A}}RX_{\tilde Z}(-\Omega) X_{\tilde Z}RX_{{\tilde B}},
\label{pend-mutation1}\\
X_{A}LX_Z\Omega X_ZRX_{A}&=&
X_{{\tilde A}}RX_{\tilde Z}(-\Omega)X_{\tilde Z}LX_{{\tilde A}},
\label{pend-mutation2}\\
X_{B}RX_Z\Omega X_ZLX_{B}&=&
X_{{\tilde B}}LX_{\tilde Z}(-\Omega)X_{\tilde Z}RX_{{\tilde B}},
\label{pend-mutation3}
\eea
where $\Omega$ is any matrix commuting with $F_\omega$; explicitly
$\Omega=\left(
         \begin{array}{cc}
           a & c \\
           -c & a-wc \\
         \end{array}
       \right)$, $a,c\in \CC$; in particular, we can take $F_\omega^k$, $k\in{\ZZ}$. \endproof

The transformation (mutation) of dual $\lambda$-lengths is depicted in Fig.~\ref{fi:clustergeneral}. It is described by the general
cluster transformations of \cite{ChSh}: $\lambda_c\lambda_d=\lambda_a^2+\lambda_b^2+\omega\lambda_a\lambda_b$. The shear coordinate
$Z$ of the edge incident to a loop is the signed geodesic distance
between perpendiculars to the third edge $c$ of the ideal triangle $abc$ through the vertex of this ideal triangle and between the hole
(orbifold point) inside the monogon and the edge $c$ (see Fig.~\ref{fi:clustergeneral}).
This signed distance is related to the cluster variables $\lambda_a$ and $\lambda_b$ by a simple formula (its proof is a nice exercise in
hyperbolic geometry),
\be
e^Z=\frac{\lambda_b}{\lambda_a}.
\label{Z-loop}
\ee

If, after a
series of morphisms, we come to a graph of the same combinatorial
type as the initial one (disregarding labeling of edges but distinguishing between
different types of orbifold points and holes with different perimeters), we
associate a {\em mapping class group} operation to this morphism
therefore passing from the groupoid of morphisms to the group of
modular transformations.

\begin{remark}\label{rm:spiraling}
In the notation adopted in this paper, the only effect of changing the decoration (spiraling direction) of a hole inside a monogon
corresponds to changing $P_i\to -P_i$ thus leaving invariant $w_i=2\cosh(P_i/2)$, so this transformation acts like the identity on the coordinates
of ${\mathfrak T}_{g,s_h+s_o}^H$.
\end{remark}

We can summarize as follows.

\begin{theorem}\label{thm:flip-m}
The whole mapping class group of $\Sigma_{g,s_h+s_o}$ is generated by morphisms described by
Lemmas~\ref{lem-abc} and~\ref{lem-pending1}.
\end{theorem}

\subsection{Quantum MCG transformations}
We now quantize a Teichm\"uller space
${\mathfrak T}_{g,s_h+s_o}$ equivariantly w.r.t.\ the mapping class group action.

Let ${\mathfrak T}^\hbar({\mathcal G}_{g,s_h+s_o})$ be a
$*$-algebra generated by the generator $Z_\alpha^\hbar$
(one generator per one unoriented edge $\alpha$)
and relations
\be
[Z^\hbar_\alpha, Z^\hbar_\beta ] = 2\pi i\hbar\{Z_\alpha, Z_\beta\}
\label{qq}
\ee
with the $*$-structure
\be
(Z^\hbar_\alpha)^*=Z^\hbar_\alpha.
\ee
Here $Z_\alpha$  and $\{\cdot,\cdot\}$ stand for the respective
coordinate functions on
the classical Teichm\"uller space and the Weil--Petersson Poisson bracket on
it. Note that according to formula (\ref{WP-PB}), the
right-hand side of (\ref{qq})
is merely a constant which may take only five values: $0$, $\pm
2\pi i \hbar$, $\pm 4 \pi i \hbar$.

 In the following two subsection we quantize the flip morphims viewed in subsections \ref{sss:mcg} and \ref{sss:pending}.
We demonstrate that in each case  the preservation of the commutation relations under quantum flip morphisms is straightforward, while verifying
the preservation of geodesic function operators requires some care.

Here and hereafter, for the rest of the paper, we assume that the ordering of
quantum operators in a product is {\em natural}, i.e.,
it is determined by the order of matrix multiplication itself.

For the notation simplicity in what follows we omit the superscript $\hbar$ for the quantum operators;
the classical or quantum nature of the object will be always clear from the context.

\subsubsection{Quantum flip morphisms for inner edges}

It was proved in \cite{ChF1} that the {\em quantum flip morphisms}
\bea
\{A,B,C,D,Z\}&\to&\{A+\phi^\hbar(Z),B-\phi^\hbar(-Z),C+\phi^\hbar(Z),
D-\phi^\hbar(-Z),-Z\}\nonumber\\
&:=&\{\tilde A,\tilde B,\tilde C,\tilde D,\tilde Z\},
\label{q-mor}
\eea
where $A$, $B$, $C$, $D$, and~$Z$ are as in Fig.~\ref{fi:flip} and
$\phi^\hbar(x)$ is the real function of one real variable,
\begin{equation} \label{phi}
\phi^\hbar(z) =
-\frac{\pi\hbar}{2}\int_{\Omega} \frac{e^{-ipz}}{\sinh(\pi p)\sinh(\pi \hbar
p)}dp,
\end{equation}
(the contour $\Omega$ goes along the real axis bypassing the
singularity at the origin from above) satisfy the standard two-, four-, and five-term relations.
The quantum dilogarithm function $\phi^\hbar(z)$ was introduced in this context by Faddeev in
\cite{Faddeev} and used in \cite{FK} for constructing quantum MCG transformations for the Liouville model,
see, e.g., \cite{ChF1} for the properties of this function.

\begin{remark}
Note that exponentiated
algebraic elements $U_i = e^{\pm Z_i}$,
which obey homogeneous commutation relations
$q^nU_iU_j=U_jU_i q^{-n}$ with $[X_i,X_j]=2ni\pi  \hbar $ and $q:=e^{i\pi\hbar}$
transform as rational functions: for example, $ e^{A+\phi^\hbar(Z)}=(1+qe^Z)e^A$.
\end{remark}

The quantum analogues of {\em matrix} relations (\ref{mutation1})--(\ref{mutation3}) were found in
\cite{ChP1}.  Then,
remarkably, {\em all four} entries of the corresponding $2\times 2$-matrices transform uniformly.

\begin{lemma}\label{lm:Ch-Pen}{\rm\cite{ChP1}}
Applying the quantum MCG transformation (\ref{q-mor}) to the curves 1,2, and 3 in Fig.~\ref{fi:flip}, we obtain
the respective quantum matrix relations:
\bea
X_D RX_ZRX_A &=&q^{1/4}X_{\tilde D}RX_{\tilde A},\label{curve1}\\
X_D RX_Z LX_B&=& X_{\tilde D} RX_{\tilde Z} LX_{\tilde B},\label{curve2}\\
X_D L X_C &=&q^{1/4}X_{\tilde D}L X_{\tilde Z} L X_{\tilde C}.\label{curve3}
\eea
\end{lemma}

\subsubsection{Quantum flip morphisms for loops}

\begin{lemma} \label{lem-pending-quantum}{\rm\cite{ChM2}}
The transformation~in Fig.~\ref{fi:interchange-p-dual}
\be
\label{morphism-pending-quantum}
\{\tilde A,\tilde B,\tilde Z\}:= \{A+\phi^\hbar(Z+i\pi/p)+\phi^\hbar(Z-i\pi/p), B-\phi^\hbar(-Z+i\pi/p)-\phi^\hbar(-Z-i\pi/p),-Z\},
\ee
with $\phi^\hbar(x)$ from (\ref{phi}) and $w=2\cos(\pi/p)$ or the transformations
\be
\label{morphism-pending-quantum2}
\{\tilde A,\tilde B,\tilde Z\}:= \{A+\phi^\hbar(Z+P/2)+\phi^\hbar(Z-P/2), B-\phi^\hbar(-Z+P/2)-\phi^\hbar(-Z-P/2),-Z\},
\ee
for $w=2\cosh(P/2)$
are morphisms of the quantum $*$-algebra
${\mathfrak T}^\hbar_{g,s_h+s_o}$.
\end{lemma}

The quantum versions of MCG transformations
(\ref{pend-mutation1})--(\ref{pend-mutation3}) are as follows:

\begin{lemma}\label{lem-quantum-geod}{\rm\cite{ChM2}}
We have the following quantum matrix relations:
\bea
X_{A}LX_Z(F_p\Omega) X_ZLX_{B}&=&q^{-1}
X_{{\tilde A}}RX_{\tilde Z}(-\Omega) X_{\tilde Z}RX_{{\tilde B}},
\label{pend-quantum-mutation1}\\
X_{A}LX_Z\Omega X_ZRX_{A}&=&
X_{{\tilde A}}RX_{\tilde Z}(-\Omega)X_{\tilde Z}LX_{{\tilde A}},
\label{pend-quantum-mutation2}\\
X_{B}RX_Z\Omega X_ZLX_{B}&=&
X_{{\tilde B}}LX_{\tilde Z}(-\Omega)X_{\tilde Z}RX_{{\tilde B}}.
\label{pend-quantum-mutation3}
\eea
\end{lemma}

\begin{remark}\label{rem-quantum}
Transformation laws (\ref{curve1})--(\ref{curve3}) and
(\ref{pend-quantum-mutation1})--(\ref{pend-quantum-mutation3}) become identities without $q$-factors if we scale
the matrices of right and left turns:
\be
L\to q^{1/4}L,\qquad R\to q^{-1/4} R.
\label{scaling-LR}
\ee
Unfortunately, this property
does not allow formulating a ``working'' ordering prescription for a general geodesic function. Indeed, we can use
these transformations for bringing \emph{any} simple closed loop geodesic function $G_\gamma$ not homeomorphic to a hole to a form
$G_\gamma=\tr R X_{\widetilde Z} L X_{\widetilde Y}$ in some transformed shear coordinates $\widetilde Z$ and $\widetilde Y$
with the natural ordering assumed. But $G_\gamma$ becomes a Hermitian operator in the Weyl ordered form, not in the naturally ordered one,
so when passing to a naturally ordered expressions we must introduce different $q$-factors for different Laurent monomials in the
expansion of $G_\gamma$, and these $q$-factors become uncontrollably spread over terms of the expansion of $G_\gamma$ in the
original shear coordinates.

Fortunately, for arcs, or
$\lambda$-lengths in Sec.~\ref{s:q}, powers of $q$ factors coincide for \emph{all} terms under the trace sign, which allows us
to solve the problem of quantum ordering completely.
\end{remark}

\section{Colliding holes: new geodesic laminations and the corresponding geodesic algebras}\label{s:limit}
\setcounter{equation}{0}

In this section, we consider degenerations of stable Riemann surfaces that correspond to colliding two holes or
two sides of the same hole. Below we show that
this results in the appearance of bordered Riemann surfaces with holes and with marked points on the
boundaries of these holes represented by \emph{bordered cusps}.

A Riemann surface $\Sigma_{g,s,n}$ of genus $g$ with $s>0$ holes/orbifold points and with $n\ge 0$ bordered cusps assigned to holes (no bordered cusps can be assigned to orbifold points) is called \emph{stable} if its hyperbolic area (with tubular domains of holes removed)
is positive. {Since every ideal triangle has area $\pi$, for a stable Riemann surface
 $\Sigma_{g,s,n}$ of genus $g$ with
$s_h>0$ holes, $s_o$ orbifiold points of respective orders $2\le k_i<\infty$, $i=1,\dots,s_o$ ($s:=s_h+s_o$),
and $n$ bordered cusps, the hyperbolic area is given by:}
$$
\hbox{Area\ of\ }\Sigma_{g,s,n}=\Bigl[4g-4+2s_h+\sum_{i=1}^{s_o}\Bigl(2-\frac{2}{k_i}\Bigr)+n\Bigr]\pi
=\Bigl[4g-4+2s+n-2\sum_{i=1}^{s_o}\frac{1}{k_i}\Bigr]\pi.
$$
So, all surfaces with $g>0$ are stable (recall that we require $s_h>0$ for any Riemann surface under consideration)
and we have only a handful of non-stable cases at $g=0$: $(s=1;n=\{0,1,2\})$, $(s=2;n=0)$, $(s_h=1,s_o=1\,(k_1=2);n=1)$,
$(s_h=1,s_o=2\,(k_1=k_2=2);n=0)$.

In our procedure of degenerations of stable Riemann surfaces, the total hyperbolic area will be preserved.

\subsection{{Qualitative description of  the collision process}}\label{ss:chewinggum}

We are interested in the process of colliding two holes with geodesic boundaries on a Riemann surface.
This means that we consider a limit in which the geodesic distance between these two holes
tends to zero. In hyperbolic geometry, this means that we obtain a ``thin'' domain between these two holes, but
the geodesic length of this domain, on the contrary, becomes infinite in this limit,
see Fig.~\ref{fi:chewinggum}.\footnote{Intuitively, it can be thought of as pulling a ``chewing gum'': the hyperbolic area of the Riemann surface
(with tubular domains of holes removed) is constant proportional
to the Euler characteristics, so pulling a chewing gum we make it thinner in the middle.}

Instead of colliding two different holes, we can as well consider  colliding two sides of the same hole (see Fig~\ref{fi:chewinggum2} and \ref{fi:chewinggum3}).
We therefore have the following three types of processes:

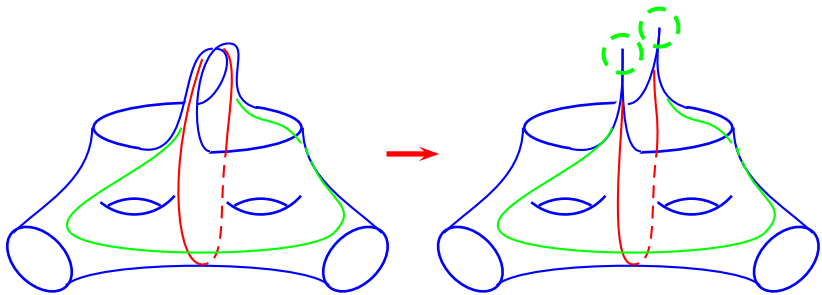
\begin{figure}[tb]
{\psset{unit=0.7}
\begin{pspicture}(-4,-2)(4,4)
\newcommand{\PATTERN}{%
{\psset{unit=1}
\rput{135}(-3,-1){\psellipse[linecolor=blue, linewidth=1pt](0,0)(.75,.5)}
\rput{45}(3,-1){\psellipse[linecolor=blue, linewidth=1pt](0,0)(.75,.5)}
\psbezier[linecolor=blue](-2.5,-1.5)(-2,-1)(2,-1)(2.5,-1.5)
\psbezier[linecolor=blue](-3.5,-.5)(-3,0)(-2,0.5)(-2,1.5)
\psbezier[linecolor=blue](3.5,-.5)(3,0)(2,0.5)(2,1.5)
\psarc[linecolor=blue, linewidth=1pt](-1.2,-.85){1}{60}{120}
\psarc[linecolor=blue, linewidth=1pt](-1.2,.85){1}{230}{320}
\psarc[linecolor=blue, linewidth=1pt](1.2,-.85){1}{60}{120}
\psarc[linecolor=blue, linewidth=1pt](1.2,.85){1}{230}{320}
\rput(0,1.5){\psellipse[linecolor=blue, linewidth=1pt](0,0)(2,0.5)}
\rput{45}(0,1.5){\psframe[linecolor=white, fillstyle=solid, fillcolor=white](-1.2,-0.5)(1.2,.5)}
}
}
\rput(0,0){\PATTERN}
\psbezier[linecolor=blue](-1.2,1.1)(-0.1,0.85)(-.5,3)(0.2,3)
\psbezier[linecolor=blue](0.15,1.05)(-0.25,1.05)(-.2,3.1)(0.5,3.1)
\psbezier[linecolor=blue](1.05,1.87)(0.25,2.07)(1,3.1)(0.5,3.1)
\psbezier[linecolor=blue](-0.1,1.95)(0.3,1.95)(.75,3)(0.25,3)
%
%
\psbezier[linecolor=red](0,2.8)(-0.7,.8)(-.5,-1.1)(0,-1.1)
\psbezier[linecolor=red, linestyle=dashed](0.45,1.05)(0.3,0)(.5,-1.1)(0,-1.1)
\psbezier[linecolor=red](0.45,1.05)(0.5,1.8)(.7,2.8)(0.4,3)
\psbezier[linecolor=green](-0.4,1.5)(-0.6,.8)(-3,0)(-2.5,-0.5)
\psbezier[linecolor=green](-2.5,-.5)(-2,-1)(2,-1)(2.5,-.5)
\psbezier[linecolor=green](2.1,.8)(2.4,0.1)(3,0)(2.5,-0.5)
\psbezier[linecolor=green, linestyle=dashed](2.1,.8)(1.8,1.4)(2.2,0.8)(1.8,1.3)
\psbezier[linecolor=green](0.65,2.05)(1,1.45)(1.4,1.8)(1.8,1.3)
\rput(4,1){\psline[linewidth=2pt,linecolor=red]{->}(-0.5,0)(0.5,0)}
\end{pspicture}
\begin{pspicture}(-4,-2)(4,4)
\newcommand{\PATTERN}{%
{\psset{unit=1}
\rput{135}(-3,-1){\psellipse[linecolor=blue, linewidth=1pt](0,0)(.75,.5)}
\rput{45}(3,-1){\psellipse[linecolor=blue, linewidth=1pt](0,0)(.75,.5)}
\psbezier[linecolor=blue](-2.5,-1.5)(-2,-1)(2,-1)(2.5,-1.5)
\psbezier[linecolor=blue](-3.5,-.5)(-3,0)(-2,0.5)(-2,1.5)
\psbezier[linecolor=blue](3.5,-.5)(3,0)(2,0.5)(2,1.5)
\psarc[linecolor=blue, linewidth=1pt](-1.2,-.85){1}{60}{120}
\psarc[linecolor=blue, linewidth=1pt](-1.2,.85){1}{230}{320}
\psarc[linecolor=blue, linewidth=1pt](1.2,-.85){1}{60}{120}
\psarc[linecolor=blue, linewidth=1pt](1.2,.85){1}{230}{320}
\rput(0,1.5){\psellipse[linecolor=blue, linewidth=1pt](0,0)(2,0.5)}
\rput{45}(0,1.5){\psframe[linecolor=white, fillstyle=solid, fillcolor=white](-1.2,-0.5)(1.2,.5)}
}
}
\rput(0,0){\PATTERN}
\psbezier[linecolor=blue](-1.2,1.1)(-0.1,0.85)(-0.2,2.5)(-0.2,3)
\psbezier[linecolor=blue](0.15,1.05)(-0.25,1.05)(-0.2,2.5)(-0.2,3)
\psbezier[linecolor=blue](1.05,1.87)(0.25,2.07)(0.5,3)(0.5,3.4)
\psbezier[linecolor=blue](-0.1,1.95)(0.3,1.95)(.5,3)(0.5,3.4)
%
%
\psbezier[linecolor=red](-0.2,2)(-0.2,1.4)(-.5,-1.1)(0,-1.1)
\psbezier[linecolor=red, linestyle=dashed](0.45,1.05)(0.3,0)(.5,-1.1)(0,-1.1)
\psbezier[linecolor=red](0.45,1.05)(0.5,1.8)(.4,1.8)(0.4,2.6)
\psbezier[linecolor=green](-0.4,1.5)(-0.6,.8)(-3,0)(-2.5,-0.5)
\psbezier[linecolor=green](-2.5,-.5)(-2,-1)(2,-1)(2.5,-.5)
\psbezier[linecolor=green](2.1,.8)(2.4,0.1)(3,0)(2.5,-0.5)
\psbezier[linecolor=green, linestyle=dashed](2.1,.8)(1.8,1.4)(2.2,0.8)(1.8,1.3)
\psbezier[linecolor=green](0.65,2.05)(1,1.45)(1.4,1.8)(1.8,1.3)
\rput(0.5,3.4){\pscircle[linecolor=green, linestyle=dashed,linewidth=1.5pt](0,0){0.4}}
\rput(-0.2,2.9){\pscircle[linecolor=green, linestyle=dashed,linewidth=1.5pt](0,0){0.4}}
\end{pspicture}
}
\caption{\small The process of colliding two holes on the Riemann surface $\Sigma_{g,s,0}$: as a result
we obtain a Riemann surface $\Sigma_{g,s-1,2}$
of the same genus with one less hole and with two new cusps on the boundary. Closed geodesics passing trough the chewing-gum become geodesic arcs after the collision.}
\label{fi:chewinggum}
\end{figure}

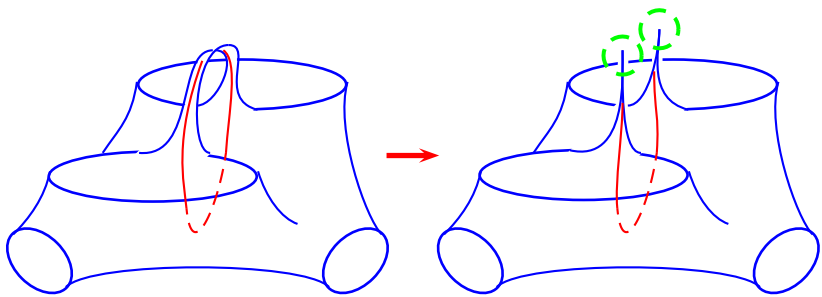
\begin{figure}[tb]
{\psset{unit=0.7}
\begin{pspicture}(-4,-2)(4,3.5)
\newcommand{\PATTERN}{%
{\psset{unit=1}
\rput{135}(-3,-1){\psellipse[linecolor=blue, linewidth=1pt](0,0)(.75,.5)}
\rput{45}(3,-1){\psellipse[linecolor=blue, linewidth=1pt](0,0)(.75,.5)}
\psbezier[linecolor=blue](-2.5,-1.5)(-2,-1)(2,-1)(2.5,-1.5)
\psbezier[linecolor=blue](-3.5,-.5)(-3.2,-0.2)(-2.9,0.2)(-2.8,.7)
\psbezier[linecolor=blue](1.9,-.3)(1.6,-0.2)(1.3,0.2)(1.15,.7)
\psbezier[linecolor=blue](3.5,-.5)(3.2,-0.2)(2.6,1.5)(2.85,2.4)
\psbezier[linecolor=blue](-1.8,1.05)(-1.5,1.5)(-1.25,1.7)(-1.15,2.4)
\rput(-.85,.6){\psellipse[linecolor=blue, linewidth=1pt](0,0)(2,0.5)}
\rput(.85,2.35){\psellipse[linecolor=blue, linewidth=1pt](0,0)(2,0.5)}
\rput{45}(0,1.5){\psframe[linecolor=white, fillstyle=solid, fillcolor=white](-1.2,-0.5)(1.2,.5)}
\rput(0.65,2.8){\psframe[linecolor=white, fillstyle=solid, fillcolor=white](-.1,-0.1)(.1,.1)}
\rput(0.1,2.8){\psframe[linecolor=white, fillstyle=solid, fillcolor=white](-.1,-0.1)(.1,.1)}
}
}
\rput(0,0){\PATTERN}
\psbezier[linecolor=blue](-1.2,1.05)(-0.1,0.95)(-.5,3)(0.2,3)
\psbezier[linecolor=blue](0.15,1.05)(-0.25,1.05)(-.2,3.1)(0.5,3.1)
\psbezier[linecolor=blue](1.05,1.87)(0.25,2.07)(1,3.1)(0.5,3.1)
\psbezier[linecolor=blue](-0.1,1.95)(0.3,1.95)(.75,3)(0.25,3)
%
%
\psbezier[linecolor=red](0,2.8)(-0.55,1.3)(-.35,0.5)(-0.3,0)
\psbezier[linecolor=red, linestyle=dashed](0.45,1.05)(0.3,0)(-.2,-1.1)(-0.3,0)
\psbezier[linecolor=red](0.45,1.05)(0.5,1.8)(.7,2.8)(0.4,3)
\rput(4,1){\psline[linewidth=2pt,linecolor=red]{->}(-0.5,0)(0.5,0)}
\end{pspicture}
\begin{pspicture}(-4,-2)(4,3.5)
\newcommand{\PATTERN}{%
{\psset{unit=1}
\rput{135}(-3,-1){\psellipse[linecolor=blue, linewidth=1pt](0,0)(.75,.5)}
\rput{45}(3,-1){\psellipse[linecolor=blue, linewidth=1pt](0,0)(.75,.5)}
\psbezier[linecolor=blue](-2.5,-1.5)(-2,-1)(2,-1)(2.5,-1.5)
\psbezier[linecolor=blue](-3.5,-.5)(-3.2,-0.2)(-2.9,0.2)(-2.8,.7)
\psbezier[linecolor=blue](1.9,-.3)(1.6,-0.2)(1.3,0.2)(1.15,.7)
\psbezier[linecolor=blue](3.5,-.5)(3.2,-0.2)(2.6,1.5)(2.85,2.4)
\psbezier[linecolor=blue](-1.8,1.05)(-1.5,1.5)(-1.25,1.7)(-1.15,2.4)
\rput(-.85,.63){\psellipse[linecolor=blue, linewidth=1pt](0,0)(2,0.5)}
\rput(.85,2.35){\psellipse[linecolor=blue, linewidth=1pt](0,0)(2,0.5)}
\rput{45}(0,1.5){\psframe[linecolor=white, fillstyle=solid, fillcolor=white](-1.2,-0.5)(1.2,.5)}
\rput(0.55,2.8){\psframe[linecolor=white, fillstyle=solid, fillcolor=white](-.1,-0.1)(.1,.1)}
\rput(-0.1,2.8){\psframe[linecolor=white, fillstyle=solid, fillcolor=white](-.1,-0.1)(.1,.1)}
}
}
\rput(0,0){\PATTERN}
\psbezier[linecolor=blue](-1.2,1.1)(-0.1,0.85)(-0.2,2.5)(-0.2,3)
\psbezier[linecolor=blue](0.15,1.05)(-0.25,1.05)(-0.2,2.5)(-0.2,3)
\psbezier[linecolor=blue](1.05,1.87)(0.25,2.07)(0.5,3)(0.5,3.4)
\psbezier[linecolor=blue](-0.1,1.95)(0.3,1.95)(.5,3)(0.5,3.4)
%
%
\psbezier[linecolor=red](-0.2,2)(-0.2,1.4)(-.35,0.5)(-0.3,0)
\psbezier[linecolor=red, linestyle=dashed](0.45,1.05)(0.3,0)(-.2,-1.1)(-0.3,0)
\psbezier[linecolor=red](0.45,1.05)(0.5,1.8)(.4,1.8)(0.4,2.6)
\rput(0.5,3.4){\pscircle[linecolor=green, linestyle=dashed,linewidth=1.5pt](0,0){0.4}}
\rput(-0.2,2.9){\pscircle[linecolor=green, linestyle=dashed,linewidth=1.5pt](0,0){0.4}}
\end{pspicture}
}
\caption{\small The process of colliding two sides of the same hole on the Riemann surface $\Sigma_{g,s,0}$
($g>0$): as a result
we obtain a Riemann surface $\Sigma_{g-1,s+1,2}$
of genus lesser by one with one more hole and two new cusps on the boundaries of
holes. A closed geodesic trough the chewing-gum becomes a geodesic arc.}
\label{fi:chewinggum2}
\end{figure}

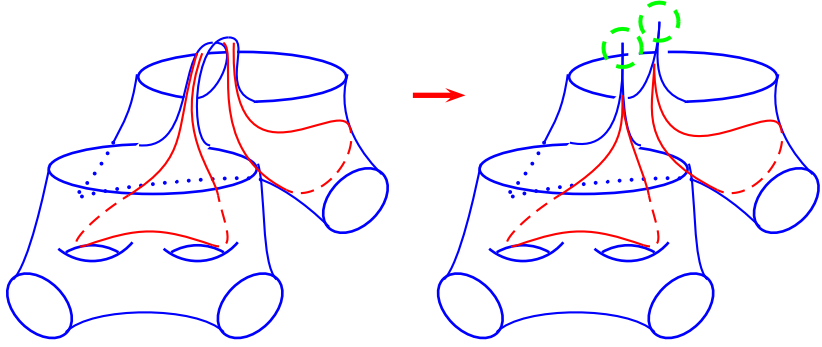
\begin{figure}[tb]
{\psset{unit=0.7}
\begin{pspicture}(-4,-3)(4,4)
\newcommand{\PATTERN}{%
{\psset{unit=1}
\rput{135}(-3,-2){\psellipse[linecolor=blue, linewidth=1pt](0,0)(.75,.5)}
\rput{45}(1,-2){\psellipse[linecolor=blue, linewidth=1pt](0,0)(.75,.5)}
\psbezier[linecolor=blue](-2.5,-2.5)(-2,-2)(0,-2)(.5,-2.5)
\psbezier[linecolor=blue](-3.5,-1.5)(-3.2,-1.2)(-2.9,0.2)(-2.8,.7)
\psbezier[linecolor=blue](1.5,-1.5)(1.2,-1.2)(1.3,0.2)(1.15,.7)
\psarc[linecolor=blue, linewidth=1pt](-2,-1.85){1}{60}{120}
\psarc[linecolor=blue, linewidth=1pt](-2,-.15){1}{230}{320}
\psarc[linecolor=blue, linewidth=1pt](0,-1.85){1}{60}{120}
\psarc[linecolor=blue, linewidth=1pt](0,-.15){1}{230}{320}
\rput{45}(3,0){\psellipse[linecolor=blue, linewidth=1pt](0,0)(.75,.5)}
\psbezier[linecolor=blue](3.5,0.5)(3.2,.8)(2.6,1.5)(2.85,2.4)
\psbezier[linecolor=blue](-1.6,1.1)(-1.4,1.5)(-1.25,1.7)(-1.15,2.4)
\psbezier[linecolor=blue](1.2,0.4)(1.5,0.3)(2,0)(2.5,-.5)
\psbezier[linecolor=blue,linestyle=dotted,linewidth=1.5pt](1.2,0.4)(.9,0.5)(-1.8,0.3)(-2.4,0)
\psbezier[linecolor=blue,linestyle=dotted,linewidth=1.5pt](-1.6,1.1)(-1.8,.7)(-2,0.4)(-2.4,0)
\rput(-.85,.6){\psellipse[linecolor=blue, linewidth=1pt](0,0)(2,0.5)}
\rput(.85,2.35){\psellipse[linecolor=blue, linewidth=1pt](0,0)(2,0.5)}
\rput{45}(0,1.5){\psframe[linecolor=white, fillstyle=solid, fillcolor=white](-1.2,-0.5)(1.2,.5)}
\rput(0.65,2.8){\psframe[linecolor=white, fillstyle=solid, fillcolor=white](-.1,-0.1)(.1,.1)}
\rput(0.1,2.8){\psframe[linecolor=white, fillstyle=solid, fillcolor=white](-.1,-0.1)(.1,.1)}
}
}
\rput(0,0){\PATTERN}
\psbezier[linecolor=blue](-1.2,1.05)(-0.1,0.95)(-.5,3)(0.2,3)
\psbezier[linecolor=blue](0.15,1.05)(-0.25,1.05)(-.2,3.1)(0.5,3.1)
\psbezier[linecolor=blue](1.05,1.87)(0.25,2.07)(1,3.1)(0.5,3.1)
\psbezier[linecolor=blue](-0.1,1.95)(0.3,1.95)(.75,3)(0.25,3)
%
%
\psbezier[linecolor=red](0,2.8)(-0.55,1.3)(0.2,0.6)(0.3,0.1)
\psbezier[linecolor=red, linestyle=dashed](0.3,0.1)(0.4,-0.4)(.7,-1)(0.2,-0.85)
\psbezier[linecolor=red](-2.2,-0.85)(-1.2,-0.5)(-.8,-0.5)(0.2,-0.85)
\psbezier[linecolor=red, linestyle=dashed](-1.4,0.1)(-2.2,-0.4)(-2.7,-1)(-2.2,-0.85)
\psbezier[linecolor=red](-0.1,2.8)(-0.55,1.3)(-.2,0.85)(-1.4,0.1)
\psbezier[linecolor=red](2.8,1.4)(2.6,2)(.5,0)(0.6,3)
\psbezier[linecolor=red, linestyle=dashed](1.6,0.2)(2.2,-0.1)(3,.8)(2.8,1.4)
\psbezier[linecolor=red](1.6,0.2)(0,1)(.7,2.8)(0.4,3)
\rput(4.5,2){\psline[linewidth=2pt,linecolor=red]{->}(-0.5,0)(0.5,0)}
\end{pspicture}
\begin{pspicture}(-4,-3)(4,4)
\newcommand{\PATTERN}{%
{\psset{unit=1}
\rput{135}(-3,-2){\psellipse[linecolor=blue, linewidth=1pt](0,0)(.75,.5)}
\rput{45}(1,-2){\psellipse[linecolor=blue, linewidth=1pt](0,0)(.75,.5)}
\psbezier[linecolor=blue](-2.5,-2.5)(-2,-2)(0,-2)(.5,-2.5)
\psbezier[linecolor=blue](-3.5,-1.5)(-3.2,-1.2)(-2.9,0.2)(-2.8,.7)
\psbezier[linecolor=blue](1.5,-1.5)(1.2,-1.2)(1.3,0.2)(1.15,.7)
\psarc[linecolor=blue, linewidth=1pt](-2,-1.85){1}{60}{120}
\psarc[linecolor=blue, linewidth=1pt](-2,-.15){1}{230}{320}
\psarc[linecolor=blue, linewidth=1pt](0,-1.85){1}{60}{120}
\psarc[linecolor=blue, linewidth=1pt](0,-.15){1}{230}{320}
\rput{45}(3,0){\psellipse[linecolor=blue, linewidth=1pt](0,0)(.75,.5)}
\psbezier[linecolor=blue](3.5,0.5)(3.2,.8)(2.6,1.5)(2.85,2.4)
\psbezier[linecolor=blue](-1.6,1.1)(-1.4,1.5)(-1.25,1.7)(-1.15,2.4)
\psbezier[linecolor=blue](1.2,0.4)(1.5,0.3)(2,0)(2.5,-.5)
\psbezier[linecolor=blue,linestyle=dotted,linewidth=1.5pt](1.2,0.4)(.9,0.5)(-1.8,0.3)(-2.4,0)
\psbezier[linecolor=blue,linestyle=dotted,linewidth=1.5pt](-1.6,1.1)(-1.8,.7)(-2,0.4)(-2.4,0)
\rput(-.85,.6){\psellipse[linecolor=blue, linewidth=1pt](0,0)(2,0.5)}
\rput(.85,2.35){\psellipse[linecolor=blue, linewidth=1pt](0,0)(2,0.5)}
\rput{45}(0,1.5){\psframe[linecolor=white, fillstyle=solid, fillcolor=white](-1.2,-0.5)(1.2,.5)}
\rput(0.65,2.8){\psframe[linecolor=white, fillstyle=solid, fillcolor=white](-.1,-0.1)(.1,.1)}
\rput(0.1,2.8){\psframe[linecolor=white, fillstyle=solid, fillcolor=white](-.1,-0.1)(.1,.1)}
}
}
\rput(0,0){\PATTERN}
\psbezier[linecolor=blue](-1.2,1.1)(-0.1,0.85)(-0.2,2.5)(-0.2,3)
\psbezier[linecolor=blue](0.15,1.05)(-0.25,1.05)(-0.2,2.5)(-0.2,3)
\psbezier[linecolor=blue](1.05,1.87)(0.25,2.07)(0.5,3)(0.5,3.4)
\psbezier[linecolor=blue](-0.1,1.95)(0.3,1.95)(.5,3)(0.5,3.4)
%
%
\psbezier[linecolor=red](-0.2,2)(-0.2,1.3)(0.2,0.6)(0.3,0.1)
\psbezier[linecolor=red, linestyle=dashed](0.3,0.1)(0.4,-0.4)(.7,-1)(0.2,-0.85)
\psbezier[linecolor=red](-2.2,-0.85)(-1.2,-0.5)(-.8,-0.5)(0.2,-0.85)
\psbezier[linecolor=red, linestyle=dashed](-1.4,0.1)(-2.2,-0.4)(-2.7,-1)(-2.2,-0.85)
\psbezier[linecolor=red](-0.2,2)(-0.2,1.3)(-.2,0.85)(-1.4,0.1)
\psbezier[linecolor=red](2.8,1.4)(2.6,2)(.4,0)(0.4,2.6)
\psbezier[linecolor=red, linestyle=dashed](1.6,0.2)(2.2,-0.1)(3,.8)(2.8,1.4)
\psbezier[linecolor=red](1.6,0.2)(0,1)(.4,1.8)(0.4,2.6)
\rput(0.5,3.4){\pscircle[linecolor=green, linestyle=dashed,linewidth=1.5pt](0,0){0.4}}
\rput(-0.2,2.9){\pscircle[linecolor=green, linestyle=dashed,linewidth=1.5pt](0,0){0.4}}
\end{pspicture}
}
\caption{\small The process of colliding two sides of the same hole on the Riemann surface $\Sigma_{g,s,0}$
($g>0$) when it results in a two-component Riemann surfaces $\Sigma_{g_1,s_1,1}$ and $\Sigma_{g_2,s_2,1}$
with $g_1+g_2=g$ and $s_1+s_2=s+1$: the hole splits into two holes on two different components, and each of the
newly generated holes contains one new bordered cusp. A closed geodesic that passed through the corresponding ``chewing gum''
before breaking it
(in this case, it must pass through it at least twice) splits into two geodesics starting and terminating at the newly created bordered cusps
on two disjoint components.}
\label{fi:chewinggum3}
\end{figure}

\vskip10pt

\noindent
{\bf1.} \ The result of colliding two holes of a Riemann surface
$\Sigma_{g,s,n}$ of genus $g$ with $s$ holes/orbifold points, and $n$ bordered cusps
is a Riemann surface of the same genus $g$, $s-1$ holes/orbifold points, and $n+2$ bordered cusps, and the hole obtained by colliding two original holes now contains two new bordered cusps (see Fig.~\ref{fi:chewinggum}).

\vskip10pt

\noindent
{\bf 2.} \ The result of colliding sides of the same hole varies
depending on whether breaking the chewing gum will result in one or two disjoint components (the latter happens unavoidably for example when the original surface has genus zero).

\begin{itemize}
\item[\bf 2a]  If breaking the chewing gum constituted by sides of the same hole in $\Sigma_{g,s,n}$
results in a one-component surface, the new surface $\Sigma_{g-1, s+1, n+2}$ has genus lesser by one (so, originally, $g>0$); the original hole then splits into two holes each containing one new bordered cusp (Fig.~\ref{fi:chewinggum2}).
\item[\bf 2b] If breaking the chewing gum constituted by sides of the same hole in $\Sigma_{g,s,n}$ results in a two-component surface, these
new connected components,
$\Sigma_{g_1,s_1,n_1}$ and $\Sigma_{g_2,s_2,n_2}$, must be stable and such that $g_1+g_2=g$, $s_1+s_2=s+1$, and
$n_1+n_2=n+2$ with $n_1>0$ and $n_2>0$ (Fig.~\ref{fi:chewinggum3}).
\end{itemize}


\subsection{{Limiting geodesic arcs and extended shear coordinates}} \label{ss:Ptolemy}

\begin{figure}[tb]
\begin{pspicture}(-4,0)(4,5)
{\psset{unit=1.5}
\psframe[linecolor=white, fillstyle=solid, fillcolor=lightgray](-3,0)(3,3)
\rput(0,0){\pswedge[linewidth=1pt,linecolor=white,fillstyle=solid,fillcolor=white](0,0){2.5}{0}{180}}
\rput(0,0){\pswedge[linewidth=1pt,linecolor=white,fillstyle=solid,fillcolor=lightgray](0,0){2}{0}{180}}
\psarc[linewidth=2pt,linecolor=red,linestyle=dashed](0.15,0){2.3}{18.5}{168}
\psarc[linewidth=2pt,linecolor=blue,linestyle=dashed](0,0){2.3}{17.5}{167}
\psline[linewidth=1.5pt]{->}(0,0)(0,3.2)
\psline[linewidth=1.5pt](-3.2,0)(3.2,0)
\psline[linewidth=2pt,linestyle=dashed](0,0)(-2.8,0.6)
\psline[linewidth=2pt,linestyle=dashed](0,0)(2.7,0.8)
\psarc[linewidth=0.7pt]{<->}(0,0){1.35}{0}{16}
\psarc[linewidth=0.7pt]{<->}(0,0){1.5}{168}{180}
\rput{245}(1.4,.8){\psline[linewidth=1.5pt,linecolor=red]{->}(0,0)(0,.45)}
\rput(1.4,0.8){\pscircle[linecolor=red,fillstyle=solid,fillcolor=white]{.2}}
\rput(1.4,0.8){\makebox(0,0)[cc]{\hbox{\tcr{\small$1$}}}}
\rput{125}(-1.5,.7){\psline[linewidth=1.5pt,linecolor=red]{->}(0,0)(0,.45)}
\rput(-1.5,.7){\pscircle[linecolor=red,fillstyle=solid,fillcolor=white]{.2}}
\rput(-1.5,.7){\makebox(0,0)[cc]{\hbox{\tcr{\small$2$}}}}
\rput{205}(-2.6,.9){\psline[linewidth=1.5pt,linecolor=red]{->}(0,0)(0,.4)}
\rput(-2.6,.9){\pscircle[linecolor=red,fillstyle=solid,fillcolor=white]{.2}}
\rput(-2.6,.9){\makebox(0,0)[cc]{\hbox{\tcr{\small$3$}}}}
\rput(0.2,1.8){\makebox(0,0){$1$}}
\rput(0.4,2.7){\makebox(0,0){$1+\varepsilon$}}
\rput(-1.75,.2){\makebox(0,0){$\varepsilon \ell_1$}}
\rput(1.6,.2){\makebox(0,0){$\varepsilon \ell_2$}}
}
\end{pspicture}
\caption{\small
{The fundamental domain of the original Riemann surface is contained in the white strip, while the grey areas represent colliding holes (or the same hole if colliding two sides of it). The collars that become boundaries of horocycles in the limit
as $\varepsilon\to 0$ are slanted dashed straight lines.}
}
\label{fi:strip}
\end{figure}
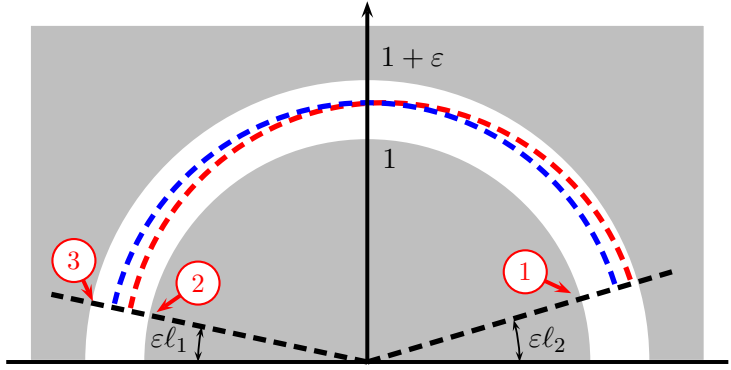

{We now give a quantitative hyperbolic description of the asymptotic process resulting from colliding two sides of the same hole in a Riemann surface in such a way that the result is disconnected. To this aim, we present this
process in the hyperbolic upper half-plane: in Fig.~\ref{fi:strip}, the two grey areas represent two sides of the hole and the white strip contains the fundamental domain of the Riemann surface. We introduce two ``collar lines'' (dashed slanted straight lines in the figure): they are loci of points equidistant from the shortest geodesic joining the boundaries of two holes (the vertical interval between
$1$ and $1+\varepsilon$ on the $y$-axis). The part of the Riemann surface which is not affected by the collision process corresponds to the part of the fundamental domain which  is contained between each collar line and the absolute. The part that is affected, namely the ``chewing-gum" is the portion of the white strip above the collars.}

{We consider parts of geodesics ``inside'' the chewing gum, (Fig.~\ref{fi:strip}). From the qualitative description discussed in sub-section \ref{ss:chewinggum}, we expect that the (total) hyperbolic length ${D_\gamma}$  of the part(s) of the geodesic function passing through the
chewing gum will blow up. Here we are going to prove this rigorously, as well as calculating the leading term in $\epsilon$ of $D_\gamma$. We will then use this information to multiplicatively renormalise the geodesic
functions such a way that when taking the limit as $\varepsilon\to 0$ the answer is finite. If a geodesic does not pass through the chewing gum, it remains non--renormalized; if it passes more than once, then we take
${D_\gamma}$ to be the total summed up lengths of all such parts.}

{Let us estimate $D_\gamma$ in terms of $\varepsilon$}. For simplicity consider the case where a geodesic passes once through the chewing gum.
A remarkable fact is that although $D_\gamma$ tends to infinity as $\varepsilon\to 0$, for any two geodesics $\gamma_e$
and $\gamma_f$, the \emph{difference} $ D_{\gamma_e}-  D_{\gamma_f}$ tends to zero.  To prove this, observe that for any such $\gamma$
this distance satisfies the inequality $D_{12}<D_\gamma<D_{13}$ where $D_{12}$ and $D_{13}$ are hyperbolic distances
between the corresponding points in Fig.~\ref{fi:strip}. By using the formula relating hyperbolic distance ${\rm d}_{\mathbb H}(z_1,z_2)$ between two points $z_1, z_2$ to the Euclidean one $|z_1-z_2|$:
$$
\left(\sinh\frac{{\rm d}_{\mathbb H}(z_1,z_2)}{2} \right)^2=\frac{|z_1-z_2|^2}{4 \Im z_1 \Im z_2},
$$
we estimate
$$
e^{D_{12}}\sim \frac{1}{\varepsilon^2 \ell_1 \ell_2} -  \frac{(\ell_1+\ell_2)^2}{ 4 \ell_1 \ell_2} +\mathcal O(\varepsilon^2),
$$
and
$$
e^{D_{13}}\sim \frac{1}{\varepsilon^2 \ell_1 \ell_2} + \frac{4-(\ell_1+\ell_2)^2}{ 4 \ell_1 \ell_2} +\mathcal O(\varepsilon),
$$
so that
$$
e^{D_{13}-D_{12}}=1+O(\varepsilon^2)
$$
and the difference $D_{12}-D_{13}$ is of order $\varepsilon^2$.

We introduce the new variables $\pi_i$,
\be
e^{\pi_i}:=\ell_i,
\label{epi}
\ee
so that
\be
e^{D_{12}/2}=(\varepsilon)^{-1}e^{-\pi_1/2-\pi_2/2}+\mathcal O(1).
\ee
{We rescale all geodesics functions by the same factor $(\varepsilon)$ and then take the limit as $\varepsilon\to 0$. In this limit the chewing-gum breaks along the vertical interval between
$1$ and $1+\varepsilon$ on the $y$-axis thus splitting the fundamental domain in two parts, or in other words, two Riemann surfaces. The vertical interval itself becomes a point (in fact two, one for each fundamenal domain) infinitely distant from the rest of the Riemann surface, thus creating two cusps and a disconnected Riemann surface.}

{In the case where the result of breaking the chewing gum is a connected Riemann surface, as in Figs. \ref{fi:chewinggum} and \ref{fi:chewinggum2}, we can always choose a connected fundamental domain in ${\mathbb H}$ whose boundary contains two copies of the interval joining the boundaries of holes. In other words, in Fig.~\ref{fi:strip}, we need to map the part of the white strip in the right quadrant to the left hand portion of white strip between the absolute and the collar of angle $\varepsilon\ell_1$ by a hyperbolic transformation in order to obtain the fundamental domain. In particular this means that
 the vertical segment between $1$ and $1+\epsilon$ splits into two, one is mapped into itself and its copy is mapped inside the left part of the strip.}

\begin{figure}[tb]
\begin{pspicture}(-2.5,0)(2.5,8)
\newcommand{\DROPLET}
{\psset{unit=2}
\psframe[linecolor=white, fillstyle=solid, fillcolor=lightgray](-1,0)(1,3)
\psbezier[linecolor=white,fillstyle=solid,fillcolor=white](0,0)(0,1)(-1.5,2.4)(0,2.4)
\psbezier[linecolor=white, fillstyle=solid,fillcolor=white](0,0)(0,1)(1.5,2.4)(0,2.4)
\psbezier[linecolor=green,fillstyle=vlines](0,0)(0,1)(-1.5,2.4)(0,2.4)
\psbezier[linecolor=green, fillstyle=vlines](0,0)(0,1)(1.5,2.4)(0,2.4)
\psline[linewidth=1pt,linecolor=red](0,0)(0,1.3)
\psline[linewidth=1pt,linecolor=blue](0,2.4)(0,2)
\rput(0,1.7){\pscircle[linecolor=green, linewidth=1pt,fillstyle=solid, fillcolor=lightgray](0,0){0.4}
\pscircle[linecolor=white, linewidth=1pt,fillstyle=solid, fillcolor=white](0,0){0.15}
{\makebox(0,0){$2$}}}
\rput(0,2.7){\pscircle[linecolor=white, linewidth=1pt,fillstyle=solid, fillcolor=white](0,0){0.15}
{\makebox(0,0){$1$}}}
\rput(0.14,2.25){\pscircle[linecolor=white, linewidth=1pt,fillstyle=solid, fillcolor=white](0,0){0.1}
\makebox(0,0){$h$}}
\rput(0.5,2){\pscircle[linecolor=white, linewidth=1pt,fillstyle=solid, fillcolor=white](0,0){0.1}
\makebox(0,0){$P$}}
\rput(0.14,1.1){\pscircle[linecolor=white, linewidth=1pt,fillstyle=solid, fillcolor=white](0,0){0.1}
\makebox(0,0){$a$}}
\pscircle[linecolor=black, linewidth=1pt,linestyle=dashed](0,0.2){0.2}
}
\rput(0,0){\DROPLET}
\rput(2.1,2.5){\makebox(0,0)[lc]{$\sim$}}
%
%
\end{pspicture}
\begin{pspicture}(-1.5,0)(7,8)
{\psset{unit=1.5}
\newcommand{\PATTERN}[1]
{\psset{unit=#1}
\psarc[linewidth=1pt,linecolor=blue](0,0){2}{20}{90}
\psarc[linewidth=1pt,linecolor=green,fillstyle=solid, fillcolor=lightgray](2.5,0){1.5}{0}{180}
\psarc[linewidth=1pt,linecolor=red](0,0){4}{0}{90}
\pscircle[linecolor=black, linewidth=1pt,linestyle=dashed](4,0.5){0.5}
\psarc[linewidth=1pt,linecolor=blue,linestyle=dotted](0,0){2}{0}{45}
}
\psframe[linecolor=white, fillstyle=solid, fillcolor=lightgray](-1,0)(0,4.5)
\pswedge[linewidth=1pt,linecolor=white,fillstyle=vlines](0,0){2}{0}{90}
\pswedge[linewidth=1pt,linecolor=white,fillstyle=solid, fillcolor=white](0,0){.5}{0}{90}
\psline[linewidth=1pt,linecolor=green](0,0)(0,4.5)
\psarc[linewidth=1pt,linecolor=green,fillstyle=solid, fillcolor=lightgray](10,0){6}{155}{205}
\psframe[linecolor=white, fillstyle=solid, fillcolor=white](3.8,0)(4.8,-3.2)
\rput(0,0){\PATTERN{1}}
\rput(0,0){\PATTERN{0.25}}
\rput(0,0){\PATTERN{0.06}}
\psarc[linewidth=1pt,linecolor=magenta,linestyle=dashed](0,-1){2.24}{60}{90}
\psarc[linewidth=1pt,linecolor=magenta,linestyle=dotted](0,-1){2.24}{27}{60}
\psarc[linewidth=1pt,linecolor=magenta,linestyle=dashed](0,0.25){.56}{10}{90}
\psarc[linewidth=1pt,linecolor=magenta,linestyle=dotted](0,0.25){.56}{-27}{10}
\rput(-0.5,2.5){\pscircle[linecolor=white, linewidth=1pt,fillstyle=solid, fillcolor=white](0,0){0.15}
{\makebox(0,0){$2$}}}
\rput(2.5,0.6){\pscircle[linecolor=white, linewidth=1pt,fillstyle=solid, fillcolor=white](0,0){0.15}
{\makebox(0,0){$1'$}}}
\rput(0.7,0.17){\pscircle[linecolor=white, linewidth=1pt,fillstyle=solid, fillcolor=white](0,0){0.1}
{\makebox(0,0){\tiny$1$}}}
\rput(1,2){\makebox(0,0){$h$}}
\rput(2,3.8){\makebox(0,0){$a$}}
\rput(0.35,0.55){\pscircle[linecolor=white, linewidth=1pt,fillstyle=solid, fillcolor=white](0,0){0.1}
\makebox(0,0){\tiny$h$}}
\rput(0.7,0.95){\pscircle[linecolor=white, linewidth=1pt,fillstyle=solid, fillcolor=white](0,0){0.15}
\makebox(0,0){\small$a$}}
\rput(-0.1,4){\makebox(0,0)[rc]{$e^P$}}
\rput(-0.1,2){\makebox(0,0)[rc]{$e^{P/2}$}}
\rput(-0.1,1){\makebox(0,0)[rc]{$1$}}
\rput(-0.1,0.5){\makebox(0,0)[rc]{\tiny$e^{-P/2}$}}
\rput(0,0.4){
\psline[linewidth=5pt]{->}(3.5,3.5)(4,3.5)
\psline[linewidth=2pt,linecolor=white]{->}(3.4,3.5)(3.9,3.5)
\rput(3.7,3.2){\makebox(0,0){$P\to\infty$}}
}
%
}
\end{pspicture}
\begin{pspicture}(-1,0)(2,8)
{\psset{unit=1.5}
\psframe[linecolor=white, fillstyle=solid, fillcolor=lightgray](-0.5,0)(0,4.5)
\psframe[linewidth=1pt,linecolor=white,fillstyle=vlines](0,0)(1,4.5)
\psframe[linecolor=white, fillstyle=solid, fillcolor=lightgray](1,0)(1.5,4.5)
\psarc[linewidth=1pt,linecolor=green,fillstyle=solid, fillcolor=lightgray](0.5,0){.5}{0}{180}
\psline[linewidth=1pt,linecolor=green](0,0)(0,4.5)
\psline[linewidth=1pt,linecolor=green](1,0)(1,4.5)
\psarc[linewidth=1pt,linecolor=red](0,0){1}{0}{90}
\pscircle[linecolor=black, linewidth=1pt,linestyle=dashed](1,0.125){0.125}
\pscircle[linewidth=1pt,linecolor=magenta,linestyle=dotted](0,0.15){.15}
\psline[linewidth=1pt,linecolor=magenta,linestyle=dotted](-0.5,3.5)(0,3.5)
\psline[linewidth=1pt,linecolor=magenta,linestyle=dashed](0,3.5)(1,3.5)
\psline[linewidth=1pt,linecolor=magenta,linestyle=dotted](1,3.5)(1.5,3.5)
\rput(-0.25,2.5){\pscircle[linecolor=white, linewidth=1pt,fillstyle=solid, fillcolor=white](0,0){0.15}
{\makebox(0,0){$2$}}}
\rput(1.25,2.5){\pscircle[linecolor=white, linewidth=1pt,fillstyle=solid, fillcolor=white](0,0){0.15}
{\makebox(0,0){$1'$}}}
\rput(0.5,0.2){\pscircle[linecolor=white, linewidth=1pt,fillstyle=solid, fillcolor=white](0,0){0.1}
{\makebox(0,0){\small$1$}}}
\rput(0.65,1.05){\pscircle[linecolor=white, linewidth=1pt,fillstyle=solid, fillcolor=white](0,0){0.15}
\makebox(0,0){\small$a$}}
\rput(-0.1,1){\makebox(0,0)[rc]{$1$}}
%
}
\end{pspicture}
\caption{\small{ On the left picture we draw the Riemann surface $\Sigma_{0,2,1}$ scematically: $P$ is the perimeter of the inner hole, $h$ is the shortest geodesic between the inner hole and the outer arc $a$; $h$ is related to $P$ via the formula $\sinh^2(h/2)=(e^P-1)^{-1}$ and tends to zero as $P\to\infty$. In the middle picture we depict the fundamental domain (the hatched area) obtained from the left picture by cutting along $h$. Rotations of the inner hole in the Riemann surface, correspond to dilations in the upper half plane, or in other words to $z\to e^P z$. We represent the collars by dashed lines. On the right picture we present the resulting fundamental domain of $\Sigma_{0,1,3}$ after we have taken the limit as
$P\to\infty$: in this limit, we obtain two new bordered cusps located at zero and infinity whereas collars transform into horocycles based at these bordered cusps.}}
\label{fi:droplet}
\end{figure}
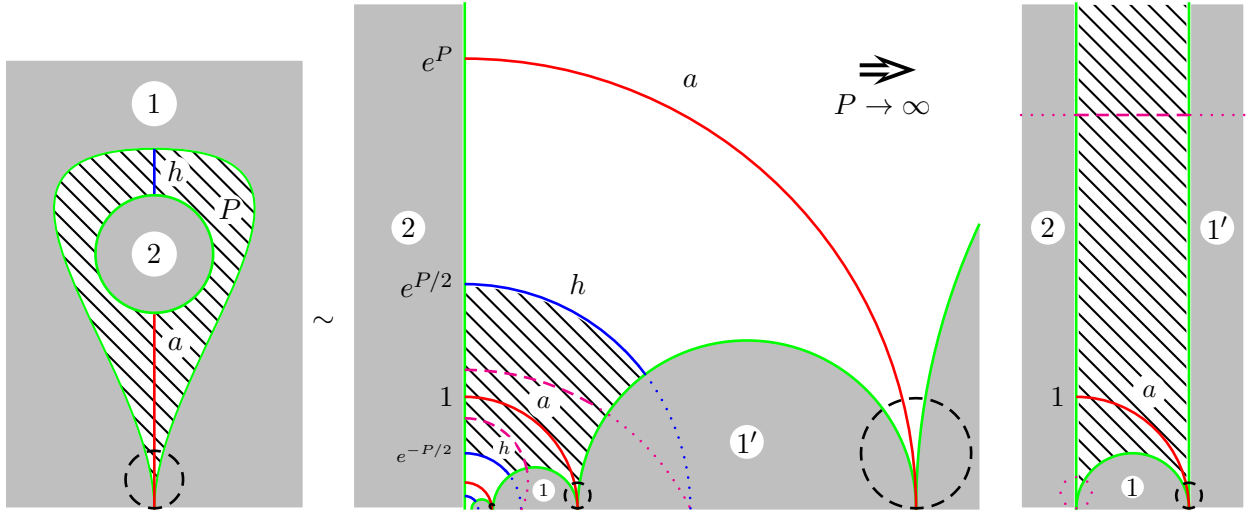

A simpler example of this procedure for $\Sigma_{0,2,1}$ is depicted in Fig.~\ref{fi:droplet}.

\begin{figure}[tb]
\begin{pspicture}(0,-2)(2,2)
{\psset{unit=1}
\rput(0,0.8){\pscircle[fillstyle=solid,fillcolor=lightgray]{.3}}
\rput(0,1.6){\pscircle[fillstyle=solid,fillcolor=lightgray]{.3}}
\rput{-30}(-0.2,0.65){\psarc[linewidth=1.5pt,linecolor=blue,linestyle=dashed](0,0){.6}{0}{180}
\psarc[linewidth=1.5pt,linecolor=blue,linestyle=dashed](0,-1.2){.6}{180}{360}
\psline[linewidth=1.5pt,linecolor=blue,linestyle=dashed](-0.6,0)(-0.6,-1.2)
\psline[linewidth=1.5pt,linecolor=blue,linestyle=dashed](0.6,0)(0.6,-1.2)
}
\rput{30}(0.2,0.65){\psarc[linewidth=1.5pt,linecolor=red,linestyle=dashed](0,0){.6}{0}{180}
\psarc[linewidth=1.5pt,linecolor=red,linestyle=dashed](0,-1.2){.6}{180}{360}
\psline[linewidth=1.5pt,linecolor=red,linestyle=dashed](-0.6,0)(-0.6,-1.2)
\psline[linewidth=1.5pt,linecolor=red,linestyle=dashed](0.6,0)(0.6,-1.2)
}
\rput(-.8,-1.4){\makebox(0,0){$\gamma_e$}}
\rput(.8,-1.4){\makebox(0,0){$\gamma_f$}}
}
\end{pspicture}
\begin{pspicture}(-2,-2)(2,2)
{\psset{unit=1}
\rput(-2,1){\makebox(0,0){${=\atop\hbox{skein}}$}}
\rput(0,0.8){\pscircle[fillstyle=solid,fillcolor=lightgray]{.3}}
\rput(0,1.6){\pscircle[fillstyle=solid,fillcolor=lightgray]{.3}}
\rput{-30}(-0.2,0.65){\psarc[linewidth=1.5pt,linecolor=green,linestyle=dashed](0,0){.6}{0}{180}
\psarc[linewidth=1.5pt,linecolor=green,linestyle=dashed](0,-1.2){.6}{180}{360}
\psline[linewidth=1.5pt,linecolor=green,linestyle=dashed](-0.6,0)(-0.6,-1.2)
\psline[linewidth=1.5pt,linecolor=green,linestyle=dashed](0.6,0)(0.6,-1.2)
}
\rput{30}(0.2,0.65){\psarc[linewidth=1.5pt,linecolor=green,linestyle=dashed](0,0){.6}{0}{180}
\psarc[linewidth=1.5pt,linecolor=green,linestyle=dashed](0,-1.2){.6}{180}{360}
\psline[linewidth=1.5pt,linecolor=green,linestyle=dashed](-0.6,0)(-0.6,-1.2)
\psline[linewidth=1.5pt,linecolor=green,linestyle=dashed](0.6,0)(0.6,-1.2)
}
\psframe[linecolor=white, fillstyle=solid, fillcolor=white](-0.4,-0.75)(0.4,0.45)
\rput(0,0){\psbezier[linewidth=1.5pt,linecolor=green,linestyle=dashed](-0.35,-.75)(0,-0.15)(0,-0.15)(0.35,-0.75)}
\rput(0,0){\psbezier[linewidth=1.5pt,linecolor=green,linestyle=dashed](-0.35,.45)(0,-0.15)(0,-0.15)(0.35,.45)}
\psframe[linecolor=white, fillstyle=solid, fillcolor=white](-0.2,1.15)(0.2,1.25)
\psline[linewidth=1.5pt,linecolor=green,linestyle=dashed](-0.2,1.15)(0.2,1.15)
\psline[linewidth=1.5pt,linecolor=green,linestyle=dashed](-0.2,1.25)(0.2,1.25)
\rput(-.4,0.1){\makebox(0,0){$\gamma_a$}}
\rput(.8,-1.4){\makebox(0,0){$\gamma_c$}}
}
\end{pspicture}
\begin{pspicture}(-1.5,-2)(1.5,2)
{\psset{unit=1}
\rput(-1.75,1){\makebox(0,0){$+$}}
\rput(1.75,1){\makebox(0,0)[lc]{$+$\hbox{\ the rest}}}
\rput(0,0.8){\pscircle[fillstyle=solid,fillcolor=lightgray]{.3}}
\rput(0,1.6){\pscircle[fillstyle=solid,fillcolor=lightgray]{.3}}
\rput{-30}(-0.2,0.65){\psarc[linewidth=1.5pt,linecolor=magenta,linestyle=dashed](0,0){.6}{0}{180}
\psarc[linewidth=1.5pt,linecolor=magenta,linestyle=dashed](0,-1.2){.6}{180}{360}
\psline[linewidth=1.5pt,linecolor=magenta,linestyle=dashed](-0.6,0)(-0.6,-1.2)
\psline[linewidth=1.5pt,linecolor=magenta,linestyle=dashed](0.6,0)(0.6,-1.2)
}
\rput{30}(0.2,0.65){\psarc[linewidth=1.5pt,linecolor=magenta,linestyle=dashed](0,0){.6}{0}{180}
\psarc[linewidth=1.5pt,linecolor=magenta,linestyle=dashed](0,-1.2){.6}{180}{360}
\psline[linewidth=1.5pt,linecolor=magenta,linestyle=dashed](-0.6,0)(-0.6,-1.2)
\psline[linewidth=1.5pt,linecolor=magenta,linestyle=dashed](0.6,0)(0.6,-1.2)
}
\psframe[linecolor=white, fillstyle=solid, fillcolor=white](-0.4,-0.75)(0.4,0.45)
\rput(0,0){\psbezier[linewidth=1.5pt,linecolor=magenta,linestyle=dashed](-0.35,-.75)(0,-0.15)(0,-0.15)(-0.35,.45)}
\rput(0,0){\psbezier[linewidth=1.5pt,linecolor=magenta,linestyle=dashed](0.35,-.75)(0,-0.15)(0,-0.15)(0.35,.45)}
\psframe[linecolor=white, fillstyle=solid, fillcolor=white](-0.2,1.15)(0.2,1.25)
\psline[linewidth=1.5pt,linecolor=magenta,linestyle=dashed](-0.2,1.15)(0.2,1.15)
\psline[linewidth=1.5pt,linecolor=magenta,linestyle=dashed](-0.2,1.25)(0.2,1.25)
\rput(-.8,-1.4){\makebox(0,0){$\gamma_b$}}
\rput(.8,-1.4){\makebox(0,0){$\gamma_d$}}
}
\end{pspicture}
\caption{\small
The skein relation applied to the geodesic functions of two geodesics ($\gamma_e$ and $\gamma_f$) having a single intersection
outside the ``chewing gum'' domain. As the result, we obtain the first term, which is a product of geodesic functions corresponding to
geodesics $\gamma_a$ and $\gamma_c$ and the geodesic function that correspond to the geodesic that splits into two geodesics,
$\gamma_b$ and $\gamma_d$, when breaking the chewing gum. ``The rest"" is a combination of geodesic  functions corresponding to
geodesics not passing through the chewing gum domain.
}
\label{fi:skein}
\end{figure}
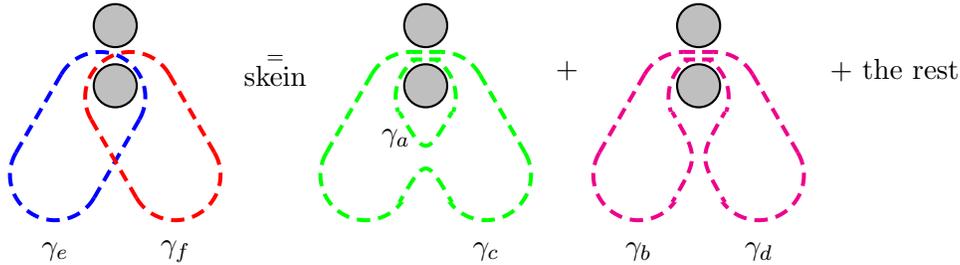

We call the new coordinates $\pi_i$ such that $e^{\pi_i}:=\ell_i$,  {\it extended shear coordinates}\/ for the decorated bordered cusps. To show that the exponentiated extended shear coordinates truly behave like $\lambda$-lengths, let us consider a useful example of a Riemann surface with at least two holes and two
geodesics, $\gamma_e$ and $\gamma_f$, passing between these two holes and having a single crossing in the rest of the Riemann surface (see Fig.~\ref{fi:skein}). The
corresponding skein relation reads
\be
G_{\gamma_e}G_{\gamma_f}=G_{\gamma_a}G_{\gamma_c}+G_{\gamma_b\gamma_d}+\hbox{ the rest},
\label{e-f}
\ee
where we let the rest denote combinations of geodesic functions for geodesics not passing through the chewing gum.

\begin{figure}[tb]
\begin{pspicture}(-4,-3)(4,2)
{\psset{unit=1}
\psarc[linewidth=1.5pt,linecolor=green,linestyle=dashed](0,2){2}{180}{360}
\psbezier[linewidth=1.5pt,linecolor=green,linestyle=dashed](-2,2)(-2,0.5)(-3.5,0.5)(-3.5,-1)
\psbezier[linewidth=1.5pt,linecolor=green,linestyle=dashed](2,2)(2,0.5)(3.5,0.5)(3.5,-1)
\psarc[linewidth=1.5pt,linecolor=green,linestyle=dashed](-1.5,-1){2}{180}{270}
\psarc[linewidth=1.5pt,linecolor=green,linestyle=dashed](1.5,-1){2}{270}{360}
\psline[linewidth=1.5pt,linecolor=green,linestyle=dashed](-1.5,-3)(1.5,-3)
\psbezier[linewidth=1.5pt,linecolor=magenta,linestyle=dashed](-2,2)(-2,0)(-2.5,0)(-2.5,-1)
\psbezier[linewidth=1.5pt,linecolor=magenta,linestyle=dashed](-2,2)(-2,0)(-1.5,0)(-1.5,-1)
\psarc[linewidth=1.5pt,linecolor=magenta,linestyle=dashed](-2,-1){0.5}{180}{360}
\psbezier[linewidth=1.5pt,linecolor=magenta,linestyle=dashed](2,2)(2,0)(2.5,0)(2.5,-1)
\psbezier[linewidth=1.5pt,linecolor=magenta,linestyle=dashed](2,2)(2,0)(1.5,0)(1.5,-1)
\psarc[linewidth=1.5pt,linecolor=magenta,linestyle=dashed](2,-1){0.5}{180}{360}
\psbezier[linewidth=1.5pt,linecolor=blue,linestyle=dashed](-2,2)(-2,0.3)(-3,0.3)(-3,-1)
\psarc[linewidth=1.5pt,linecolor=blue,linestyle=dashed](-1.7,-1){1.3}{180}{270}
\psbezier[linewidth=1.5pt,linecolor=blue,linestyle=dashed](2,2)(2,0.5)(1,-2.3)(-1.7,-2.3)
\psbezier[linewidth=1.5pt,linecolor=red,linestyle=dashed](2,2)(2,0.3)(3,0.3)(3,-1)
\psarc[linewidth=1.5pt,linecolor=red,linestyle=dashed](1.7,-1){1.3}{270}{360}
\psbezier[linewidth=1.5pt,linecolor=red,linestyle=dashed](-2,2)(-2,0.5)(-1,-2.3)(1.7,-2.3)
\rput(-2,1.5){\pscircle[linecolor=white,fillstyle=solid,fillcolor=white]{0.5}}
\rput(2,1.4){\pscircle[linecolor=white,fillstyle=solid,fillcolor=white]{0.6}}
\psclip{\pscircle[linecolor=white](-2,1.5){0.5}}
\psarc[linewidth=.5pt,linecolor=green](0,2){2}{180}{360}
\psbezier[linewidth=.5pt,linecolor=green](-2,2)(-2,0.5)(-3.5,0.5)(-3.5,-1)
\psbezier[linewidth=.5pt,linecolor=magenta](-2,2)(-2,0)(-2.5,0)(-2.5,-1)
\psbezier[linewidth=.5pt,linecolor=magenta](-2,2)(-2,0)(-1.5,0)(-1.5,-1)
\psbezier[linewidth=.5pt,linecolor=blue](-2,2)(-2,0.3)(-3,0.3)(-3,-1)
\psbezier[linewidth=.5pt,linecolor=red](-2,2)(-2,0.5)(-1,-2.3)(1.7,-2.3)
\endpsclip
\psclip{\pscircle[linecolor=white](2,1.4){0.6}}
\psarc[linewidth=.5pt,linecolor=green](0,2){2}{180}{360}
\psbezier[linewidth=.5pt,linecolor=green](2,2)(2,0.5)(3.5,0.5)(3.5,-1)
\psbezier[linewidth=.5pt,linecolor=magenta](2,2)(2,0)(2.5,0)(2.5,-1)
\psbezier[linewidth=.5pt,linecolor=magenta](2,2)(2,0)(1.5,0)(1.5,-1)
\psbezier[linewidth=.5pt,linecolor=blue](2,2)(2,0.5)(1,-2.3)(-1.7,-2.3)
\psbezier[linewidth=.5pt,linecolor=red](2,2)(2,0.3)(3,0.3)(3,-1)
\endpsclip
\rput(-2,1.5){\pscircle[linestyle=dashed,linewidth=1.5pt]{0.5}}
\rput(2,1.4){\pscircle[linestyle=dashed,linewidth=1.5pt]{0.6}}
\rput(0,.3){\makebox(0,0){$\tilde\gamma_a$}}
\rput(0,-2.7){\makebox(0,0){$\tilde\gamma_c$}}
\rput(-2,-1){\makebox(0,0){$\tilde\gamma_b$}}
\rput(2,-1){\makebox(0,0){$\tilde\gamma_d$}}
\rput(1,-0.4){\makebox(0,0){$\tilde\gamma_e$}}
\rput(-1,-0.4){\makebox(0,0){$\tilde\gamma_f$}}
}
\end{pspicture}
\caption{\small
In the limit as $\varepsilon\to 0$ we obtain the Ptolemy relation for the new geodesic functions (corresponding to arcs).
}
\label{fi:Ptolemy}
\end{figure}
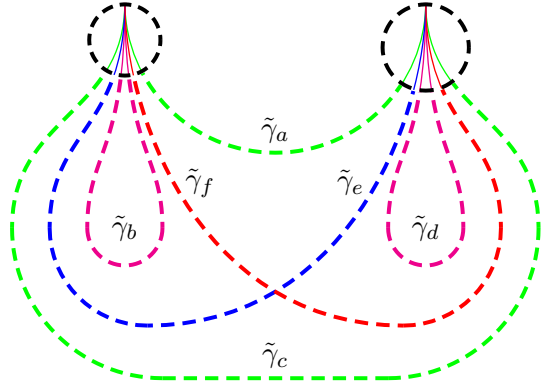

We now consider the limit as $\varepsilon\to 0$
of the skein relations  (\ref{e-f}) for two geodesics, $\gamma_e$ and $\gamma_f$ in Fig.~\ref{fi:skein}. For this, we multiply this relation
by $\varepsilon^{ 2} $ (because we have two geodesic lines passing through the chewing gum) and
take the limit as $\varepsilon\to 0$. All terms corresponding to geodesics not passing through the chewing gum vanish and
only the three terms in the figure contribute to the limiting relation. The chewing gum breaks into two \emph{bordered cusps} whereas
collar lines transform into \emph{horocycles}
decorating the corresponding cusps (see Fig.~\ref{fi:Ptolemy}). Then, for the renormalised geodesic functions,
\be
G_{\tilde\gamma}:=
\lim_{\varepsilon\to0}  \varepsilon G_{\gamma} ,
\label{renorm}
\ee
we obtain just the Ptolemy relation:
\be
G_{\tilde\gamma_e}G_{\tilde\gamma_f}=G_{\tilde\gamma_a}G_{\tilde\gamma_c}+G_{\tilde\gamma_b}G_{\tilde\gamma_d},
\label{Ptolemy}
\ee
in which the geodesic in the third term in (\ref{e-f}) now splits into two new geodesics. Note that the thus obtained geodesic
functions $G_{\tilde\gamma}$ are exponentials of halves of lengths of geodesics $\tilde\gamma$ confined between two horocycles
(which can be the same horocycle):
\be
\ell_{\tilde\gamma}=e^{D_{\tilde\gamma}/2}.
\ee
That we have obtained the Ptolemy relation should not be surprising: indeed, since collars transform into horocycles in the above limit, the
closed geodesics transform into $\lambda$-paths bounded by horocycles. What is more surprising is that we have an explicit way of doing this transition on the level of original shear coordinates.

\subsection{{Limiting matrix decomposition}}\label{ss:cusped-matrices}

Let us consider the shear coordinates $P_\alpha$ of the original fat graph ${\mathcal G}_{g,s}$ corresponding to an edge whose sides are incident to the two holes or the two sides of the same hole that we are going to merge in the collision. As discussed in sub-section \ref{ss:Ptolemy}, in the limit $\varepsilon\to 0$, $e^{P_\alpha/2}=\varepsilon e^{\pi_1/2+\pi_2/2}+\mathcal O(\varepsilon^2)$. As a result, the edge matrix associated to that edge is affected by the collision process in the following way:
\be
\varepsilon X_{P_\alpha}=\varepsilon
\biggl[
\begin{array}{cc}
0  &  -e^{P_\alpha/2}   \\
 e^{-P_\alpha/2} &   0
\end{array}
\biggr]
\sim \biggl[
\begin{array}{cc}
0  &  -\frac{e^{\pi_1/2+\pi_2/2}}{\varepsilon} \\
 \varepsilon e^{-\pi_1/2-\pi_2/2} &   0
\end{array}
\biggr] \varepsilon \mapsto\biggl[
\begin{array}{cc}
0  &  -e^{\pi_1/2+\pi_2/2}   \\
0 &   0
\end{array}
\biggr]=X_{\pi_1}K X_{\pi_2},
\label{scaling}
\ee
where
\be
K=\left(
\begin{array}{cc}
0  &  0   \\
-1 &   0
\end{array}
\right)
\label{K}
\ee
is the new matrix appearing because of the collision.
So, we must merely replace $X_{P_\alpha}$ by $X_{\pi_1} K X_{\pi_2}$.

Note that, for arbitrary $2\times 2$-matrices $F_1,\dots, F_k$,
\be
\tr \bigl( F_1 K F_2 K\cdots F_n K\bigr)=\prod_{j=1}^n \tr(F_i K),
\label{FFF}
\ee
so if an original geodesic was passing $n$ times through the chewing gum, in the limit as $\varepsilon\to 0$
 it will be partitioned into $n$ disjoint geodesics
each endowed with its own limiting geodesic function $\tr(F_i K)$.

\section{{Bordered cusped Teichm\"uller space and its combinatorial description}}\label{s:graph-bordered}
\setcounter{equation}{0}

In this section we  introduce the notion of bordered cusped Teichm\"uller space for Riemann surfaces with at least one decorated bordered cusp. We first provide a fat graph description of such Riemann surfaces. For this purpose, we need to distinguish between the holes that have a cusp on them, we denote their number by $s_c$, and the holes without cusps on them. We treat the latter on the same footing as orbifold points.

\subsection{Fat graph description for Riemann surfaces with holes, orbifold points, and decorated bordered cusps}\label{ss:fatgraph}

In this subsection, we are going to use the insights from subsection \ref{ss:cusped-matrices} to introduce the correct notion of fat graph for a Riemann surface with decorated bordered cusps.

\begin{definition}\label{def-graph-cusp}
We call {\it cusped fat graph ${\mathcal G}_{g,s,n}$} a special type of {\em spine of the Riemann surface} $\Sigma_{g,s,n}$
with genus $g$, $s$ of holes or orbifold points (at least one hole), and  $n\ge1$ decorated bordered cusps, such that it is a graph
 with a prescribed cyclic ordering of edges
entering each vertex and the following properties are satisfied:
\begin{itemize}
\item[(a)] this graph can be embedded  without self-intersections in $\Sigma_{g,s,n}$;
\item[(b)] all vertices of ${\mathcal G}_{g,s,n}$ are three-valent except exactly $n$
one-valent vertices (endpoints of the open edges), which are placed at the corresponding
bordered cusps;
\item[(c)] upon cutting along all \emph{nonopen} edges of ${\mathcal G}_{g,s,n}$ the Riemann surface
$\Sigma_{g,s,n}$ splits into $s$ polygons each containing exactly one hole or orbifold point and being
simply connected upon contracting  this hole or erasing the orbifold point.
\end{itemize}
We denote by $\Gamma_{g,s,n}$ the set of all such cusped fat graphs associated to  $\Sigma_{g,s,n}$.

We call  {\it good cusped fat graph} $\widehat \Gamma_{g,s,n}$ a cusped fat-graph such that  the polygons described in property (c)
are actually monogons for \emph{all} orbifold points and \emph{all} holes to which
no bordered cusps are associated. We denote by $\widehat \Gamma_{g,s,n}$ the set of all such good cusped fat graphs associated to  $\Sigma_{g,s,n}$.
 \end{definition}

According to this definition, for every {good cusped fat graph} $\widehat{\mathcal G}_{g,s,n}\in\widehat \Gamma_{g,s,n}$,
 every orbifold point and every hole with no associated bordered cusps is contained inside a closed loop, which is an edge
starting and terminating at the same three-valent vertex. Vice versa, every such closed loop corresponds either to an orbifold point or
to a hole with no associated bordered cusps. {We shall see that the good cusped fat graphs present good properties under the action of the Mapping Class Group, therefore we
shall mainly stick to these, distinguishing them by the ``hat'' symbol.}

Because every open edge  corresponding to a bordered cusp
``protrudes'' towards the interior of some face of the graph, and we have exactly one hole contained inside this face,
every fat graph ${\mathcal G}_{g,s,n}\in \Gamma_{g,s,n}$
determines a natural partition of the set of bordered cusps into nonintersecting (maybe empty)
subsets $\delta_k$, $k=1,\dots,s_{c}$ of cusps incident to the corresponding holes, and in every such set we have
the natural cyclic ordering coming from the orientation of the Riemann surface. We therefore have a marking on the space of
bordered cusps.

Using the above marking, we prescribe the following labelling for the  edges of a good cusped fat graph $\widehat{\mathcal G}_{g,s,n}$: to every edge that is neither open, nor a loop we set into
the correspondence a real number $Z_\alpha$; to every open edge we set into correspondence a real number $\pi_i$,
to every loop we set into the correspondence the number $\omega_j$ equal to $2\cosh (P_j/2)$ for
a hole with the perimeter $P_j\ge 0$ (the vanishing perimeter corresponds to a puncture) or equal to
$2\cos(\pi/r_j)$ for a ${\mathbb Z}_{r_j}$-orbifold point.

{We call the set of variables $\{Z_\alpha,\pi_i\}$ {\it extended shear coordinates.}\/ In subsection \ref{suse:CC-MCG} we show that they coordinatise metrics on bordered cusped Riemann surfaces and in subsection \ref{ss:Goldman} we give the Goldman bracket for them.}

\subsection{Cusped laminations}

In this section we introduce the concept of cusped laminations, loosely speaking collections of closed geodesics and geodesic arcs that do not intersect in the interior of the Riemann surface. We show that when we have at least one bordered cusp, there always exist a {\it complete} cusped geodesic lamination, namely a lamination such that any geodesic function associated to a closed geodesic or any $\lambda$--length of a geodesic arc in the Riemann surface is obtained as a Laurent polynomial in terms of the geodesic functions associated to closed geodesics and $\lambda$--lengths of geodesic arcs belonging to the complete cusped geodesic lamination.

\begin{definition}\label{def:geom-laminations}
We call geometric
\emph{cusped geodesic lamination} (CGL) on a bordered cusped Riemann surface a set of non-directed curves up to a homotopy equivalence such that
\begin{itemize}
\item[(a)] these curves are either closed curves ($\gamma$) or \emph{arcs} ($\mathfrak a$) that start and terminate at bordered cusps
(which can be the same cusp);
\item[(b)] these curves have no (self)intersections inside the Riemann surface (but can be incident to the same bordered cusp);
\item[(c)] these curves are contractible neither to a point in the interior of the Riemann surfaces nor to a cusp.
\end{itemize}
\end{definition}

Note that in each thus defined CGL sets of ends of arcs entering the same bordered cusp are \emph{linearly ordered} w.r.t. the orientation of the
Riemann surface.

We now set an algebraic cusped geodesic lamination into correspondence to its geometric counterpart.

\begin{definition}\label{def:alg-lamination}
The algebraic CGL corresponding to a geometric CGL is
\be
\prod_{\gamma\in CGL} (2\cosh (l_\gamma/2))\prod_{\mathfrak a\in CGL}e^{l_{\mathfrak a}/2}:=\prod_{\gamma\in CGL}G_\gamma
\prod_{\mathfrak a\in CGL} G_{\mathfrak a}
\label{GL}
\ee
where $l_\gamma$ are the geodesic lengths of the corresponding closed curves and $l_{\mathfrak a}$ are the signed
geodesic lengths of the parts of arcs $\mathfrak a$ contained between two horocycles decorating the corresponding bordered
cusps (or between the same horocycle if the arc starts and ends in the same cusp); the sign is negative when these horocycles intersect. The geodesic functions $G_\gamma=2\cosh (l_\gamma/2)$ for closed
curves, as before, and $G_{\mathfrak a}=e^{l_{\mathfrak a}/2}$ for arcs.
\end{definition}

\begin{remark}\label{lambda-length}
We stress that the functions $e^{l_{\mathfrak a}/2}$, associated to the arcs $\mathfrak a$ in the CGL are nothing but $\lambda$-lengths on the corresponding bordered cusped Riemann surfaces (see \cite{FST,FT}).
\end{remark}

\begin{definition}\label{def:maximumarc}
We call a \emph{maximum arc CGL,} denoted by CGL$_{\mathfrak a}^{\text{max}}$,
of a bordered cusped Riemann surface $\Sigma_{g,s,n}$ with $s>0$ and $n>0$ the collection of
all edges of the following ideal-triangle decomposition of $\Sigma_{g,s,n}$: (a) all edges of ideal triangles terminate at bordered cusps; (b)
all holes/orbifold points that do not contain bordered cusps are contained in monogons; (c) the rest of the surface obtained by eliminating
all monogons containing holes/orbifold points without bordered cusps is partitioned into ideal triangles (all three sides of every such triangle are necessarily distinct).
\end{definition}

\begin{proposition}\label{def:dual}
Given a fat graph $\widehat{\mathcal G}_{g,s,n}\in \widehat\Gamma_{g,s,n}$ with $n>0$ we have a unique CGL$_{\mathfrak a}^{\text{max}}$
\emph{dual} to $\widehat{\mathcal G}_{g,s,n}$.
\end{proposition}

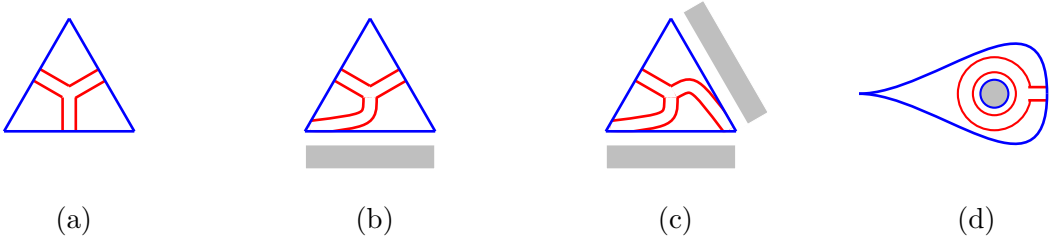
\begin{figure}[tb]
\begin{pspicture}(-6,-2)(6,2){
\newcommand{\TRI}{%
{\psset{unit=1}
\rput{0}(0,0){\psline[linewidth=5pt,linecolor=white,linestyle=solid](-0.866,-0.6)(0.866,-0.6)
\psline[linewidth=1pt,linecolor=blue,linestyle=solid](-0.866,-0.5)(0.866,-0.5)
}
\rput{120}(0,0){\psline[linewidth=5pt,linecolor=white,linestyle=solid](-0.866,-0.6)(0.866,-0.6)
\psline[linewidth=1pt,linecolor=blue,linestyle=solid](-0.866,-0.5)(0.866,-0.5)
}
\rput{240}(0,0){\psline[linewidth=5pt,linecolor=white,linestyle=solid](-0.866,-0.6)(0.866,-0.6)
\psline[linewidth=1pt,linecolor=blue,linestyle=solid](-0.866,-0.5)(0.866,-0.5)
}
\pscircle[linecolor=white,fillstyle=solid,fillcolor=white](0,0){0.05}
}
}
\newcommand{\LINE}{%
{\psset{unit=1}
\rput(0,0){\psline[linewidth=6pt,linecolor=red](0,-0.05)(0,-0.5)
\psline[linewidth=4pt,linecolor=white](0,0)(0,-0.55)}
}
}
\newcommand{\LINEBORD}{%
{\psset{unit=1}
\rput(0,0){\psbezier[linewidth=6pt,linecolor=red](0,-0.05)(0,-0.35)(0,-0.35)(-0.8,-0.45)
\psbezier[linewidth=4pt,linecolor=white](0,-0.05)(0,-0.35)(0,-0.35)(-0.8,-0.45)
\psframe[linewidth=0pt, linecolor=white, fillstyle=solid, fillcolor=lightgray](-0.86,-1)(0.86,-0.6)
}
}
}
\newcommand{\MONO}{%
{\psset{unit=1}
\pscircle[linecolor=red](0.3,0){0.3}
\pscircle[linecolor=red](0.3,0){0.5}
\pscircle[linecolor=blue,fillstyle=solid,fillcolor=lightgray](0.3,0){0.2}
\psline[linewidth=6pt,linecolor=red](0.75,0)(1,0)
\psline[linewidth=4pt,linecolor=white](0.7,0)(1.05,0)
\psbezier[linewidth=1pt,linecolor=blue](-1.5,0)(-0.5,0)(1,1.5)(1,0)
\psbezier[linewidth=1pt,linecolor=blue](-1.5,0)(-0.5,0)(1,-1.5)(1,0)
}
}
\rput(-6,0){\LINE}
\rput(-6.1,0.05){\rput{120}(0,0){\LINE}}
\rput(-6.1,-0.05){\rput{240}(0,0){\LINE}}
\rput(-6,0){\TRI}
\rput(-6,-1.5){\makebox(0,0)[ct]{(a)}}
\rput(-2,0){\LINEBORD}
\rput(-2.1,0.05){\rput{120}(0,0){\LINE}}
\rput(-2.1,-0.05){\rput{240}(0,0){\LINE}}
\rput(-2,0){\TRI}
\rput(-2,-1.5){\makebox(0,0)[ct]{(b)}}
\rput(2,0){\LINEBORD}
\rput(1.9,0.05){\rput{120}(0,0){\LINEBORD}}
\rput(1.9,-0.05){\rput{240}(0,0){\LINE}}
\rput(2,0){\TRI}
\rput(2,-1.5){\makebox(0,0)[ct]{(c)}}
\rput(6,0){\MONO}
\rput(6,-1.5){\makebox(0,0)[ct]{(d)}}
}
\end{pspicture}
\caption{\small Four types of tiles used for tessellating a Riemann surface $\Sigma_{g,s,n}$. Every vertex is situated at a bordered cusp (these cusps may coincide for a single ideal triangle); for simplicity we do not draw the horocycles decorating the bordered cusps. 
Every inner side is glued to an inner side of another ideal triangle. We then glue the half-edges of a fat graph incident to these two inner sides into a single edge. An ideal triangle may have zero (case (a)), one (case (b)), or two (case (c)) outer sides marked gray in the picture. Outer ends of the corresponding edges of the fat graph slip to the nearest-to-the-right bordered cusp (cases (b) and (c)). Case (d) represents a monogon containing a hole/orbifold point without bordered cusps. The edge bordering a monogon is always inner. In such a tesselation, any arc (with or without self-intersections) is homotopic to a unique path in the thus constructed fat graph.}
\label{fi:tiles}
\end{figure}

Before proving this proposition, we give a brief combinatorial description of how we pass from ideal triangle decomposition to fat graphs and vce-versa. We exclude the exceptional cases of non stable hyperbolic Riemann surfaces $\Sigma_{0,2,1}$ (single monogon) and $\Sigma_{0,1,3}$ (single ideal triangle).
We tessellate a Riemann surface $\Sigma_{g,s,n}$ with $n\ge 1$ by tiles of four sorts depicted in Fig.~\ref{fi:tiles}: ideal triangles with distinct sides among which zero, one, or two sides may be outer sides bordering cusped holes (Cases (a), (b), and (c)) and monogons (Case (d)) each containing exactly one hole/orbifold point without bordered cusps on it.  Outer ends of edges of the fat graph incident to outer sides slip to the nearest-to-the-right bordered cusp (Cases (b) and (c)). Then, for any arc with ends at bordered cusps we have a unique path in the fat graph that is homotopic to this arc.

\proof Given a good cusped fat-graph $\widehat{\mathcal G}_{g,s,n}$, we  embed it
into a Riemann surface $\Sigma_{g,s,n}$. We then partition $\Sigma_{g,s,n}$
into ideal triangles and monogons (containing holes/orbifold points without bordered cusps) in such a way that every internal edge of $\widehat{\mathcal G}_{g,s,n}$ that is not a loop intersects with exactly one (internal) arc from
the triangulation, every outer edge of $\widehat{\mathcal G}_{g,s,n}$ terminates at its own bordered cusp (and we set the bordered arc to the left of this cusp in correspondence to this edge), and every loop is homeomorphic to its $\omega$-cycle from this triangulation. The elements of this triangulation constitute the
CGL$_{\mathfrak a}^{\text{max}}$ that is dual to $\widehat{\mathcal G}_{g,s,n}$ .
Vice versa, we obtain the fat graph $\widehat{\mathcal G}_{g,s,n}\in\widehat\Gamma_{g,s,n}$ (with $s>0$ and $n>0$)
dual to a maximum arc CGL$_{\mathfrak a}^{\text{max}}$
as follows: we set three-valent vertices into correspondence to every monogon with
a hole/orbifold point and to every ideal triangle placing these vertices inside the corresponding monogons (but outside
holes contained inside monogons) and triangles. We then draw loops (edges starting and terminating at the same vertex) around holes/orbifold points
inside monogons; all other internal edges of $\widehat{\mathcal G}_{g,s,n}$
joint pairwise distinct neighboring three-valent vertices; each edge crosses exactly one arc from
CGL$_{\mathfrak a}^{\text{max}}$ and we have exactly $n$ ``outer'' edges starting at three-valent vertices of
$\widehat{\mathcal G}_{g,s,n}$ and
terminating at the bordered cusps; these edges correspond to $n$ bordering arcs (framing holes with bordered cusps), and
for each such arc the corresponding edge terminates at the ``right'' bordered cusp incident to this arc (when looking from inside the
Riemann surface). All edges of this fat graph are endowed with real numbers (shear coordinates):
edges that are neither loops nor ``outer'' edges (terminating at bordered cusps) are endowed with $Z_\alpha\in {\mathbb R}$, ``outer'' edges are endowed with $\pi_j\in {\mathbb R}$, and loops are endowed with $\omega_i=2\cosh(P_i/2)$, $P_i\in {\mathbb R}$,
for holes and $\omega_i=2\cos(\pi/p_i)$, $p_i\in {\mathbb Z}_{\ge 2}$, for orbifold points. \endproof

We now describe the geometric meaning of the new shear coordinates $\pi_i$ associated with decorated bordered cusps.
In the ideal triangle decomposition dual to a fat graph ${\widehat{\mathcal G}}_{g,s,n}\in\widehat\Gamma_{g,s,n}$,
we establish a 1-1 correspondence between arcs and all edges of the
graph that are not loops. For inner edges and edges adjacent to loops, the correspondence is as in
Sec.~\ref{s:preliminaries} while we set into correspondence to
an open edge terminating at a decorated bordered cusp the edge of the ideal triangle that borders the surface to the left (if looking from inside the surface) from this bordered cusp. We therefore allow arcs between neighbouring cusps to be included into laminations; these arcs actually play important role in our construction.

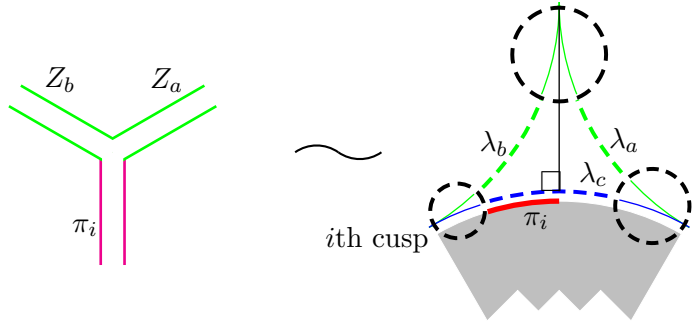
\begin{figure}[tb]
\begin{pspicture}(-4,-2.5)(4,2.5){
\newcommand{\DISC}{%
{\psset{unit=1.3}
\rput{30}(0,0){\psarc[linewidth=1.5pt,linecolor=green,linestyle=dashed](3,0){2.6}{150}{210}}
\rput{150}(0,0){\psarc[linewidth=1.5pt,linecolor=green,linestyle=dashed](3,0){2.6}{150}{210}}
\rput{270}(0,0){\psarc[linewidth=1.5pt,linecolor=blue,linestyle=dashed](3,0){2.6}{150}{210}}
\rput{90}(0,0){\pscircle[linecolor=white,fillstyle=solid,fillcolor=white](1,0){0.5}}
\rput{-30}(0,0){\pscircle[linecolor=white,fillstyle=solid,fillcolor=white](1.1,0){0.4}}
\rput{210}(0,0){\pscircle[linecolor=white,fillstyle=solid,fillcolor=white](1.2,0){0.3}}
\rput{90}(0,0){
\psclip{\pscircle[linecolor=white](1,0){0.5}}
\rput{-60}(0,0){\psarc[linewidth=.5pt,linecolor=green](3,0){2.6}{150}{180}}
\rput{60}(0,0){\psarc[linewidth=.5pt,linecolor=green](3,0){2.6}{180}{240}}
\endpsclip
}
\rput{-30}(0,0){
\psclip{\pscircle[linecolor=white](1.1,0){0.4}}
\rput{60}(0,0){\psarc[linewidth=.5pt,linecolor=green](3,0){2.6}{180}{210}}
\rput{-60}(0,0){\psarc[linewidth=.5pt,linecolor=blue](3,0){2.6}{120}{180}}
\endpsclip
}
\rput{-150}(0,0){
\psclip{\pscircle[linecolor=white](1.2,0){0.3}}
\rput{-60}(0,0){\psarc[linewidth=.5pt,linecolor=green](3,0){2.6}{150}{180}}
\rput{60}(0,0){\psarc[linewidth=.5pt,linecolor=blue](3,0){2.6}{180}{240}}
\endpsclip
}
\rput{-90}(0,0){\pswedge[linewidth=0pt,linecolor=white,fillstyle=solid, fillcolor=lightgray](3,0){2.5}{150}{210}}
\rput{270}(0,0){\psarc[linewidth=2pt,linecolor=red](3,0){2.5}{180}{197}}
\rput{-45}(0,0){
\psframe[linewidth=0pt, linecolor=white, fillstyle=solid, fillcolor=white](0.7,-2.9)(2.3,-1.3)
\psframe[linewidth=0pt, linecolor=white, fillstyle=solid, fillcolor=white](1,-2.6)(2.6,-1)
\psframe[linewidth=0pt, linecolor=white, fillstyle=solid, fillcolor=white](1.3,-2.3)(2.9,-0.7)
}
\rput{90}(0,0){\pscircle[linecolor=black,linewidth=1.5pt,linestyle=dashed](1,0){0.5}}
\rput{-30}(0,0){\pscircle[linecolor=black,linewidth=1.5pt,linestyle=dashed](1.1,0){0.4}}
\rput{-150}(0,0){\pscircle[linecolor=black,linewidth=1.5pt,linestyle=dashed](1.2,0){0.3}}
\rput(0,0){\psline[linewidth=0.5pt](0,1.5)(0,-0.4)}
\rput(0,0){\rput(0.03,-0.4){\makebox(0,0)[rb]{$\square$}}}
}
}
\rput(-3,0){
\rput{-90}(0,0){\psline[linewidth=10pt,linecolor=magenta](0,0)(1.5,0)}
\rput{30}(0,0){\psline[linewidth=10pt,linecolor=green](0,0)(1.5,0)}
\rput{150}(0,0){\psline[linewidth=10pt,linecolor=green](0,0)(1.5,0)}
\rput{-90}(0,0){\psline[linewidth=8pt,linecolor=white](0,0)(1.6,0)}
\rput{30}(0,0){\psline[linewidth=8pt,linecolor=white](0,0)(1.6,0)}
\rput{150}(0,0){\psline[linewidth=8pt,linecolor=white](0,0)(1.6,0)}
}
\rput{55}(0,0){
\psarc(0.5,0){0.5}{180}{250}
\psarc(-0.5,0){0.5}{0}{70}
}
\rput(3,0){\DISC
}
\rput(-3,0){
\rput(-0.5,.8){\makebox(0,0)[rb]{$Z_b$}}
\rput(0.5,0.8){\makebox(0,0)[lb]{$Z_a$}}
\rput(-0.2,-1){\makebox(0,0)[rc]{$\pi_i$}}
}
\rput(3,0){
\rput(.6,0){\makebox(0,0)[lb]{$\lambda_a$}}
\rput(-0.75,-0.05){\makebox(0,0)[rb]{$\lambda_b$}}
\rput(0.2,-0.45){\makebox(0,0)[lb]{$\lambda_c$}}
\rput(-0.2,-0.8){\makebox(0,0)[rt]{$\pi_i$}}
\rput(-1.8,-1){\makebox(0,0)[rt]{$i$th cusp}}
}
}
\end{pspicture}
\caption{\small The shear coordinate of a open edge corresponding to the $i$th decorated bordered cusp.
On the left-hand side we present a part of a fat graph with the open edge endowed with the
variable $\pi_i$; on the right-hand side we present
the corresponding ideal triangle whose side corresponding to $\pi_i$ borders a hole (other sides can also border holes or can be
adjacent to loops), the shear coordinate of this edge is $e^{\pi_i}=(\lambda_c\lambda_b/\lambda_a)$ and
it is sensitive only to the decoration at the $i$th bordered cusp.}
\label{fi:cusp}
\end{figure}

Explicitly, $\pi_i$ is the geodesic distance between the perpendicular to the ``outer'' side of the
triangle through the vertex opposite to this side and horocycle decorating the
$i$th cusp (see Fig.~\ref{fi:cusp}). The relation to $\lambda$-lengths of edges of this triangle is
\be
e^{\pi_i}=\frac{\lambda_c\lambda_b}{\lambda_a},
\label{pi-lambda}
\ee
where the edges $a$, $b$, and $c$ are as in the figure. The new shear coordinates are therefore hybrids of genuine shear coordinates and
$\lambda$-lengths being distances between perpendiculars and horocycles.

\subsection{Bordered cusped Teichm\"uller space and Mapping Class Group}\label{suse:CC-MCG}
We are now ready to define the bordered cusped Teichm\"uller space:

\begin{definition}\label{def-Teich}
The {\it bordered cusped Teichm\"uller space} $\widehat{\mathfrak T}_{g,s,n}$ of Riemann surfaces of genus $g$, $s$ of holes or orbifold points (at least one hole), and  $n\ge1$ decorated bordered cusps is defined by
$$
\widehat{\mathfrak T}_{g,s,n}:={\mathbb R}^{6g-6+2 s+ s_{c}+2n}\times \Omega^{s-s_c},
$$
where $s_{c}$ is the number of holes with cusps on them.
\end{definition}

The following result links points in the  bordered cusped Teichm\"uller space to conformal classes of metrics on a Riemann surface with bordered cusps:

\begin{proposition}
For any point in $\{\pi_i,Z_\alpha,\omega_j\}\in\widehat{\mathfrak T}_{g,s,n}$ there exists a Riemann surface $\Sigma_{g,s,n}$ that is smooth everywhere except a finite set of orbifold points, is endowed with a metric of constant curvature $-1$, and such that the distribution of decorated bordered cusps into boundary components is fixed,  and, vice versa, for any Riemann surface $\Sigma_{g,s,n}$ having the constant-curvature metric everywhere except a finite set of orbifold points we have a (nonunique) point in
$\widehat{\mathfrak T}_{g,s,n}$  such that all the lengths of all geodesics and (decorated) arcs are given in terms of
extended shear coordinates $\{\pi_i,Z_\alpha,\omega_j\}\in \widehat{\mathfrak T}_{g,s,n}$.
\end{proposition}

\proof
The proof is constructive and its idea is the same as in the one for Riemann surfaces without bordered cusps in \cite{ChSh}. We begin with an obvious observation that every surface $\Sigma_{g,s,n}$ (still without decorations at the bordered cusps) can be glued out of ideal triangles and monogons containing holes without boundary cusps and orbifold points based at the bordered cusps (the boundaries of these triangles and monogons are elements of a CGL$_{\mathfrak a}^{\text{max}}$ from Definition~\ref{def:maximumarc}); exactly $n$ sides of triangles remain unpaired; all gluings along internal edges are completely fixed by the choice of standard shear coordinates $Z_\alpha$ (which can be arbitrary real numbers), and, vice versa, for any set of shear coordinates we have exactly one (up to overall automorphisms by Denh twists) way to glue the surface. All ideal triangles and monogons are endowed with the hyperbolic metric, and all gluings along sides of triangles/monogons are smooth w.r.t. this metric. Introducing decorations we introduce simultaneously $n$ extended shear coordinates $\pi_j$ and, vice versa, for any set of the extended shear coordinates we have exactly one decoration by horocycles.
 \endproof

The Mapping Class Group (MCG) acts on the fat-graphs by flipping edges. We consider only MCG transformations on $\Gamma_{g,s,n}$ that are dual to generalized cluster, or Ptolemy, transformations on the set of CGL$_{\mathfrak a}^{\text{max}}$. The elementary move, or flip, of such a transformation corresponds either to flipping the fat-graph inner edge that is not a loop (see Fig.~\ref{fi:flip}) or to a special flip of an edge that is a loop (see Fig.~\ref{fi:interchange-p-dual}). {We never flip an open edge - this means that we do not flip bordered cusps.} If after a series of such flips we obtain a graph of the same combinatorial type as the starting one and with the same labelling of bordered cusps, we can always associate this series of MCG transformations with the Dehn twist transformation along some closed geodesic curve.

\begin{remark}
{We can relax the condition that the polygons containing holes must be monogons,
enlarging therefore the set of mapping class group
transformations. }This would be in line with \cite{FST,FT} where tagged cluster varieties were introduced.
The aim to impose this restriction in this paper is twofold: first,
it allows considering holes without cusps and orbifold points on an absolutely equal footing based on generalised cluster transformations of \cite{ChSh}. Second, in this case, we are able to establish an isomorphism between the sets of shear coordinates
of $\widehat{\mathcal G}\in\widehat\Gamma_{g,s,n}$
and cluster variables in an explicit and simple way, which enables us to quantize the corresponding cluster variables in Sec.~\ref{s:q}.
\end{remark}

In the case where we consider the set of fat graphs
$\widehat\Gamma_{g,s,n}$ with $n>0$
all morphisms of shear coordinates on $\widehat{\mathfrak T}_{g,s,n}$
are generated by flips (mutations) of inner edges (not adjacent to loops and not open) described by formula (\ref{abc}),
by flips (mutations) of edges adjacent to loops described by Lemma~\ref{lem-pending1},
and neither loop edges nor open edges are allowed to mutate.
This set of morphisms obviously acts inside the set
$\widehat\Gamma_{g,s,n}$ provided we have at least one bordered cusp.

\begin{remark}\label{notching}
Restricting the set of admissible spines to that of $\widehat\Gamma_{g,s,n}$
we (intentionally) impose restrictions on the set of morphisms; one
reason is that in this presentation we want to avoid further complications (and inflating the text volume)
related to introducing \emph{notching} of edges (see \cite{FST}, \cite{FT}).
thus postponing developing corresponding cluster structures to subsequent publications. Another important
reason is that only for $\widehat{\mathcal G}_{g,s,n}\in\widehat\Gamma_{g,s,n}$ all dual cluster variables correspond to arcs between decorated
bordered cusps (\emph{regular arcs} in terminology of \cite{FT}). Although the extended shear coordinates $\{\pi_i,Z_\alpha\}$
are well defined for any ${\mathcal G}_{g,s,n}\in \Gamma_{g,s,n}$ and admit the Poisson algebras and quantization, only $\lambda$-lenghts
of ordinary arcs, not those of tagged arcs, can be expressed in terms of these extended shear coordinates (for any spine from
$\widehat\Gamma_{g,s,n}$) enabling us to derive Poisson and quantum algebras for these $\lambda$-lengths.
On the other hand, notching the edges terminating at holes without bordered cusps corresponds (see \cite{FST}, \cite{FT}) to
enlarging the lamination sets including geodesic lines winding to the geodesic boundaries of the corresponding holes (the notching then corresponds to choosing the winding direction). But because we obtain our system of lambda-lengths from colliding holes of an original
Riemann surface endowed with a set of closed geodesic lines (which degenerate into arcs in the bordered cusped Riemann surfaces), we do not have geodesic lines winding to holes in the original formulation and we do not expect their appearance in the final expressions.
\end{remark}

\begin{remark}
The tropical (projective) limit of hole confluence process was considered by Fock and Goncharov in \cite{FG3}. However, because no decoration is needed in the projective limit, in the approach of \cite{FG3}, actual dimension of laminations decreases in the process of confluence; on the contrary, in our approach it increases, not decreases. On the level of laminations, this means that we also allow curves in CGLs that go parallel to boundary curves joining neighbor borderer cusps.
\end{remark}

\subsection{Quantitative description of CGL$_{\mathfrak a}^{\text{max}}$ and properties of the bordered Teichm\"uller space}\label{suse-mat-dec}
In this subsection, we provide a quantitative description of the algebraic geodesic laminations that allows to characterise the  bordered Teichm\"uller space as certain quotient of a representation space (similar to the noncusped case).

As in Sec.~\ref{s:preliminaries} we evaluate
geodesic functions as traces of products of $2\times 2$-matrices by the following rules.

\begin{itemize}
\item We first choose the direction on a path or on an arc (the final result does not depend on this choice). We also choose the starting edge: for a closed path it can be any edge, for an arc we choose it to be an open edge of the ``starting'' bordered cusp.
\item We take the product of $2\times 2$-matrices from right to left accordingly to the chosen direction: {if we start  at an open edge labelled by $i$, we insert the \emph{edge matrix} $X_{\pi_i}$ (\ref{XZ}), where $\pi_i$ is the extended shear
coordinate associated to the $i$-th cusp. If we start at a closed edge, and
every time the path goes through
an internal edge (labeled $\alpha$) not incident to a loop we insert the \emph{edge matrix} $X_{Z_\alpha}$ (\ref{XZ}), where $Z_\alpha$ is the shear
coordinate of the corresponding edge;}
every time it makes a right or left
turn at a three-valent vertex not incident to a loop we insert the corresponding matrices $R$ and $L$ (\ref{R}); every time it goes
along the edge (with the shear coordinate $Z_\beta$) incident to a loop, then along the loop with the parameter $\omega$, then back along the same edge, we introduce the product of matrices
$X_{Z_\beta}F_\omega X_{Z_\beta}$ if the path goes (once) clockwise along the loop or the product of matrices
$X_{Z_\beta}(-F_\omega^{-1}) X_{Z_\beta}$ if it goes (once) counterclockwise $F_\omega$ is defined in (\ref{F-p});
if a path makes more than one tour along the loop, it intersects itself and cannot enter a CGL.
\item When a path corresponding to an arc terminates at a decorated bordered cusp we insert the matrix $K$ (\ref{K}). We obtain the
geodesic functions $G_\gamma$ and $G_{\mathfrak a}$ by taking the traces of the thus constructed products of matrices.
\end{itemize}

{We are now going to use this description to show that the notion of bordered cusped Teichm\"uller space is the same as the real slice of the {\it decorated character variety} introduced in \cite{CMR}. Let us remind here that construction: topologically speaking a Riemann surface  $\Sigma_{g,s,n}$  of genus $g$ with $s$ holes/orbifold points and $n$ bordered cusps is equivalent to a Riemann surface $\tilde\Sigma_{g,s,n}$ of the same genus $g$, with the same number $s$ of holes/orbifold points and $n$ marked points $m_1,\dots,m_n$ on the boundaries. On $\tilde\Sigma_{g,s,n}$  the  {\it fundamental groupoid of arcs}\/ $\pi_{\mathfrak a}(\tilde \Sigma_{g,s,n})$ is well defined as the set of all directed paths $\mathfrak a_{ij}:[0,1]\to\tilde \Sigma_{g,s,n}$ such that $\mathfrak a_{ij}(0)=m_i$ and $\mathfrak a_{ij}(1)=m_j$ modulo homotopy. The groupoid structure is dictated by the usual path--composition rules. }

{\begin{proposition}
The {\it bordered cusped Teichm\"uller space} $\widehat{\mathfrak T}_{g,s,n}$ of Riemann surfaces of genus $g$, $s$ of holes or orbifold points (at least one hole), and  $n>1$ decorated bordered cusps is the  real slice of the {\it decorated character variety} \cite{CMR}:
$$
{\rm Hom}\left(
\pi_{\mathfrak a}( \tilde\Sigma_{g,s,n}),SL_2(\mathbb R)
\right)\Bigm\slash {\prod_{j=1}^n U_j},
$$
where every $U_j$ is a unipotent Borel subgroup in $SL_2(\mathbb R)$ (one one-dimensional Borel subgroup for each cusp).
\end{proposition}}

\proof
{First of all let us observe that the real dimension of the real slice of the  decorated character variety is $6g-6+3s+2n$. In fact, let us fix a bordered cusp as the base point $c_0$, we have $2g$ matrices for the usual $A$- and $B$-cycles starting and terminating at $c_0$, $s-1$ matrices corresponding to going around all holes except the one to
which the cusp $c_0$ belongs, $n-1$ matrices corresponding to paths starting at $c_0$ and terminating at other
cusps. Each matrix in $SL_2(\mathbb R)$ depends on three independent complex coordinates, giving
$3(2g+s-1+n-1)$, by taking the quotient by ${\prod_{j=1}^n B_j}$, which eliminates $n$ degrees of freedom, we obtain the final result. }

{Let us show that the extended shear coordinates and $\omega$--cycles in the good cusped fat-graph associated to $\Sigma_{g,s,n}$ are coordinates of points in the  real slice of the decorated character variety. Using the decoration by a horocycle at each cusp, one can associate to each arc $\mathfrak a_{ij}$ in the fundamental groupoid of arcs $\pi_{\mathfrak a}( \Sigma_{g,s,n})$ a matrix  $\gamma_{ij}\in SL_2(\mathbb R)$ according to the rules outlined at the beginning of this section. These matrices are by construction matrix functions of the extended shear coordinates and $\omega$--cycles in the good cusped fat-graph associated to $\Sigma_{g,s,n}$ as we wanted to prove.
Since the extended shear coordinates and $\omega$--cycles  are independent and there are $6g-6+2 s+ s_{c}+2n+s-s_c$ of them, the result follows.}
\endproof

Another consequence of the matrix decomposition rules is the following inversion formula expressing $\lambda$-lengths of arcs from CGL$_{\mathfrak a}^{\text{max}}$ in terms of the
extended shear coordinates of the dual fat graph:

\begin{theorem}\label{prop:monoidal}
Given a CGL$_{\mathfrak a}^{\text{max}}$ and its dual fat graph $\widehat{\mathcal G}_{g,s,n}$,
every arc ${\mathfrak a}_{\alpha}\in $ CGL$_{\mathfrak a}^{\text{max}}$ intersects exactly one edge of
$\widehat{\mathcal G}_{g,s,n}$ labeled $\alpha$ and carrying the shear coordinate $Z_\alpha$; denote the shear coordinate of the cusp at which the arc
${\mathfrak a}_{\alpha}$ starts by $\pi_1^{(\alpha)}$ and the one at which it ends by $\pi_2^{(\alpha)}$ (note that $\pi_1^{(\alpha)}$ and $\pi_2^{(\alpha)}$ may coincide), then
this arc is obtained by starting at the cusp  $\pi_1^{(\alpha)}$, turning left to an edge denoted by $Z_{i_1}$, then left again to an edge denoted by $Z_{i_2}$ and again until the edge $Z_\alpha$ is reached, then turning right to an edge denoted by $Z_{j_1}$, then turning right again to an edge denoted by $Z_{j_2}$ and again and again until the final cusp with coordinate $\pi_2^{(\alpha)}$ is reached. Correspondingly the $\lambda$-length $\lambda_{\alpha}$ of the arc ${\mathfrak a}_{\alpha}$  is given by:
\bea
\lambda_{\alpha}&=&\tr \Bigl[ K X_{\pi_2^{(\alpha)}} R X_{Z_{j_n}} \cdots R X_{Z_{j_k}} F_{\omega_{j_k}} X_{Z_{j_k}} R \cdots
R X_{Z_{j_1}} R X_{Z_\alpha}  L\cdots L X_{Z_{i_r}} F_{\omega_{i_r}}X_{Z_{i_r}}L \cdots  L X_{Z_{i_1}} L X_{\pi_1^{(\alpha)}}\Bigr]
\nonumber\\
&=& \exp \bigl[  (\pi_1^{(\alpha)} + \pi_2^{(\alpha)}  +Z_{i_1}+Z_{i_2}+\cdots + 2Z_{i_r}+\cdots  +Z_\alpha+ Z_{j_1}+Z_{j_2}+\cdots +2Z_{j_k} +\cdots +Z_{j_n})/2\bigr]
\nonumber
\eea
if the $\alpha$ edge is not incident to a loop and by
\bea
\lambda_{\alpha}&=&\tr \Bigl[ K X_{\pi_1^{(\alpha)}} R X_{Z_1} R X_{Z_2} R \cdots R  X_{Z_{\alpha-1}} R X_{Z_\alpha} F_{\omega_\alpha}
X_{Z_\alpha} L X_{Z_{\alpha-1}} L\cdots L X_{Z_1}  L X_{\pi_1^{(\alpha)}}\Bigr]
\nonumber\\
&=&{ \exp \bigl[ (2 \pi_1^{(\alpha)}   + Z_1+\cdots + 2 Z_\alpha)/2 \bigr]},
\nonumber
\eea
if  the $\alpha$ edge is incident to a loop.

So in the both cases, the corresponding $\lambda$-lengths are monomial products of the exponentiated extended shear coordinates corresponding to all the edges (with multiplicities) passed by the corresponding arc. The inverse transformations expressing $\{Z_\alpha,\pi_j\}$ through $\lambda_{\mathfrak a}$ are (\ref{cross-l}), (\ref{Z-loop}),  and  (\ref{pi-lambda}).
 \end{theorem}

An example of such lamination is given in Fig. \ref{fi:face}.

\begin{remark}\label{rm:parity}
Our construction is obviously nonsymmetric w.r.t. changing the orientation of the surface.
Instead of taking the limit as $P_\alpha\to +\infty$
in (\ref{scaling}) we may take the limit as $P_\alpha\to -\infty$ in the same expression. This will then result in the insertion of the matrix
$K'=\left(
\begin{array}{cc}
0  &  1   \\
0 &   0
\end{array}
\right) $ instead of the matrix (\ref{K}) in the proper places, and the shear variable $\pi_i$ will then be based on the ideal triangle that is
to the right, not to the left, from the corresponding bordered cusp. Nevertheless, components of geodesic laminations remain to be
$G_{\mathfrak a}=e^{+l_{\mathfrak a}/2}$ with the same definition of the signed length (with the plus sign for nonintersecting horocycles).
\end{remark}

\subsection{The skein relation for CGLs}\label{ss:skein}

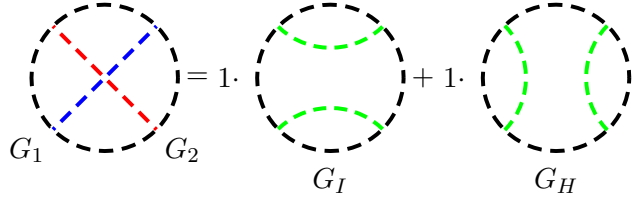
\begin{figure}[tb]
\begin{pspicture}(-4,-1.5)(4,1.5){
\rput(-3,0){
\psclip{\pscircle[linewidth=1.5pt, linestyle=dashed](0,0){1}}
\rput(0,0){\psline[linewidth=1.5pt,linecolor=red, linestyle=dashed](1,-1)(-1,1)}
\rput(0,0){\psline[linewidth=3pt,linecolor=white](-1,-1)(1,1)}
\rput(0,0){\psline[linewidth=1.5pt,linecolor=blue, linestyle=dashed](-1,-1)(1,1)}
\endpsclip
\rput(-0.8,-.8){\makebox(0,0)[rt]{$G_1$}}
\rput(0.8,-.8){\makebox(0,0)[lt]{$G_2$}}
}
\rput(0,0){
\psclip{\pscircle[linewidth=1.5pt, linestyle=dashed](0,0){1}}
\rput(0,-1.4){\psarc[linewidth=1.5pt,linecolor=green, linestyle=dashed](0,0){1}{45}{135}}
\rput(0,1.4){\psarc[linewidth=1.5pt,linecolor=green, linestyle=dashed](0,0){1}{225}{315}}
\endpsclip
\rput(0,-1.2){\makebox(0,0)[ct]{$G_I$}}
\rput(-1.2,0){\makebox(0,0)[rc]{$1\cdot$}}
}
\rput(3,0){
\psclip{\pscircle[linewidth=1.5pt, linestyle=dashed](0,0){1}}
\rput(-1.4,0){\psarc[linewidth=1.5pt,linecolor=green, linestyle=dashed](0,0){1}{-45}{45}}
\rput(1.4,0){\psarc[linewidth=1.5pt,linecolor=green, linestyle=dashed](0,0){1}{135}{225}}
\endpsclip
\rput(0,-1.2){\makebox(0,0)[ct]{$G_H$}}
\rput(-1.2,0){\makebox(0,0)[rc]{$1\cdot$}}
}
\rput(-1.7,0){
\rput(0,0){\makebox(0,0){$=$}}}
\rput(1.3,0){
\rput(0,0){\makebox(0,0){$+$}}}
}
\end{pspicture}
\caption{\small The classical skein relation: for an inner-point crossing of two curves $\gamma_1$ and $\gamma_2$, the corresponding
geodesic functions satisfy (\ref{skein1}) where in the left-hand side we have two CGLs (comprising one curve each if at least one of $\gamma_i$
is a closed curve and two arcs each if the both $\gamma_i$ are arcs) obtained by two possible resolutions of the crossing.}
\label{fi:skein1}
\end{figure}

In this section we introduce the skein relation for elements of new CGLs. Let us recall that the standard skein relation between
two closed curves corresponds to the following trace relation valid for any two matrices in $SL_2$:
$$
\tr A\tr B=\tr (AB)+\tr(AB^{-1}).
$$
We can still use this formula when the matrix $A$ is no longer invertible, namely we can trivially extend the skein relation to the case
of an arc and a closed curve. This means that by choosing $G_1=\tr A$ and
$G_2=\tr B$ where $B$ corresponds to a closed curve, we obtain that we can resolve their intersection in the standard way depicted in
Fig.~\ref{fi:skein1}.

However, when both geodesic functions correspond to arcs, the above formula is no longer valid and we must use a more ``refined'' version which will turn out to be useful also when we want to quantise. To this aim, we first approach the skein relation from a purely algebraic view point. Let us consider the permutation matrix
$$
P_{12}:=\sum_{i,j} \stackrel{1}{e}_{i,j}\otimes \stackrel{2}{e}_{j,i},
$$
where we use the standard notation for the matrix $e_{i,j}$ that has a unity at the intersection of $i$th row and $j$th column with all other elements
equal to zero.  It is not difficult to prove that for any two matrices $A$ and $B$
$$
\tr (AB )= \tr_{12}\left( \stackrel {1}{A} P_{12} \stackrel {2}{B}\right).
$$
Let us now consider the transposition of the
permutation matrix in one of the matrix spaces (does not matter in which as the total transposition leaves $P_{12}$ invariant):
$$
P^{{\mathrm T}_1}_{12}=\sum_{i,j} \stackrel{1}{e}_{i,j}\otimes \stackrel{2}{e}_{i,j},
$$
and introduce
$$
{\widetilde P}_{12}:=\bigl(\stackrel{1} F\otimes \stackrel{2}{\mathbb I} \bigr)P^{{\mathrm T}_1}_{12}
\bigl(\stackrel{1} F\otimes \stackrel{2}{\mathbb I} \bigr),
$$
where
$$
 F=
\left(
\begin{array}{cc}
0  &  1   \\
 -1 &  0
\end{array}
\right).
$$
Again it is not difficult to prove that for any matrix $A$ and any matrix $B\in SL_2$, thanks to the fact that $B^{-1}=-F B^T F$, one has:
$$
\tr (AB^{-1} )=- \tr_{12}\left( \stackrel {1}{A}{\widetilde P}_{12} \stackrel {2}{B}\right),
$$
so that the skein relation can be written as follows:
\begin{equation}\label{refined-skein}
\tr A \tr B= \tr_{12}\left( \stackrel {1}{A} P_{12} \stackrel {2}{B}\right)-\tr_{12}\left( \stackrel {1}{A}{\widetilde P}_{12} \stackrel {2}{B}\right)
\end{equation}
This new version (\ref{refined-skein}) of the skein relation is valid for any $2\times 2$ matrices, non necessarily in $SL_2$. Indeed it is a simple consequence of the fact that
\be\label{eq:trick}
 \tr(A)\tr(B)=\tr_{12}( \stackrel{1}{A} \stackrel{2}{B})=\tr_{12}( \stackrel{1}{A}\stackrel{1}{\mathbb I}\otimes \stackrel{2}{\mathbb I} \stackrel {2}{B})
 \ee
and
$$
\stackrel{1}{\mathbb I}\otimes \stackrel{2}{\mathbb I}=P_{12}-\widetilde P_{12}.
$$
Now we match this algebraic explanation to the geometric picture. Since the matrices $A$ and $B$ describe arcs or geodesics, they will generically be given by some product of left, right, edge matrices and possibly a cusp matrix $K$ as explained in sub-section \ref{suse-mat-dec}. In particular for the skein relation to make sense geometrically,  $A$ and $B$ must intersect, or in other words  they must  contain at least one edge matrix with the same coordinate, and the two right hand side terms in (\ref{refined-skein}) must also define arcs or geodesics. For example, assume:
$$
A=K \cdots  L X_Z R\cdots  \qquad\hbox{and}\qquad
B=\cdots K\cdots R X_Z L\cdots
$$
then the traces
\bea
&{}&\tr(A)=\tr (\stackrel{1}{K} \cdots \stackrel {1}{L}\stackrel{1}{X}_Z\stackrel{1}{R}\cdots )\nonumber\\
&{}&\tr(B)=\tr (\cdots\stackrel{2}{K} \cdots \stackrel {2}{R}\stackrel{2}{X}_Z\stackrel{2}{L}\cdots ),
\label{XXX}
\eea
 are invariant with respect to cyclic permutation. When using the relation (\ref{eq:trick}), we must cyclically permute the building blocks in $A$ and $B$ as follows:
 \bea
&{}&\tr(A)=\tr (\stackrel{1}{R}\cdots \stackrel{1}{K} \cdots \stackrel {1}{L}\stackrel{1}{X}_Z )\nonumber\\
&{}&\tr(B)=\tr (\stackrel{2}{L}\cdots \stackrel{2}{K} \cdots \stackrel {2}{R}\stackrel{2}{X}_Z),
\eea
so that now the two right hand side terms in (\ref{refined-skein}) become:
$$
\tr_{12}\left(\stackrel{1}{R}\cdots \stackrel{1}{K} \cdots \stackrel {1}{L}\stackrel{1}{X}_ZP_{12}\stackrel{2}{L}\cdots \stackrel{2}{K} \cdots \stackrel {2}{R}\stackrel{2}{X}_Z\right)-\tr_{12}\left(\stackrel{1}{R}\cdots \stackrel{1}{K} \cdots \stackrel {1}{L}\stackrel{1}{X}_Z{\widetilde P}_{12}\stackrel{2}{L}\cdots \stackrel{2}{K} \cdots \stackrel {2}{R}\stackrel{2}{X}_Z\right).
$$
To convince oneself that both these terms describe arcs we need to use the following two properties of $P_{12}$ and ${\widetilde P}_{12}$: for any  matrix $S$:
$$
P_{12} \stackrel{1}{S}=\stackrel{2}{S}P_{12}
$$
and
$$
\widetilde P_{12}\stackrel{2}{S}=  \stackrel{1}{F} \stackrel{1}{S}^T \stackrel{1}{F}\widetilde P_{12},
$$
so that
$$
\tr_{12}\left(\stackrel{1}{R}\cdots \stackrel{1}{K} \cdots \stackrel {1}{L}\stackrel{1}{X}_ZP_{12}\stackrel{2}{L}\cdots \stackrel{2}{K} \cdots \stackrel {2}{R}\stackrel{2}{X}_Z\right)=\tr_{12}\left(\stackrel{1}{R}\cdots \stackrel{1}{K} \cdots \stackrel {1}{L}\stackrel{1}{X}_Z\stackrel{1}{L}\cdots \stackrel{1}{K} \cdots \stackrel {1}{R}\stackrel{1}{X}_ZP_{12}\right),
$$
which defines an ark by construction. Analogously:
$$
\tr_{12}\left(\stackrel{1}{R}\cdots \stackrel{1}{K} \cdots \stackrel {1}{L}\stackrel{1}{X}_Z{\widetilde P}_{12}\stackrel{2}{L}\cdots \stackrel{2}{K} \cdots \stackrel {2}{R}\stackrel{2}{X}_Z\right)=\tr_{12}\left(\stackrel{1}{R}\cdots \stackrel{1}{K} \cdots \stackrel {1}{L}\stackrel{1}{X}_Z
\stackrel{1}F\stackrel{1}{X}_Z^T\stackrel{1}{L}\cdots\stackrel{1}{K}^T \cdots \stackrel{1}{R}\stackrel{1}{F}{\widetilde P}_{12}\right),
$$
that, thanks to the fact that $X_Z^2=F^2=-{\mathbb I}$  and $F K^T F=K$, again defines an arc.

Let us now show that (\ref{refined-skein})  implies the Ptolemy relation when both $A$ and $B$ correspond to arcs. Let us again proceed first by a purely algebraic point of view. Thanks to the results of section \ref{s:graph-bordered}, we have:
$$
A=A_1 K A_2,\qquad B=B_1 K B_2,
$$
where $A_1,A_2,B_1,B_2$ will be given by some products of left, right and edge matrices, or in other words they are elements of $SL_2(\mathbb R)$ while $K$ is defined in (\ref{K}) and satisfies $F K^T F=K$. We then have
$$
\tr_{12}\left( \stackrel {1}{A} P_{12} \stackrel {2}{B}\right)= \tr (A_1 K A_2 B_1 K B_2)=\tr (B_2 A_1 K A_2 B_1 K)= \tr (B_2 A_1 K)\tr( A_2 B_1 K),
$$
due to the nice property of $K$ that $\tr(S K T K)=\tr(S K)\tr(T K)$ for any two matrices $S$ and $T$.
Analogously:
\bea
-\tr_{12}\left( \stackrel {1}{A}{\widetilde P}_{12} \stackrel {2}{B}\right)&=&\tr( A F B^T F)=\tr( A_1 K A_2 F B_2^T K^T B_1^T F)=\tr( A_1 K A_2 F B_2^T F^2 K^T F^2 B_1^T F)=\nn\\
&=&
\tr( A_1 K A_2  B_2^{-1} K  B_1^{-1})=\tr ( B_1^{-1} A_1 K )\tr (A_2  B_2^{-1} K),\nn
\eea
so that in the end we obtain the Ptolemy relation:
\begin{equation}\label{refined-skein-pt}
\tr (A_1KA_2) \tr(B_1 K B_2)= \tr (B_2 A_1 K)\tr( A_2 B_1 K)+\tr ( B_1^{-1} A_1 K )\tr (A_2  B_2^{-1} K).
\end{equation}
Here again to match this algebraic explanation to the geometric picture we assume that $A$ and $B$ contain at least one edge matrix with the same coordinate and we need to cyclically permute the building blocks in $A$ and $B$ in such a way that all terms on the right hand side of (\ref{refined-skein-pt}) define arcs. We leave this to the reader as it is analogous to the previous case.

So, in all cases we can still present the skein relations as in Fig.~\ref{fi:skein1}: for two curves $\gamma_1$ and $\gamma_2$ having a
single crossing inside the Riemann surface, the corresponding geodesic functions $G_1=\tr(A)$ and $G_2=\tr(B)$ satisfy the relation
\be
G_1G_2=G_I+G_H,
\label{skein1}
\ee
where any of $G_1$ and $G_2$ can be either closed curves or arcs, and we obtain the geodesic or arcs $G_I$ and $G_H$
by resolving the crossing locally in two ways shown in the figure. In the case of multiple crossings, we resolve them one at the time and it is straightforward to prove that the order in which we resolve the crossings does not change the final result.

We can then extend the skein relation to laminations: the skein relation between two CGLs, call them CGL$_1$ and CGL$_2$ reads
\be
{\mathrm {CGL}}_1 {\mathrm {CGL}}_2=\sum_{{\mathrm {resolutions}}} {\mathrm {CGL}}_{HIIHHI\cdots},
\label{skein-CGL}
\ee
where in the left-hand side we have the sum of CGLs obtained by applying resolutions to all crossings of CGL$_1$ and CGL$_2$ (for $m$ crossings
the left-hand side contains $2^m$ terms). If, in the resolution process, we obtain a closed empty loop,
we assign the factor $-2$ to this loop (so not all
terms come with plus sign in the left-hand side of (\ref{skein-CGL})). If, in the resolution process, we obtain an empty loop starting and terminating
at a bordered cusp, we assign zero to this curve thus killing the whole corresponding CGL$_{HIHII\cdots}$. If a loop homeomorphic to going
around hole/orbifold point appears, we substitute its parameter $\omega_i$.

\subsection{Open/closed string diagrammatics as a projective limit of $\lambda$-lengths}\label{ss:KP}

We now establish a correspondence between our description of Riemann surfaces with bordered cusps and the approach of \emph{windowed surfaces} by Kaufmann and Penner \cite{KP}.  The authors of \cite{KP} proposed to consider laminations on Riemann surfaces with marked points on boundary components determining \emph{windows}: the domains between neighboring marked points; elements of laminations are allowed to escape through these windows. In our approach, we naturally identify these windows with parts of horocycles confined between two bordered arcs. We then have the following \emph{correspondence principle}:

\vskip10pt

\noindent
Given a Riemann surface $\Sigma_{g,s,n}$ with $n>0$, a fat graph $\widehat{\mathcal G}_{g,s,n}\in \widehat\Gamma_{g,s,n}$, and a lamination, which is a finite set of nonintersecting curves that are either closed or start and terminate at windows (see, e.g., Fig.~\ref{fi:Ptolemy} for examples of such curves), we can always collapse this lamination to
$\widehat{\mathcal G}_{g,s,n}$ in such a way that all lines of the lamination terminating at a window will terminate at the corresponding bordered cusp.
Then the parameters $\ell_\alpha\in {\mathbb Z}_{(+,0)}$ indicating how many lines of the lamination pass through the given ($\alpha$th) edge are determined uniquely. We identify these parameters with the \emph{projective limit} of $\lambda$-lengths of arcs: specifically, $\ell_\alpha$ is the projective limit of $\log \lambda_\alpha$ where $\lambda_\alpha$ is the $\lambda$-length
of the arc that is dual to the $\alpha$th edge and belongs to a unique CGL$_{\mathfrak a}^{\text{max}}$ dual to $\widehat{\mathcal G}_{g,s,n}$.

\vskip10pt

The above identification is based on the fact that the \emph{tropical limit} (or the projective limit) of mutations describes transformations of the variables $\ell_\alpha$ upon flips; indeed, when flipping an inner edge as in Fig.~\ref{fi:flip}, we obtain
\be
\ell_e+\ell_f=\max[\ell_a+\ell_c,\ell_b+\ell_d],
\label{proj1}
\ee
and when flipping an edge incident to a loop as in Fig.~\ref{fi:interchange-p-dual} we have
\be
\ell_e+\ell_f=\max[ 2\ell_a, 2\ell_b].
\label{proj2}
\ee
Here $\ell_e$ and $\ell_f$ are parameters of lamination for the original and transformed edges and in the right-hand sides of
(\ref{proj1}) and (\ref{proj2}) we can easily recognize projective limits of the corresponding mutation formulas
\bea
&{}&\hbox{\cite{Penn1}}\qquad \lambda_e\lambda_f=\lambda_a\lambda_c+\lambda_b\lambda_d,\\
&{}& \hbox{\cite{ChSh}}\qquad \lambda_e\lambda_f=\lambda_a^2+\omega\lambda_a\lambda_b +\lambda_b^2
\eea
obtained by taking the scaling limit $\lambda_\alpha\to e^{N \ell_\alpha/2}$ with the same $N\to +\infty$ for all $\alpha$.

We therefore identify windows by Kaufmann and Penner with asymptotic domains (a decoration becomes irrelevant in the projective limit), which correspond in the open/closed string terminology to incoming/outgoing \emph{open strings}; we thus have a convenient paramteterization of an open/closed string worldsheet in terms of the extended shear coordinates provided we have at least one open string asymptotic state. The open/closed string worldsheet corresponding to $\Sigma_{g,s,n}$ then has genus $g$, has exactly $n$ open string asymptotic states, and exactly
$s_{h_o}$ closed string asymptotic states (in the absence of conical singularities corresponding to orbifold points).

\subsection{Comparing to the theory of bordered surfaces by Fomin, M. Shapiro, and D. Thurston}

In two nice papers by Fomin, M. Shapiro, and D. Thurston \cite{FST} and by Fomin and D. Thurston \cite{FT}, the authors developed a theory of bordered Riemann surfaces. Riemann surfaces with bordered cusps we consider in the present paper are in fact bordered Riemann surfaces of \cite{FST,FT} with cusp decorations by horocycles (also introduced in \cite{FT}). However, our description of $\lambda$-lengths in terms of the extended shear coordinates that enables us to quantize the formers seems to be new. So, let us present the list of similarities/differences between our approach and that of Fomin, Shapiro, and Thurston:
\begin{itemize}
\item[(i)] Our arcs are \emph{ordinary arcs} between decorated bordered cusps in the terminology of \cite{FT}; related ideal triangulations comprising only compatible ordinary arcs (with all punctures enclosed in monogons) are our CGL$_{\mathfrak a}^{\text{max}}$.
\item[(ii)] Our (exponentiated) extended shear coordinate $e^{\pi_j}$ associated with the $j$th cusps is reciprocal to the $L_r$ from \cite{FT}, which is the length of the horocycle segment cut out by the corresponding ideal triangle.
\item[(iii)] We always add to an ideal triangulation system arcs between neighboring cusps (excluded in \cite{FST,FT}). These arcs never mutate, correspond to frozen variables in the quantum cluster algebra case, but they are not central in the sense of Poisson or quantum algebra having nontrivial commutation relations with ordinary arcs and between themselves.
\item[(iv)] In our treatment, we consider \emph{only} triangulations by ordinary arcs (thus avoiding tagging and notching issues, which were crucial in \cite{FST,FT}). In fact, we can introduce shear coordinates for \emph{any} fat graph ${\mathcal G}_{g,s,n}\in \Gamma_{g,s,n}$ dual to the corresponding partition of $\Sigma_{g,s,n}$ into ideal triangles whose sides are both tagged and ordinary arcs in the terminology of \cite{FST,FT}. Then the Poisson and quantum algebras of the shear coordinates on ${\mathcal G}_{g,s,n}$ will be given by the same formulas (Theorems
\ref{th-WP-cusp} and \ref{th-q-cusp} below) as for any fat graph $\widehat{\mathcal G}_{g,s,n}\in \widehat{\Gamma}_{g,s,n}$ and we can again express both $\lambda$-lengths of ordinary arcs and geodesic functions of closed curves in terms of these shear coordinates (using exactly the same combinatorial rules as before) thus obtaining the corresponding Poisson and quantum algebras (which, of course, retain their forms).  We cannot however express $\lambda$-lengths of tagged arcs in terms of shear coordinates of ${\mathcal G}_{g,s,n}$ because these shear coordinates are insensitive to the tagging and to horocycle decorations corresponding to the tagging. We are therefore lacking Poisson and quantum algebras of tagged arcs. A possible reason hindering the very existence of Poisson and quantum algebras of tagged arcs compatible with the surface orientation is that, unlike ordinary arcs, we have only cyclic, not linear, ordering of tagged arcs winding to a hole/approaching a puncture, so, presumably, no decoration-free notion of a Poisson or quantum algebra exists for tagged arcs. In what follows, we thus consider only a sub-groupoid of MCG transformations that preserve the ``monogon'' property and are described by the generalized cluster algebra mutations of \cite{ChSh}; in reward we can explicitly quantize $\lambda$-lengths of the \emph{ordinary arcs} from CGL$_{\mathfrak a}^{\text{max}}$ dual to corresponding fat graphs
$\widehat{\mathcal G}_{g,s,n}\in\widehat{\Gamma}_{g,s,n}$
thus obtaining quantum cluster algebras of geometric type (see Sec.~\ref{s:q}).
\end{itemize}

\subsection{Goldman brackets for CGLs}\label{ss:Goldman}

We now introduce the Goldman bracket on the CGLs comprising closed curves and arcs ($\lambda$-lengths). For this, we introduce the Poisson relations for intersecting curves entering CGLs.
Curves (either closed curves or arcs) can intersect either in the interior of the Riemann surface or at bordered cusps (if they are arcs incident to the same cusp(s)).

\begin{figure}[tb]
\begin{pspicture}(-4,-1.5)(4,1.5){
\rput(-3,0){
\psclip{\pscircle[linewidth=1.5pt, linestyle=dashed](0,0){1}}
\rput(0,0){\psline[linewidth=1.5pt,linecolor=red, linestyle=dashed](1,-1)(-1,1)}
\rput(0,0){\psline[linewidth=3pt,linecolor=white](-1,-1)(1,1)}
\rput(0,0){\psline[linewidth=1.5pt,linecolor=blue,  linestyle=dashed](-1,-1)(1,1)}
\endpsclip
\rput(0,-1.2){\makebox(0,0)[ct]{$\{G_1,G_2\}_k$}}
\rput(0.8,0.8){\makebox(0,0)[lb]{$\gamma_1$}}
\rput(0.8,-0.8){\makebox(0,0)[lt]{$\gamma_2$}}
}
\rput(0,0){
\psclip{\pscircle[linewidth=1.5pt, linestyle=dashed](0,0){1}}
\rput(0,-1.4){\psarc[linewidth=1.5pt,linecolor=green, linestyle=dashed](0,0){1}{45}{135}}
\rput(0,1.4){\psarc[linewidth=1.5pt,linecolor=green, linestyle=dashed](0,0){1}{225}{315}}
\endpsclip
\rput(0,-1.2){\makebox(0,0)[ct]{$G_I$}}
\rput(-1.2,0){\makebox(0,0)[rc]{$\dfrac12$}}
}
\rput(3,0){
\psclip{\pscircle[linewidth=1.5pt, linestyle=dashed](0,0){1}}
\rput(-1.4,0){\psarc[linewidth=1.5pt,linecolor=green, linestyle=dashed](0,0){1}{-45}{45}}
\rput(1.4,0){\psarc[linewidth=1.5pt,linecolor=green, linestyle=dashed](0,0){1}{135}{225}}
\endpsclip
\rput(0,-1.2){\makebox(0,0)[ct]{$G_H$}}
\rput(-1.2,0){\makebox(0,0)[rc]{$\dfrac12$}}
}
\rput(-1.7,0){
\rput(0,0){\makebox(0,0){$=$}}}
\rput(1.3,0){
\rput(0,0){\makebox(0,0){$-$}}}
}
\end{pspicture}
\caption{\small The ``elementary'' Poisson bracket (the Goldman bracket) $\{G_1,G_2\}_k$ (\ref{Gold-inner}) between two
geodesic functions of the two corresponding curves $\gamma_1$ and $\gamma_2$ at their $k$th
intersection point inside a Riemann surface: the curves and CGLs here are the same as in
Fig.~\ref{fi:skein1}.}
\label{fi:Goldman1}
\end{figure}
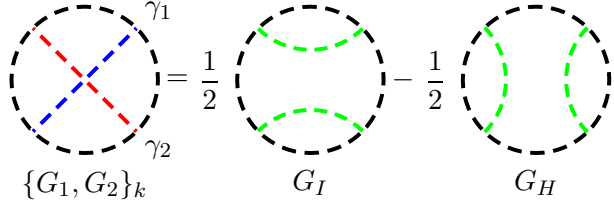

Let us define the \emph{local resolution} $\{G_1,G_2\}_k$ at the $k$th intersection point $p_k$
of two curves $\gamma_1$ and $\gamma_2$. When $p_k$ is an internal point of the surface, we set (see Fig.~\ref{fi:Goldman1})
\be
\{G_1,G_2\}_k=\frac 12 G_I -\frac 12 G_H,
\label{Gold-inner}
\ee
where $G_I$ and $G_H$ are the same resolutions of the crossing as in the skein relation in Fig.~\ref{fi:skein1}

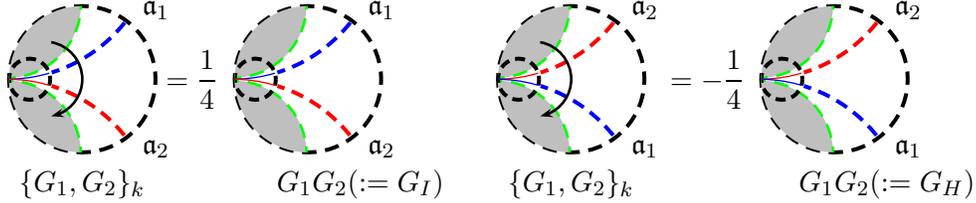
\begin{figure}[tb]
\begin{pspicture}(-4,-1.5)(4,1.5){
\rput(-5,0){
\psclip{\pscircle[linewidth=1.5pt, linestyle=dashed](0,0){1}}
\rput(-1,2){\psarc[linewidth=1.5pt,linecolor=blue, linestyle=dashed](0,0){2}{270}{360}}
\rput(-1,-2){\psarc[linewidth=1.5pt,linecolor=red, linestyle=dashed](0,0){2}{0}{90}}
\rput(-0.7,0){\pscircle[linewidth=0pt, linecolor=white,fillstyle=solid, fillcolor=white](0,0){0.3}}
\rput(-1,1){\pscircle[linewidth=1pt,linecolor=green, linestyle=dashed, fillstyle=solid, fillcolor=lightgray](0,0){1}}
\rput(-1,-1){\pscircle[linewidth=1pt,linecolor=green, linestyle=dashed, fillstyle=solid, fillcolor=lightgray](0,0){1}}
\psclip{\pscircle[linestyle=dashed,linewidth=1.5pt](-0.7,0){0.3}}
\rput(-1,2){\psarc[linewidth=.5pt,linecolor=blue](0,0){2}{270}{320}}
\rput(-1,-2){\psarc[linewidth=.5pt,linecolor=red](0,0){2}{60}{90}}
\endpsclip
\psarc[linewidth=1pt]{<-}(-0.5,0){.5}{-80}{80}
\endpsclip
\rput(0.8,0.8){\makebox(0,0)[lb]{$\mathfrak a_1$}}
\rput(0.8,-0.8){\makebox(0,0)[lt]{$\mathfrak a_2$}}
\rput(0,-1.2){\makebox(0,0)[ct]{$\{G_1,G_2\}_k$}}
}
\rput(-2,0){
\psclip{\pscircle[linewidth=1.5pt, linestyle=dashed](0,0){1}}
\rput(-1,2){\psarc[linewidth=1.5pt,linecolor=blue, linestyle=dashed](0,0){2}{270}{360}}
\rput(-1,-2){\psarc[linewidth=1.5pt,linecolor=red, linestyle=dashed](0,0){2}{0}{90}}
\rput(-0.7,0){\pscircle[linewidth=0pt, linecolor=white,fillstyle=solid, fillcolor=white](0,0){0.3}}
\rput(-1,1){\pscircle[linewidth=1pt,linecolor=green, linestyle=dashed, fillstyle=solid, fillcolor=lightgray](0,0){1}}
\rput(-1,-1){\pscircle[linewidth=1pt,linecolor=green, linestyle=dashed, fillstyle=solid, fillcolor=lightgray](0,0){1}}
\psclip{\pscircle[linestyle=dashed,linewidth=1.5pt](-0.7,0){0.3}}
\rput(-1,2){\psarc[linewidth=.5pt,linecolor=blue](0,0){2}{270}{320}}
\rput(-1,-2){\psarc[linewidth=.5pt,linecolor=red](0,0){2}{60}{90}}
\endpsclip
\endpsclip
\rput(0.8,0.8){\makebox(0,0)[lb]{$\mathfrak a_1$}}
\rput(0.8,-0.8){\makebox(0,0)[lt]{$\mathfrak a_2$}}
\rput(0.7,-1.2){\makebox(0,0)[ct]{$G_1G_2(:=G_I)$}}
\rput(-1.2,0){\makebox(0,0)[rc]{$=\dfrac 14$}}
}
\rput(1.5,0){
\psclip{\pscircle[linewidth=1.5pt, linestyle=dashed](0,0){1}}
\rput(-1,2){\psarc[linewidth=1.5pt,linecolor=red, linestyle=dashed](0,0){2}{270}{360}}
\rput(-1,-2){\psarc[linewidth=1.5pt,linecolor=blue, linestyle=dashed](0,0){2}{0}{90}}
\rput(-0.7,0){\pscircle[linewidth=0pt, linecolor=white,fillstyle=solid, fillcolor=white](0,0){0.3}}
\rput(-1,1){\pscircle[linewidth=1pt,linecolor=green, linestyle=dashed, fillstyle=solid, fillcolor=lightgray](0,0){1}}
\rput(-1,-1){\pscircle[linewidth=1pt,linecolor=green, linestyle=dashed, fillstyle=solid, fillcolor=lightgray](0,0){1}}
\psclip{\pscircle[linestyle=dashed,linewidth=1.5pt](-0.7,0){0.3}}
\rput(-1,2){\psarc[linewidth=.5pt,linecolor=red](0,0){2}{270}{320}}
\rput(-1,-2){\psarc[linewidth=.5pt,linecolor=blue](0,0){2}{60}{90}}
\endpsclip
\psarc[linewidth=1pt]{<-}(-0.5,0){.5}{-80}{80}
\endpsclip
\rput(0.8,-0.8){\makebox(0,0)[lt]{$\mathfrak a_1$}}
\rput(0.8,0.8){\makebox(0,0)[lb]{$\mathfrak a_2$}}
\rput(0,-1.2){\makebox(0,0)[ct]{$\{G_1,G_2\}_k$}}
}
\rput(5,0){
\psclip{\pscircle[linewidth=1.5pt, linestyle=dashed](0,0){1}}
\rput(-1,2){\psarc[linewidth=1.5pt,linecolor=red, linestyle=dashed](0,0){2}{270}{360}}
\rput(-1,-2){\psarc[linewidth=1.5pt,linecolor=blue, linestyle=dashed](0,0){2}{0}{90}}
\rput(-0.7,0){\pscircle[linewidth=0pt, linecolor=white,fillstyle=solid, fillcolor=white](0,0){0.3}}
\rput(-1,1){\pscircle[linewidth=1pt,linecolor=green, linestyle=dashed, fillstyle=solid, fillcolor=lightgray](0,0){1}}
\rput(-1,-1){\pscircle[linewidth=1pt,linecolor=green, linestyle=dashed, fillstyle=solid, fillcolor=lightgray](0,0){1}}
\psclip{\pscircle[linestyle=dashed,linewidth=1.5pt](-0.7,0){0.3}}
\rput(-1,2){\psarc[linewidth=.5pt,linecolor=red](0,0){2}{270}{320}}
\rput(-1,-2){\psarc[linewidth=.5pt,linecolor=blue](0,0){2}{60}{90}}
\endpsclip
\endpsclip
\rput(0.8,-0.8){\makebox(0,0)[lt]{$\mathfrak a_1$}}
\rput(0.8,0.8){\makebox(0,0)[lb]{$\mathfrak a_2$}}
\rput(0.7,-1.2){\makebox(0,0)[ct]{$G_1G_2(:=G_H)$}}
\rput(-1.2,0){\makebox(0,0)[rc]{$=-\dfrac 14$}}
}
}
\end{pspicture}
\caption{\small The ``elementary'' Poisson bracket (the Goldman bracket) $\{G_1,G_2\}_k$ (\ref{Gold-inner}) between two
geodesic functions of the two corresponding arcs ${\mathfrak a}_1$ and ${\mathfrak a}_2$ coming to the same bordered cusp
of a Riemann surface: the sign depends on the ordering of ends of the corresponding curves w.r.t. the orientation of the
Riemann surface (indicated by an arrow.}
\label{fi:Goldman2}
\end{figure}

When two arcs meet at the same bordered cusp, the Goldman bracket between their geodesic functions
at this cusp depends on the ordering of the corresponding arcs w.r.t. the orientation of the Riemann surface (see Fig.~\ref{fi:Goldman2})
($G_1=G_{\mathfrak a_1}$ and $G_2=G_{\mathfrak a_2}$)
\be
\{G_{1},G_{2}\}_k=\pm \frac 14 G_{1}G_{2}:=\left\{\begin{array}{cl}
\frac14 G_I & \hbox{if\ } \mathfrak a_1\ \hbox{is to the right of}\ \mathfrak a_2\\
-\frac14 G_H & \hbox{if\ } \mathfrak a_1\ \hbox{is to the left of}\ \mathfrak a_2
 \end{array}\right.
\label{Goldman2}
\ee
where we have the plus sign if the arc $\mathfrak a_1$ lies to the right from the arc $\mathfrak a_2$ when looking ``from inside'' the Riemann
surface and minus sign if the arc  $\mathfrak a_1$ lies to the left from the arc $\mathfrak a_2$.
Note that, since every arc has two ends, we must evaluate the brackets (\ref{Goldman2}) for all four combinations of these ends (ends at different cusps Poisson commute); for instance, in the case where all four ends are at the same cusp, and the both ends of $\mathfrak a_1$ are to the right of both ends of $\mathfrak a_2$ (provided these arcs has no intersections inside the Riemann surface), the total bracket will be
$\{G_{\mathfrak a_1},G_{\mathfrak a_2}\}=G_{\mathfrak a_1}G_{\mathfrak a_2}$.

The Poisson bracket (the Goldman bracket) between two geodesic laminations CGL$_1$ and CGL$_2$
(which may comprise both closed curves and arcs) with the set $Q$ of intersection points $p_k$ (for two arcs
$\mathfrak a^{(1)}_l\in \mathrm{CGL}_1$ and $\mathfrak a^{(2)}_m\in \mathrm{CGL}_2$,
we count intersections separately for every pair of endpoints of
$\mathfrak a^{(1)}_l$ and $\mathfrak a^{(2)}_m$)
 is then geometrically defined to be
\be
\{ \mathrm{CGL}_1, \mathrm{CGL}_2\}=\sum_{k} \sum_{Q\setminus p_k}c_{\{I,H\}}
\mathrm{CGL}_{HIIH\stackrel{k\mathrm{th}}{\{I,H\}}IHH\cdots},
\label{Goldman-cusp}
\ee
where $c_{\{I\}}=1/2\hbox{ or }1/4$ and  $c_{\{H\}}=-1/2\hbox{ or }-1/4$ depending on whether the point $p_k$ is an inner point
or a bordered cusp and we take the sum over all resolutions of the corresponding intersection of CGLs. Again, if in the process of
resolution we obtain an empty closed loop, we assign the factor $-2$ to it; if we obtain an empty loop starting and terminating at the bordered cusp,
we assign zero to it killing the corresponding CGL$_{HII\{I,H\}I\cdots}$.

\begin{lemma}\label{lem-CGL-Poisson}
The semiclassical algebra of CGLs with the product defined in (\ref{skein-CGL}) and the Poisson bracket defined by
(\ref{Goldman-cusp}) satisfies the classical Whitehead moves and semiclassical Jacobi relations.
\end{lemma}

We postpone the \emph{proof} till Sec.~\ref{s:q} where the above two cases will be corollaries of the quantum skein relations
(they arise as the respective terms of orders $\hbar^0$ and $\hbar^1$ in the $\hbar$-expansion of the quantum Whitehead moves).

\subsection{Poisson brackets for shear variables of bordered cusped Riemann surfaces}

We now introduce the Poisson bivector field (the Poisson bracket) on $\widehat{\mathfrak T}_{g,s,n}$ that is invariant w.r.t.
morphisms of $\widehat{\mathfrak T}_{g,s,n}$. {In what follows we denote by $Y_J$ the both the usual shear coordinates  $Z_\alpha$ and the cusp shear coordinates $\pi_i$. The following theorem states that the cusp shear coordinates behave in the Poisson relations exactly as usual shear coordinates - in particular they are not central:}

\begin{theorem}\label{th-WP-cusp} In {the extended shear coordinates $Y_J$ (including the standard shear coordinates and the bordered cusp shear coordinates)} of $\widehat{\mathfrak T}_{g,s,n}$ on any fixed spine
$\widehat{\mathcal G}_{g,s,n}\in\widehat\Gamma_{g,s,n}$
corresponding to a surface with at least one bordered cusp,
the Weil--Petersson bracket $B_{{\mbox{\tiny WP}}}$ reads
\be
\label{WP-cusp}
\bigl\{f({\mathbf Y}),g({\mathbf Y})\bigr\}=\sum_{{\hbox{\small 3-valent} \atop \hbox{\small vertices $J=1$} }}^{4g+2s+{n}-4}
\,\sum_{i=1}^{3 \mod 3}
\left(\frac{\partial f}{\partial Y_{J_i}} \frac{\partial g}{\partial Y_{J_{i+1}}}
- \frac{\partial g}{\partial Y_{J_i}} \frac{\partial f}{\partial Y_{J_{i+1}}}\right),
\ee
where the sum ranges all three-valent vertices of a graph that are not adjacent to loops
and $J_i$ are the labels of the cyclically (clockwise)
ordered ($J_4\equiv J_1 $) edges (irrespectively whether inner or outer) incident to the vertex
with the label $J$. This bracket
\begin{itemize}
\item[(1)] is equivariant w.r.t. the morphisms generated by flips (mutations) of inner edges
described by formula (\ref{abc}) and
by flips (mutations) of edges adjacent to loops described by Lemma~\ref{lem-pending1};
\item[(2)]
gives rise to the {\em Goldman
bracket} {\rm(\ref{Goldman-cusp})} on the space of CGLs \cite{Gold}.
\end{itemize}
The centre of this Poisson algebra is a linear span of $\sum_{J\in I}Y_J$ where we take the sum (with proper multiplicities)
over indices of edges bounding a cusped hole (labeled $I$) and of the coefficients $\omega_j$ corresponding to monogons.
These coefficients are either $2\cosh(P_j/2)$, where $P_j$ are perimeters of holes that do not contain bordered cusps, or $2\cos(\pi/p_j)$,
where $p_j\in {\mathbb Z}_{\ge 2}$ are orders of the orbifold points.
The dimension of the centre is  $s$, and the total dimension
of any Poisson leaf of  $\widehat{\mathfrak T}_{g,s,n}$ is $6g-6+2s+2n$.
\end{theorem}

\proof We just outline the proof because we can consider it a corollary of the corresponding statement in the quantum case Theorem \ref{th-q-cusp}. Proving the preservation
of Poisson brackets is easy
(and in fact was already done in Sec.~\ref{s:preliminaries} because we do not enlarge the set of mutations: we are not allowed to
mutate open edges). The strategy of proving that the brackets (\ref{WP-cusp}) imply the Goldman brackets is based on the invariance of
products of matrices under the trace signs under the flip morphisms (formulas (\ref{mutation1})-- (\ref{mutation3}) and
(\ref{pend-mutation1})--(\ref{pend-mutation3}). Using MCG transformations, we can then reduce the intersection pattern between
two curves in two CGLs to a handful of cases, each of which admits a local (quantum) resolution presented in the next section.\endproof

{We now show that the Poisson relations between $\lambda$-lengths of arcs belonging to the same CGL are completely combinatorial. In order to describe the combinatorial nature of the Poisson brackets we need to introduce some more notation: given a CGL$_{\mathfrak a}^{\text{max}}$,  let us fix an orientation of the fat graph and of each open edge so that  we can  enumerate all bordered cusps in $\Sigma_{g,s,n}$ once for ever and at each cusp, and we can prescribe a linear ordering on the set of the ends of arcs coming into the cusp. This means that all arcs in the lamination CGL$_{\mathfrak a}^{\text{max}}$ are uniquely determined by 2 indices and two sub-indices:  $s_i$ and $t_j$, where $s$ determines the cusp at which the arc originates, $i$ determines which arc in the $s$ cusp we pick, $t$ gives the cusp where the arc ends and $j$ determines which arc in the $t$ cusp we pick.
We denote this arc by $\mathfrak a_{s_i,t_j}$ and its $\lambda$-length by $\lambda_{s_i,t_j}$.}

\begin{corollary}\label{cor:arcs}
{The Poisson relations between any two $\lambda$-lengths $\lambda_{s_i,t_j}$ and $\lambda_{p_l,q_k}$ of two arcs in the same CGL  read:
\be\label{eq:comb-p}
\{\lambda_{s_i,t_j},\lambda_{p_l,q_k}\}= \frac{1}{4}I({s_i,t_j}; {p_l,q_k}) \lambda_{s_i,t_j}\lambda_{p_l,q_k},
\ee
where $I(({s_i,t_j}; {p_l,q_k}))/4$ is called  \emph{incidence index}  between the two arcs $\mathfrak a_{s_i,t_j}$ and $\mathfrak a_{p_l,q_k}$ and is defined by
\be\label{eq:comb-p1}
 I({s_i,t_j}; {p_l,q_k}):={\rm sign}({i-l})\delta_{s,p}+{\rm sign}({j-l})\delta_{t,p}+{\rm sign}({i-k})\delta_{s,q}+{\rm sign}({j-k})\delta_{t,q}.
\ee}
\end{corollary}

\proof
This is a  consequence of the fact that two arcs in the same lamination never intersect inside the Riemann surface, but can only meet in a cusp. Since the
 $\lambda$-lengths $\lambda_{s_i,t_j}$ and $\lambda_{p_l,q_k}$
of arbitrary two arcs from the same CGL$_{\mathfrak a}^{\text{max}}$ admit monomial representations in terms of the extended shear coordinates of
the fat graph $\widehat{\mathcal G}_{g,s,n}$ dual to CGL$_{\mathfrak a}^{\text{max}}$ (see Proposition~\ref{prop:monoidal}),
the Possion relations for these shear coordinates (\ref{WP-cusp}) become homogeneous.
\endproof

\subsection{Poisson algebras of geodesic functions in the case of no bordered cusps.}\label{se:no-cusp}

Let us explain here how to fully characterise the Poisson algebra of geodesic functions on a Riemann surface $\Sigma_{g,s}$ for any genus g and any number $s_h>1$ of holes and any number of $s_o$ of orbifold points ($s = s_o + s_h$) as a specific Poisson sub-algebra of the set of geodesics functions and arcs on $\tilde\Sigma_{g,s ,1}$ i.e. a Riemann surface with the same genus $g$, the same number $s_h>1$ of holes and the same number of $s_o$ of orbifold points with at least one bordered cusp on one of the holes.

For simplicity let us restrict to the case when there are no orbifold points, so that $s_h=s$. The general case can be done in the same way.

The Teichm\"uller space for $\Sigma_{g,s}$ is $\mathbb R^{6g-6+2 s}\times\Omega^s$, while for  $\tilde\Sigma_{g,s ,1}$  is $\mathbb R^{6g-6+2 s+3}\times\Omega^{s-1}$ because we have $s-1$ holes with no cusps and $s_{h_1}=1$ holes with $1$ cusp on it.

The Riemann surface $\tilde\Sigma_{g,s,1}$ is laminated by $s-1$ closed geodesics around the non-cusped holes and by $6g-6+2s+3$ arcs. The Poisson algebra is therefore of dimension  $6g-6+3s+2$ and admits $s$ central elements - the
$s-1$ parameters $\omega_1,\dots,\omega_{s-1}$ corresponding to the lengths of the closed geodesics around the non-cusped holes and the $\lambda$-length of the arc that follows the fat graph starting from the cusp, going always left until it ends at the cusp again.

Let us now consider the closed geodesic $g$ around the cusped hole (homeomorphic to the closed path going exactly around the hole and separating its part with the cusp from the rest of the surface) and take the set $\mathcal F_g$ of all functions of the lamination that Poisson commute with it. This forms a closed Poisson algebra due to the following simple lemma.

\begin{lemma}
Given a Poisson algebra $(\mathcal A,\{\cdot,\cdot\})$ and any element $g\in\mathcal A$, the set $\mathcal F=\{f\in\mathcal A|\{f,g\}=0\}$ is a Poisson sub-algebra with the induced Poisson bracket.
\end{lemma}

\proof
The statement is a trivial consequence of the Jacobi identity.
\endproof

The Poisson algebra $\mathcal F_g$ coincides with the Poisson algebra of geodesic functions on the Riemann surface $\Sigma_{g,s}$ by construction and has dimension $6g-6+3s+1$ with $s+1$ Casimirs.

\begin{example}
Let us illustrate the procedure in the case of a torus with one hole $\Sigma_{1,1}$.
In this case the fat-graph is given by a prezzle (see Fig. \ref{prezzle}) and the Poisson algebra is generated by the lengths of the three simple closed geodesics going along two edjes:
$G_{Z_1 Z_0}$, $G_{Z_2 Z_1}$,  $G_{Z_0 Z_2}$ which satisfy the following Poisson relations:
$$
\{G_{Z_2 Z_1},G_{Z_1 Z_0}\} = \frac{1}{2} G_{Z_2 Z_1}G_{Z_1 Z_0} - G_{Z_0Z_2},
$$
$$
\{G_{Z_1 Z_0},G_{Z_0Z_2},\} = \frac{1}{2} G_{Z_1 Z_0} G_{Z_0Z_2}-G_{Z_2 Z_1},
$$
$$
\{G_{Z_0Z_2},G_{Z_2 Z_1}\} = \frac{1}{2}G_{Z_0Z_2} G_{Z_2 Z_1}- G_{Z_1 Z_0}  ,
$$
with central element:
$$
G_{Z_2 Z_1}^2 +G_{Z_1 Z_0}^2 + G_{Z_0Z_2}^2-G_{Z_2 Z_1}G_{Z_1 Z_0} G_{Z_0Z_2}.
$$
\begin{figure}[h]
\begin{center}
\begin{pspicture}(4,-3.5)(12,2)
 \psarc(8,0){3}{195}{340}
 \psarc(8,0){2.5}{220}{319}
 \psarc(8,0){2.5}{336}{360}
 \psarc(8,0){2.5}{180}{195}
  \rput(8,-3.5){{${Z_1}$}}
 \psarc(6.7,0){1.7}{0}{180}
  \psarc(8,0){3}{180}{195}
  \psarc(6.7,0){1.2}{0}{180}  \psarc(9.6,0){1.7}{180}{281}
  \psarc(9.6,0){1.2}{180}{303}
  \rput(6,2){{${Z_2}$}}
   \psarc(8,0){3}{340}{360}
      \psarc(9.3,0){1.7}{0}{140}
  \psarc(9.3,0){1.2}{0}{146}
     \psarc(9.3,0){1.7}{159}{180}
\psarc(6.5,0){1.6}{253}{340}
  \psarc(6.5,0){1.1}{212}{360}
 \rput(10,2){\textcolor{black}{${Z_0}$}}
    \psarc[linecolor=blue,linewidth=2pt](8,0){2.7}{215}{360}
         \psarc[linecolor=blue,linewidth=2pt](9.3,0){1.4}{0}{144}
   \psarc[linecolor=blue,linewidth=2pt](6.5,0){1.4}{246}{348}
      \psarc[linecolor=blue,linewidth=2pt](6.0,-1.5){0.2}{90}{192}
\end{pspicture}
\caption{Fat-graph of $\Sigma_{1,1}$ with the geodesic $G_{Z_1 Z_0}$ in blue.}\label{pic:gammab}
\end{center}\label{prezzle}
\end{figure}
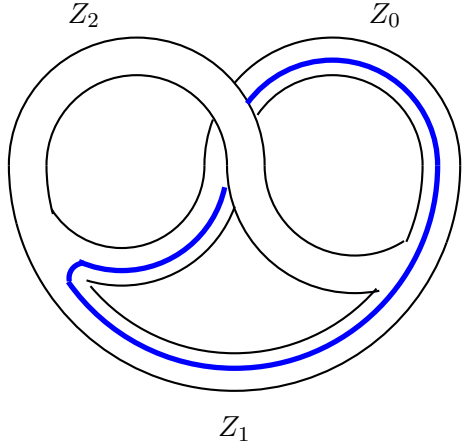

Let us now characterise this algebra as a sub-algebra of the algebra given by the lamination of a torus with one hole and one cusp on the hole $\tilde\Sigma_{1,1,1}$. The fat-graph in this case is given in  Fig.~\ref{fi:face}.
\begin{figure}[h]
{\psset{unit=1}
\begin{pspicture}(-4,-5.5)(4,5.5)
\newcommand{\PATTERN}[1]{
{\psset{unit=#1}
\psarc[linecolor=black, linewidth=1pt](0,0){4}{180}{360}
\psarc[linecolor=black, linewidth=1pt](0,0){2}{180}{360}
\rput(0,-4){\psframe[linecolor=white, fillstyle=solid, fillcolor=white](-1,0.2)(1,-1)}
\pcline[linewidth=1pt,linecolor=black](-1,-5)(-1,-3.875)
\pcline[linewidth=1pt,linecolor=black](1,-5)(1,-3.875)
\pcline[linewidth=1pt,linecolor=black](-4,0)(-4,3)
\pcline[linewidth=1pt,linecolor=black](-2,2)(-2,3)
\pcline[linewidth=1pt,linecolor=black](2,2)(2,3)
\pcline[linewidth=1pt,linecolor=black](4,0)(4,3)
\psarc[linecolor=black, linewidth=1pt](-1.5,3){0.5}{0}{180}
\psarc[linecolor=black, linewidth=1pt](-1.5,3){2.5}{0}{180}
\psarc[linecolor=black, linewidth=1pt](1.5,3){0.5}{0}{180}
\psarc[linecolor=black, linewidth=1pt](1.5,3){2.5}{0}{124}
\psarc[linecolor=black, linewidth=1pt](2,3){1}{180}{270}
\psarc[linecolor=black, linewidth=1pt](2,3){3}{180}{270}
\psarc[linecolor=black, linewidth=1pt](-2,3){1}{270}{360}
\psarc[linecolor=black, linewidth=1pt](-2,3){3}{270}{310}
\rput(-1.3,-4.5){$\pi$}
\rput(3.1,-3){$Z_1$}
\rput(-3.1,-3){$Z_2$}
\rput(-4.3,3){$Z_4$}
\rput(4.3,3){$Z_3$}
}
}
\newcommand{\TAIL}[1]{%
{\psset{unit=#1}
\psarc[linecolor=blue,  linestyle=dashed, linewidth=1pt](0,0){0.1}{0}{75}
\psarc[linecolor=green, linewidth=1pt](0,0){0.25}{0}{75}
\psarc[linecolor=red, linewidth=1pt](0,0){0.4}{0}{75}
\psarc[linecolor=blue,  linestyle=dashed, linewidth=1pt](0,0){0.55}{0}{75}
\psarc[linecolor=magenta, linestyle=dashed, linewidth=1pt](0,0){0.7}{0}{75}
\psarc[linecolor=green, linewidth=1pt](0,0){0.85}{0}{75}
\psarc[linecolor=red, linewidth=1pt](0,0){1}{0}{75}
\psarc[linecolor=magenta, linestyle=dashed, linewidth=1pt](0,0){1.15}{0}{75}
\pcline[linecolor=blue,  linestyle=dashed, linewidth=1pt](0.1,0)(0.1,-1)
\pcline[linecolor=green, linewidth=1pt](0.25,0)(0.25,-1)
\pcline[linecolor=red, linewidth=1pt](0.4,0)(0.4,-1)
\pcline[linecolor=blue,  linestyle=dashed, linewidth=1pt](0.55,0)(0.55,-1)
\pcline[linecolor=magenta, linestyle=dashed, linewidth=1pt](0.7,0)(0.7,-1)
\pcline[linecolor=green, linewidth=1pt](0.85,0)(0.85,-1)
\pcline[linecolor=red, linewidth=1pt](1,0)(1,-1)
\pcline[linecolor=magenta, linestyle=dashed, linewidth=1pt](1.15,0)(1.15,-1)
\pcline[linewidth=0.5pt,linecolor=black]{->}(0.15,-1.5)(0.05,-1.1)
\rput(.2,-1.65){$a_0$}
\pcline[linewidth=0.5pt,linecolor=black]{->}(0.65,-1.5)(0.12,-1)
\rput(.7,-1.65){$a_1$}
\pcline[linewidth=0.5pt,linecolor=black]{->}(1.15,-1.5)(0.28,-0.9)
\rput(1.2,-1.65){$a_2$}
\pcline[linewidth=0.5pt,linecolor=black]{->}(1.65,-1.5)(0.43,-0.85)
\rput(1.7,-1.65){$a_3$}
\pcline[linewidth=0.5pt,linecolor=black]{->}(2.15,-1.5)(0.73,-0.85)
\rput(2.2,-1.65){$a_4$}
}
}
\newcommand{\LOOPS}[1]{%
{\psset{unit=#1}
\psarc[linecolor=blue,  linestyle=dashed, linewidth=1pt](0,0){3.9}{180}{270}
\psarc[linecolor=green, linewidth=1pt](0,0){3.75}{180}{270}
\psarc[linecolor=red, linewidth=1pt](0,0){3.6}{180}{270}
\psarc[linecolor=blue,  linestyle=dashed, linewidth=1pt](0,0){3.45}{180}{270}
\psarc[linecolor=magenta, linestyle=dashed, linewidth=1pt](0,0){3.3}{180}{270}
\psarc[linecolor=green, linewidth=1pt](0,0){3.15}{180}{270}
\psarc[linecolor=red, linewidth=1pt](0,0){3}{180}{270}
\psarc[linecolor=magenta, linestyle=dashed, linewidth=1pt](0,0){2.85}{180}{270}
\rput{-15}(0,-2){\psframe[linecolor=white, fillstyle=solid, fillcolor=white](-0.5,2)(0.5,-2)}
\psarc[linecolor=blue,  linestyle=dashed, linewidth=1pt](0,0){2.55}{180}{270}
\psarc[linecolor=blue,  linestyle=dashed, linewidth=1pt](0,1.35){3.9}{270}{360}
\psarc[linecolor=red, linewidth=1pt](0,0){2.4}{180}{360}
\psarc[linecolor=green, linewidth=1pt](0,0){2.25}{180}{360}
\psarc[linecolor=blue,  linestyle=dashed, linewidth=1pt](0,0){2.1}{180}{360}
\pcline[linecolor=blue,  linestyle=dashed, linewidth=1pt](-3.9,0)(-3.9,3)
\pcline[linecolor=green, linewidth=1pt](-3.75,0)(-3.75,3)
\pcline[linecolor=red, linewidth=1pt](-3.6,0)(-3.6,3)
\pcline[linecolor=blue,  linestyle=dashed, linewidth=1pt](-3.45,0)(-3.45,3)
\pcline[linecolor=magenta, linestyle=dashed, linewidth=1pt](-3.3,0)(-3.3,3)
\pcline[linecolor=green, linewidth=1pt](-3.15,0)(-3.15,3)
\pcline[linecolor=red, linewidth=1pt](-3,0)(-3,3)
\pcline[linecolor=green, linewidth=1pt](-2.55,2)(-2.55,3)
\pcline[linecolor=blue,  linestyle=dashed, linewidth=1pt](-2.1,2)(-2.1,3)
\psarc[linecolor=green, linewidth=1pt](-2,2){0.55}{180}{270}
\psarc[linecolor=blue,  linestyle=dashed, linewidth=1pt](-2,2){0.1}{180}{270}
\psarc[linecolor=green, linewidth=1pt](-2,3.3){1.85}{270}{310}
\psarc[linecolor=blue,  linestyle=dashed, linewidth=1pt](-2,3){1.1}{270}{340}
\psarc[linecolor=magenta, linestyle=dashed, linewidth=1pt](-2,3.07){2.22}{270}{310}
\psarc[linecolor=blue,  linestyle=dashed, linewidth=1pt](-2,3){2.45}{270}{310}
\psarc[linecolor=red, linewidth=1pt](-2,3){2.6}{270}{310}
\psarc[linecolor=green, linewidth=1pt](-2,3){2.75}{270}{310}
\psarc[linecolor=blue,  linestyle=dashed, linewidth=1pt](-2,3){2.9}{270}{310}
\psarc[linecolor=magenta, linestyle=dashed, linewidth=1pt](-2,0){0.85}{90}{180}
\psarc[linecolor=blue,  linestyle=dashed, linewidth=1pt](-2,0){0.55}{90}{180}
\psarc[linecolor=red, linewidth=1pt](-2,0){.4}{90}{180}
\psarc[linecolor=green, linewidth=1pt](-2,0){0.25}{90}{180}
\psarc[linecolor=blue,  linestyle=dashed, linewidth=1pt](-2,0){0.1}{90}{180}
\psarc[linecolor=red, linewidth=1pt](2,0){.4}{0}{90}
\psarc[linecolor=green, linewidth=1pt](2,0){0.25}{0}{90}
\psarc[linecolor=blue,  linestyle=dashed, linewidth=1pt](2,0){0.1}{0}{90}
\psarc[linecolor=blue,  linestyle=dashed, linewidth=1pt](-1.5,3){2.4}{0}{180}
\psarc[linecolor=green, linewidth=1pt](-1.5,3){2.25}{0}{180}
\psarc[linecolor=red, linewidth=1pt](-1.5,3){2.1}{0}{180}
\psarc[linecolor=blue,  linestyle=dashed, linewidth=1pt](-1.5,3){1.95}{0}{180}
\psarc[linecolor=magenta, linestyle=dashed, linewidth=1pt](-1.5,3){1.8}{0}{180}
\psarc[linecolor=green, linewidth=1pt](-1.5,3){1.65}{0}{180}
\psarc[linecolor=red, linewidth=1pt](-1.8,3){1.2}{0}{180}
\psarc[linecolor=green, linewidth=1pt](-1.65,3){.9}{0}{180}
\psarc[linecolor=blue,  linestyle=dashed, linewidth=1pt](-1.5,3){0.6}{0}{180}
\psarc[linecolor=blue,  linestyle=dashed, linewidth=1pt](2,3){1.1}{180}{270}
\psarc[linecolor=green, linewidth=1pt](2,3){1.25}{180}{270}
\psarc[linecolor=red, linewidth=1pt](2,3){1.4}{180}{270}
\psarc[linecolor=blue,  linestyle=dashed, linewidth=1pt](2,3){1.55}{180}{270}
\psarc[linecolor=magenta, linestyle=dashed, linewidth=1pt](2,3){1.7}{180}{270}
\psarc[linecolor=green, linewidth=1pt](2,3){1.85}{180}{270}
\psarc[linecolor=red, linewidth=1pt](2,3){2.6}{180}{270}
\psarc[linecolor=green, linewidth=1pt](2,3){2.75}{180}{270}
\psarc[linecolor=blue,  linestyle=dashed, linewidth=1pt](2,3){2.9}{180}{270}
\psarc[linecolor=blue,  linestyle=dashed, linewidth=1pt](2,2){.1}{270}{360}
\psarc[linecolor=green, linewidth=1pt](2,2){.25}{270}{360}
\psarc[linecolor=red, linewidth=1pt](2,2){.4}{270}{360}
\psarc[linecolor=blue,  linestyle=dashed, linewidth=1pt](2,2){.55}{270}{360}
\psarc[linecolor=magenta, linestyle=dashed, linewidth=1pt](2,2){.7}{270}{360}
\psarc[linecolor=green, linewidth=1pt](2,2){.85}{270}{360}
\pcline[linecolor=blue,  linestyle=dashed, linewidth=1pt](2.1,2)(2.1,3)
\pcline[linecolor=green, linewidth=1pt](2.25,2)(2.25,3)
\pcline[linecolor=red, linewidth=1pt](2.4,2)(2.4,3)
\pcline[linecolor=blue,  linestyle=dashed, linewidth=1pt](2.55,2)(2.55,3)
\pcline[linecolor=magenta, linestyle=dashed, linewidth=1pt](2.7,2)(2.7,3)
\pcline[linecolor=green, linewidth=1pt](2.85,2)(2.85,3)
\pcline[linecolor=blue, linestyle=dashed, linewidth=1pt](3.9,1.35)(3.9,3)
\psarc[linecolor=blue,  linestyle=dashed, linewidth=1pt](1.5,3){.6}{0}{140}
\psarc[linecolor=green, linewidth=1pt](1.5,3){.75}{0}{132}
\psarc[linecolor=red, linewidth=1pt](1.5,3){.9}{0}{128}
\psarc[linecolor=blue,  linestyle=dashed, linewidth=1pt](1.5,3){1.05}{0}{125}
\psarc[linecolor=magenta, linestyle=dashed, linewidth=1pt](1.5,3){1.2}{0}{123}
\psarc[linecolor=green, linewidth=1pt](1.35,3){1.5}{0}{119.5}
\psarc[linecolor=blue,  linestyle=dashed, linewidth=1pt](1.5,3){2.4}{0}{124}
}
}
\rput(0,0){\LOOPS{1}}
\rput(0,0){\PATTERN{1}}
\rput(-1.05,-3.86){\TAIL{1}}
\end{pspicture}
}
\caption{\small The (canonical) system of arcs for $\Sigma_{1,1,1}$ .}
\label{fi:face}
\end{figure}
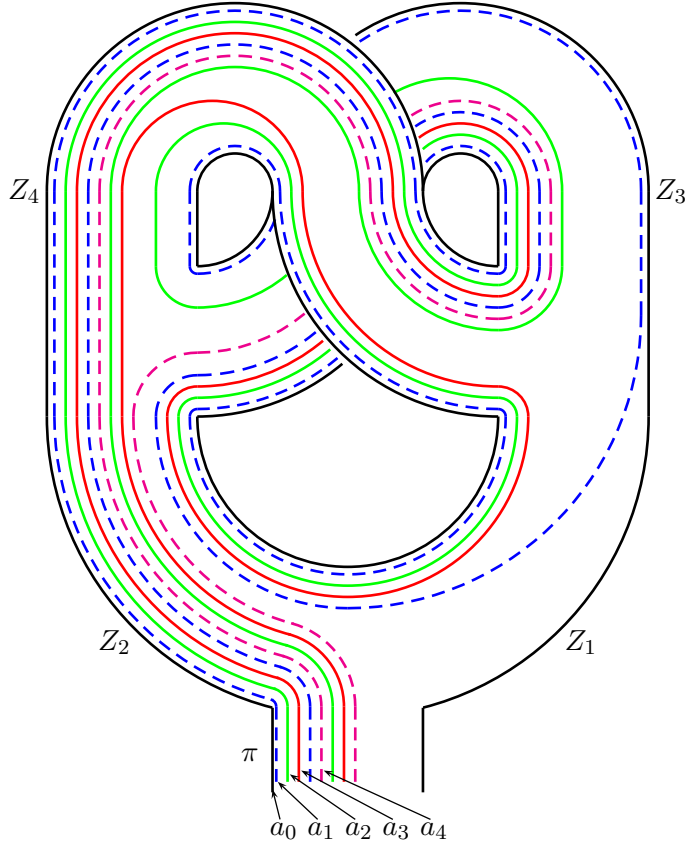

We choose the lamination in the canonical form from Proposition~\ref{prop:monoidal}  (in this form, all arcs are monomials in the exponentiated shear coordinates).

The elements constituting the lamination are:
\bea
&&
a_0= e^{\pi+Z_1+Z_2+Z_3+Z_4},\qquad a_1= e^{\pi+Z_1+2Z_2+\frac32 Z_3+\frac32 Z_4},\qquad
a_2=e^{\pi+\frac12 Z_1+\frac32 Z_2+Z_3+\frac32 Z_4},\nonumber\\
&&
a_3= e^{\pi+\frac12 Z_1+\frac32 Z_2+\frac12 Z_3+Z_4},\qquad
a_4=e^{\pi+Z_2+\frac12 Z_3+\frac12 Z_4},\label{Z2a}
\eea
where $a_0$ is central. Note that because the above relations are invertible, we can equivalently express all shear coordinates in terms of arcs using formulas  (\ref{cross-l}) and (\ref{pi-lambda}):
\bea
&&e^{\pi}=\frac{a_0a_4}{a_1},\qquad e^{Z_1}=\frac{a_0a_3}{a_2a_4}, \qquad e^{Z_2}=\frac{a_1a_3}{a_0a_2},\nonumber\\
&&e^{Z_3}=\frac{a_1a_4}{a_3^2},\qquad e^{Z_4}=\frac{a_2^2}{a_1a_4}.\label{a2Z}
\eea

Let us now consider the closed geodesic around the hole:
$$
g={\rm tr}\left(RX_{Z_1}LX_{Z_3}LX_{Z_4}LX_{Z_1}LX_{Z_2}LX_{Z_3}LX_{Z_4}LX_{Z_2}\right).
$$
It is straightforward to verify that the set of functions that Poisson commute with $g$ is generated by $g,a_0,\frac{a_1}{a_2},\frac{a_2}{a_3},\frac{a_3}{a_4}$. The three
simple closed geodesics of the uncusped case now correspond to:
$$
G_1={\rm tr}\left(LX_{Z_2}RX_{Z_4}LX_{Z_1}\right),\quad G_2={\rm tr}\left(LX_{Z_2}LX_{Z_3}RX_{Z_1}\right),\quad
G_3={\rm tr}\left(LX_{Z_4}RX_{Z_3}\right).
$$
We can express these in terms of the lamination as follows:
\bea
&&
G_1= \frac{a_4}{a_3}+\frac{a_3}{a_4}+ \frac{a_2^2}{a_1 a_3}+\frac{a_0 a_2}{a_1 a_4},\nn\\
&&
G_2= \frac{a_2}{a_1}+\frac{a_1}{a_2}+ \frac{a_3^2}{a_2 a_4}+\frac{a_0 a_3}{a_1 a_4},\nn\\
&&
G_3= \frac{a_2}{a_3}+\frac{a_3}{a_2}+\frac{a_1 a_4}{a_2 a_3}.\nn
\eea
It is straightforward to see that $G_1,G_2,G_3$ and $g$
generate the sub-algebra of all functions of $a_0,\dots,a_4$ that Poisson commute with $g$ and that these geodesic functions
satisfy the Poisson relations
$$
\{G_1,G_2\}=\frac{1}{2} G_1 G_2-G_3,\qquad
\{G_2,G_3\}=\frac{1}{2} G_2 G_3-G_1,\qquad
\{G_3,G_1\}=\frac{1}{2} G_3 G_1-G_2,
$$
with the central element
$$
G_1 G_2 G_3-G_1^2-G_2^2 -G_3^2.
$$
This central element is equal to $2-g$.
\end{example}


\section{Quantum algebras}\label{s:q}

\subsection{Quantum algebras of arcs}\label{ss:q-arcs}

We first start with quantising the Poisson relations for the generalised shear coordinates on Riemann surfaces with bordered cusps. We have the quantum analogue of
Theorem~\ref{th-WP-cusp} - note that here again $Y^\hbar_J$ denotes both the operator corresponding to inner edges and to cusps.

\begin{theorem}\label{th-q-cusp} Let $Y^\hbar_J$ denote the Hermitian operators  corresponding to the extended shear coordinates coordinates $Y_J$
of $\widehat{\mathfrak T}_{g,s,n}$ on any fixed spine $\widehat{\mathcal G}_{g,s,n}\in\widehat\Gamma_{g,s,n}$,
the commutation relations between these operators are given by the formula
\be
[Y^\hbar_{J_1},Y^\hbar_{J_2}]=2\pi i \hbar\{Y_{J_1},Y_{J_2}\},\label{qqq}
\ee
where $\{Y_{J_1},Y_{J_2}\}$ are the Poisson brackets given by the formula (\ref{WP-cusp}).
These commutation relations
\begin{itemize}
\item[(1)] are equivariant w.r.t. the quantum flip morphisms generated by flips (mutations) of inner edges
(\ref{q-mor}) and by flips (mutations) of edges adjacent to loops (\ref{morphism-pending-quantum}) and (\ref{morphism-pending-quantum2}).
\item[(2)]
gives rise to the {\em quantum skein relations} on the space of CGLs \cite{Gold}.
\end{itemize}
The Casimirs of these quantum algebras are again $\sum_{J\in I}Y^\hbar_J$ where we take the sum (with proper multiplicities)
over indices of edges bounding a cusped hole (labeled $I$).
\end{theorem}

Whereas no obvious natural ordering of quantum shear coordinates entering a quantum geodesic function for a closed geodesic
exists, it appears that we have one for quantum shear coordinates of arcs.

\begin{lemma}\label{lm:ordering}
The quantum ordering that
\begin{itemize}
\item[(1)] is preserved by the quantum flip morphisms in (\ref{q-mor}),  (\ref{morphism-pending-quantum}) and (\ref{morphism-pending-quantum2}),
\item[(2)] ensures that \emph{all} geodesic arcs are Hermitian operators,
\end{itemize}
is the natural quantum ordering (coinciding with the ordering of matrix product) provided we replace $R\to q^{-1/4}R$ and
$L\to q^{1/4}L$ at all the formulae.
\end{lemma}

\proof The { proof} is based on formulas (\ref{curve1})--(\ref{curve3})  and (\ref{pend-quantum-mutation1})--(\ref{pend-quantum-mutation3})
using which we can reduce \emph{any} arc to one of the following  cases:
\begin{itemize}
\item[(1)]
if an arc starts and terminates at different bordered cusps labeled $1$ and $2$, then we have either
$\Tr \bigl[ K X_{\pi_2}Rq^{-1/4} X_{\pi_1} \bigr]$ with $[\pi_2,\pi_1]=2\pi i \hbar$
or $\Tr \bigl[ K X_{\pi_2}L X_{Y} R X_{\pi_1}\bigr]$ with $[Y,\pi_2]=[Y,\pi_1]=2\pi i \hbar$, $[\pi_1,\pi_2]=0$;
a direct calculation in the both cases demonstrate that these expressions are Hermitian operators.
\item[(2)] if an arc starts and terminates at the same bordered cusp, then we have either
$$
\Tr \bigl[K X_{\pi} R X_{Y} F_{\omega} X_{Y} L X_{\pi}\bigr],\quad [Y,\pi]=2\pi i \hbar,
$$
or
$$
q^{-1/4} \Tr \bigl[K X_{\pi} R X_{Y_1} L X_{Y_2} R X_{\pi}\bigr],\quad [Y_1,\pi]=[Y_1,Y_2]=[\pi,Y_2]=2\pi i \hbar
$$
or
$$
q^{-1/2}\Tr \bigl[K X_{\pi} R X_{Y_1} R X_{Y_2} L X_{Y_3} R X_{\pi}\bigr] ,\quad [Y_1,\pi]=[Y_2,Y_1]=[Y_2,Y_3]=[\pi,Y_3]=2\pi i \hbar
$$
All these expressions with the natural ordering of quantum entries are Hermitian operators.
\end{itemize}\endproof

\begin{lemma}\label{lm:arcs}
All quantum arcs from the same quantum CGL have homogeneous ($q$-commutation) relations:
\be
q^{I({\mathfrak a}_1,{\mathfrak a}_2)/4}G^\hbar_{{\mathfrak a}_1}G^\hbar_{{\mathfrak a}_2}
=q^{-I({\mathfrak a}_1,{\mathfrak a}_2)/4}G^\hbar_{{\mathfrak a}_2}G^\hbar_{{\mathfrak a}_1},
\label{homog}
\ee
where $I({\mathfrak a}_1,{\mathfrak a}_2)=-I({\mathfrak a}_2,{\mathfrak a}_1)$ is the ``incidence index'' (\ref{eq:comb-p1})
of two arcs ${\mathfrak a}_1$ and ${\mathfrak a}_2$ that have no intersections inside the Riemann surface. Recall that  this index can take values $-4,-2,-1,0,1,2,4$.
\end{lemma}

\proof The {proof} again uses the invariance of quantum arcs w.r.t. quantum flip morphisms. Using this invariance we can again reduce the
pattern to one of a finite number of cases. We can then verify the quantum skein relations (\ref{homog}) at each case separately.\endproof

\subsection{Quantum skein relations for arcs}\label{ss:q-skein}

As we have seen in sub-section \ref{ss:skein}, in order for the skein relation (\ref{refined-skein}) to make sense geometrically, we need to cyclically permute the factors that form the matrices $A$ and $B$ in such a way that the matrices on the right hand side of  (\ref{refined-skein})  indeed correspond to geodeic arcs or closed geodesics. When we are dealing with the quantum case, the entries of these matrices no longer commute, so that cyclic permutations bring in some $q$-factors. In this section we explain how to control these factors in a way to define a quantum analogue to the skein relation.

\begin{figure}[tb]
\begin{pspicture}(-6,-2)(6,2){
{\psset{unit=0.7}
\newcommand{\DISCMIX}{%
{\psset{unit=1}
\rput(0,0){\psline[linewidth=30pt,linecolor=blue](-1,0)(1,0)}
\rput(0,0){\psline[linewidth=30pt,linecolor=magenta](1,0)(2,1.5)}
\rput(0,0){\psline[linewidth=30pt,linecolor=magenta](1,0)(2,-1.5)}
\rput(0,0){\psline[linewidth=30pt,linecolor=magenta](-1,0)(-2,1.5)}
\rput(0,0){\psline[linewidth=30pt,linecolor=magenta](-1,0)(-2,-1.5)}
\rput(0,0){\psline[linewidth=27pt,linecolor=white](-1,0)(1,0)}
\rput(0,0){\psline[linewidth=27pt,linecolor=white](1,0)(2,1.5)}
\rput(0,0){\psline[linewidth=27pt,linecolor=white](1,0)(2,-1.5)}
\rput(0,0){\psline[linewidth=27pt,linecolor=white](-1,0)(-2,1.5)}
\rput(0,0){\psline[linewidth=27pt,linecolor=white](-1,0)(-2,-1.5)}
\rput(0,0){\psline[linewidth=1.5pt,linecolor=red, linestyle=dashed](1.4,0.6)(2,1.5)}
\rput(0,0){\psline[linewidth=1.5pt,linecolor=blue, linestyle=dashed](1.4,-0.6)(2,-1.5)}
\rput(0,0){\psline[linewidth=1.5pt,linecolor=blue, linestyle=dashed](-1.4,0.6)(-2,1.5)}
\rput(0,0){\psline[linewidth=1.5pt,linecolor=red, linestyle=dashed](-1.4,-0.6)(-2,-1.5)}
\psbezier[linewidth=1.5pt,linecolor=red,linestyle=dashed](-1.4,-0.6)(-0.8,0.3)(.8,-0.3)(1.4,0.6)
\psbezier[linewidth=3pt,linecolor=white](-1.4,0.6)(-0.8,-0.3)(.8,0.3)(1.4,-0.6)
\psbezier[linewidth=1.5pt,linecolor=blue,linestyle=dashed](-1.4,0.6)(-0.8,-0.3)(.8,0.3)(1.4,-0.6)
}
}
\newcommand{\DISCBLUE}{%
{\psset{unit=1}
\rput(0,0){\psline[linewidth=30pt,linecolor=magenta](-1,0)(1,0)}
\rput(0,0){\psline[linewidth=30pt,linecolor=magenta](1,0)(2,1.5)}
\rput(0,0){\psline[linewidth=30pt,linecolor=magenta](1,0)(2,-1.5)}
\rput(0,0){\psline[linewidth=30pt,linecolor=magenta](-1,0)(-2,1.5)}
\rput(0,0){\psline[linewidth=30pt,linecolor=magenta](-1,0)(-2,-1.5)}
\rput(0,0){\psline[linewidth=27pt,linecolor=white](-1,0)(1,0)}
\rput(0,0){\psline[linewidth=27pt,linecolor=white](1,0)(2,1.5)}
\rput(0,0){\psline[linewidth=27pt,linecolor=white](1,0)(2,-1.5)}
\rput(0,0){\psline[linewidth=27pt,linecolor=white](-1,0)(-2,1.5)}
\rput(0,0){\psline[linewidth=27pt,linecolor=white](-1,0)(-2,-1.5)}
\rput(0,0){\psline[linewidth=1.5pt,linecolor=blue, linestyle=dashed](1.4,-0.6)(2,-1.5)}
\rput(0,0){\psline[linewidth=1.5pt,linecolor=blue, linestyle=dashed](-1.4,0.6)(-2,1.5)}
\psbezier[linewidth=3pt,linecolor=white](-1.4,0.6)(-0.8,-0.3)(.8,0.3)(1.4,-0.6)
\psbezier[linewidth=1.5pt,linecolor=blue,linestyle=dashed](-1.4,0.6)(-0.8,-0.3)(.8,0.3)(1.4,-0.6)
}
}
\newcommand{\DISCRED}{%
{\psset{unit=1}
\rput(0,0){\psline[linewidth=30pt,linecolor=magenta](-1,0)(1,0)}
\rput(0,0){\psline[linewidth=30pt,linecolor=magenta](1,0)(2,1.5)}
\rput(0,0){\psline[linewidth=30pt,linecolor=magenta](1,0)(2,-1.5)}
\rput(0,0){\psline[linewidth=30pt,linecolor=magenta](-1,0)(-2,1.5)}
\rput(0,0){\psline[linewidth=30pt,linecolor=magenta](-1,0)(-2,-1.5)}
\rput(0,0){\psline[linewidth=27pt,linecolor=white](-1,0)(1,0)}
\rput(0,0){\psline[linewidth=27pt,linecolor=white](1,0)(2,1.5)}
\rput(0,0){\psline[linewidth=27pt,linecolor=white](1,0)(2,-1.5)}
\rput(0,0){\psline[linewidth=27pt,linecolor=white](-1,0)(-2,1.5)}
\rput(0,0){\psline[linewidth=27pt,linecolor=white](-1,0)(-2,-1.5)}
\rput(0,0){\psline[linewidth=1.5pt,linecolor=red, linestyle=dashed](1.4,0.6)(2,1.5)}
\rput(0,0){\psline[linewidth=1.5pt,linecolor=red, linestyle=dashed](-1.4,-0.6)(-2,-1.5)}
\psbezier[linewidth=1.5pt,linecolor=red,linestyle=dashed](-1.4,-0.6)(-0.8,0.3)(.8,-0.3)(1.4,0.6)
}
}
\rput{-40}(0,0){
\rput(0,0){\DISCMIX}
\rput(-4,-3){\DISCRED}
\rput(4,3){\DISCRED}
\rput(0,1){\makebox(0,0)[cb]{$Z$}}
\rput(2.7,-1){\makebox(0,0)[lc]{$Y$}}
\rput(-2.7,-0.8){\makebox(0,0)[rc]{$T$}}
\rput(2.7,1){\makebox(0,0)[lc]{$X$}}
\rput(-3,1){\makebox(0,0)[rc]{$P$}}
\rput(4,4){\makebox(0,0)[cb]{$X_1$}}
\rput(6.7,4){\makebox(0,0)[lc]{$X_2$}}
\rput(-4,-2){\makebox(0,0)[cb]{$T_1$}}
\rput(-7.5,-4){\makebox(0,0)[lc]{$T_2$}}
\rput(0,0){\psline[linewidth=2pt,linestyle=dashed,linecolor=green](0.6,1)(0.6,-1)}
}
}
}
\end{pspicture}
\caption{\small Example of a single arc intersection. The vertical dashed line indicates the position of insertions.}
\label{fi:q-skein}
\end{figure}
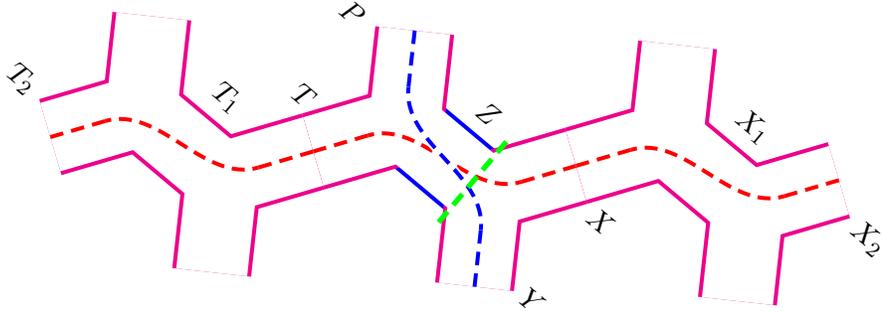

Let us consider a specific example (see  Fig.~\ref{fi:q-skein}) with two quantum arcs intersecting once:
\be
G_1^\hbar=\tr\left(\cdots X_{T_2} L X_{T_1} R X_T L X_Z R X_X L X_{X_1} R  X_{X_2} \cdots\right) \ \hbox{and}\ G_2^\hbar=\tr\left( \cdots X_P R X_Z L X_Y \cdots\right).
\label{twoarcs}
\ee

We use that
$$
\stackrel{1} X_X \stackrel{2} X_Y=\stackrel{2} X_Y \stackrel{1} X_X Q = \stackrel{2} X_Y  Q^{-1} \stackrel{1} X_X,
$$
where $Q$ is a diagonal matrix acting in the space product:
\beq
Q=\sum_{i,j}\stackrel{1}e_{i,i}\otimes\stackrel{2}e_{j,j} q^{(-1)^{|i-j|}/2}.
\label{Q}
\eeq
We then push the second arc in (\ref{twoarcs}) from the right through the first arc until two insertions of $X_Z$ will become neighbour (this is to make sure that all quantities involved in our quantum skein relation indeed describe quantum arcs). We obtain
\be\label{eq:bullet-q}
G_1^\hbar G_2^\hbar=\tr_{12}\left(\cdots \stackrel{1}X_{T_2} \stackrel{1} L \stackrel{1} X_{T_1} \stackrel{1} R  \stackrel{1}X_T  \stackrel{1}L
 \stackrel{2} X_P \stackrel{1}X_Z  Q^{-1} \stackrel{2} R\stackrel{2} X_Z \bullet \stackrel{1} R \stackrel{1} X_X Q^{-1} \stackrel{2} L \stackrel{2}X_Y
 \stackrel{1} L\stackrel{1} X_{X_1}\stackrel{1} R \stackrel{1} X_{X_2} \cdots\right)
\ee
where the bullet is the place where we are going to insert $\mathbb I\times \mathbb I$ like in the classical case.  We now however have to replace the classical matrices $P_{12}$ and $\widetilde{P}_{12}$ by their quantum analogues.

Let us start from the quantum analogue $\widetilde r_{12}$ of $-\widetilde P_{12}$:
\be
\widetilde r_{12}:=q \stackrel{1} e_{22}\otimes \stackrel{2} e_{11}+q^{-1} \stackrel{1} e_{11}\otimes \stackrel{2} e_{22}
-\stackrel{1} e_{12}\otimes \stackrel{2} e_{21}- \stackrel{1} e_{21}\otimes \stackrel{2} e_{12},
\label{wtr12}
\ee
or
$$
\widetilde r_{12}=
\left[
\begin{array}{cccc}
 0 & 0  & 0 & 0 \\
 0 & q^{-1}  &-1  &0 \\
 0 & -1  & q & 0 \\
 0 & 0  & 0 & 0
\end{array}
\right].
$$
It is straightforward to verify that
\bea
&{}&
\widetilde r_{12}\bigl( \stackrel{1} {RX_S}\otimes \stackrel{2} {\mathbb E}\bigr)Q^{-1}=q^{1/2}
\widetilde r_{12} \bigl( \stackrel{1} {\mathbb E}\otimes \stackrel{2}  {X_S L}\bigr),
\nonumber\\
&{}&
\widetilde r_{12}\bigl( \stackrel{1} {LX_S}\otimes \stackrel{2} {\mathbb E}\bigr)Q=q^{-1/2}
\widetilde r_{12} \bigl( \stackrel{1} {\mathbb E}\otimes \stackrel{2}  {X_S R}\bigr),
\nonumber\\
&{}&
Q^{-1}\bigl( \stackrel{1} {X_S R}\otimes \stackrel{2} {\mathbb E}\bigr)\widetilde r_{12}
=q^{1/2} \bigl( \stackrel{1} {\mathbb E}\otimes \stackrel{2}  {L X_S }\bigr)\widetilde r_{12},
\nonumber\\
&{}&
Q\bigl( \stackrel{1} {X_S L}\otimes \stackrel{2} {\mathbb E}\bigr)\widetilde r_{12}
=q^{-1/2} \bigl( \stackrel{1} {\mathbb E}\otimes \stackrel{2}  {R X_S }\bigr)\widetilde r_{12},
\nonumber
\eea
that is, $\widetilde r_{12}$ is indeed the quantum analogue of $-\widetilde P_{12}$.

We now define the quantum analogue $r_{12}$ of  $P_{12}$. This is defined as
\be
r_{12}=q \stackrel{1}{\mathbb I}\otimes \stackrel{2}{\mathbb I}-\widetilde r_{12},
\label{SRP}
\ee
so that
$$
r_{12}=\left[
\begin{array}{cccc}
 q & 0  & 0 & 0 \\
 0 & q-q^{-1}  &1  &0 \\
 0 & 1  & 0 & 0 \\
 0 & 0  & 0 & q
\end{array}
\right].
$$
Observe that $r_{12}=- q^{\frac{1}{2}} s_{12} P_{12}^q$ where
\be
P^q_{12}=
\left[
\begin{array}{cccc}
 1 & 0  & 0 & 0 \\
 q-1 & q-q^{-1}  &1  &0 \\
 0 & 1  & 0 & 0 \\
 0 & q^{-1}-1  & 0 & 1
\end{array}
\right],
\ee
and
\be
 s_{12} =\stackrel{1} L Q^{-1} \stackrel{1} R
=\left[
\begin{array}{cccc}
 -q^{1/2} & 0  & 0 & 0 \\
 q^{1/2}-q^{-1/2} & -q^{-1/2}  &0  &0 \\
 0 & 0  & -q^{-1/2} & 0 \\
 0 & 0  & q^{-1/2}-q^{1/2} & -q^{1/2}
\end{array}
\right].
\label{s12}
\ee
These two matrices satisfy the following useful properties:
$$
P^q_{12}\stackrel{1} R Q\stackrel{2} L Q = P_{12} \stackrel{1} R \otimes \stackrel{2} L,
$$
and
$$
\stackrel{2} X_Z   s_{12} \stackrel{1} L \stackrel{1} X_Y= \stackrel{1} L \stackrel{1} X_Y\otimes \stackrel{2} X_Z=
 \stackrel{1} L \stackrel{2} X_Z Q^{-1} \stackrel{1} X_Y,
$$
so that $s_{12}$ to effectively permutes $\stackrel{2} X_Z$ and $\stackrel{1} X_Y$.

Let us now insert $q \mathbb I\times \mathbb I=r_{12}+\widetilde r_{12}$ in (\ref{eq:bullet-q}) and see the effect of $\widetilde r_{12}$ (the case of $r_{12}$ is easier and we leave it to the reader: one just have to check that all matrices $Q$ appearing when pushing matrices $X$ one through another are indeed killed by $r_{12}$). On the left of $\widetilde r_{12}$, we then obtain that
\bea
&{}&
\widetilde r_{12} \stackrel{1} R \stackrel{1} X_X Q^{-1} \stackrel{2} L \stackrel{2}X_Y
 \stackrel{1} L\stackrel{1} X_{X_1}\stackrel{1} R \stackrel{1} X_{X_2} \cdots
 =q^{1/2}\widetilde r_{12} \stackrel{2} X_X \stackrel{2} L \stackrel{2} L \stackrel{2}X_Y
 \stackrel{1} L\stackrel{1} X_{X_1}\stackrel{1} R \stackrel{1} X_{X_2} \cdots\nonumber\\
 &{}&=-q^{1/2}\widetilde r_{12}\stackrel{1} L\stackrel{1} X_{X_1}Q \stackrel{2} X_X \stackrel{2} R  \stackrel{2}X_Y \stackrel{1} R \stackrel{1} X_{X_2} \cdots
 =-\widetilde r_{12}\stackrel{2} X_{X_1}\stackrel{2} R \stackrel{1} R \stackrel{1} X_{X_2} \stackrel{2} X_X \stackrel{2} R  \stackrel{2}X_Y \cdots
 \nonumber\\
&{}& =-\widetilde r_{12}\stackrel{1} R \stackrel{1} X_{X_2} Q^{-1}\stackrel{2} X_{X_1}\stackrel{2} R \stackrel{2} X_X \stackrel{2} R  \stackrel{2}X_Y \cdots
  =- q^{1/2}\widetilde  r_{12}\stackrel{2} X_{X_2} \stackrel{2} L \stackrel{2} X_{X_1}\stackrel{2} R \stackrel{2} X_X \stackrel{2} R  \stackrel{2}X_Y \cdots,
  \nonumber
\eea
so the action of $\widetilde r_{12}$ inverts the order of quantum operators entering a quantum arc. This happens on the left side of $\widetilde r_{12}$ as well: the first action however happens in ``opposite'' order, we use that
$$
\stackrel{1}{LX_Z}Q^{-1}\stackrel{2}{RX_Z}\widetilde r_{12}=q^{12}\stackrel{1}{LX_Z}\stackrel{1}{X_ZL}\widetilde r_{12}=q^{12}\stackrel{1}R \widetilde r_{12}
$$
to present the expression to the left from $\widetilde r_{12}$ as
\bea
&{}&
q^{1/2} \cdots \stackrel{1}X_{T_2} \stackrel{1} L \stackrel{1} X_{T_1} \stackrel{1} R  \stackrel{1}X_T  \stackrel{1}R
 \stackrel{2} X_P\widetilde  r_{12}
 = q^{1/2} \cdots \stackrel{1}X_{T_2} \stackrel{1} L \stackrel{1} X_{T_1} \stackrel{1} R  \stackrel{2} X_P Q^{-1}\stackrel{1}X_T  \stackrel{1}R
\widetilde  r_{12}\nonumber\\
&{}&=q \cdots \stackrel{1}X_{T_2} \stackrel{1} L \stackrel{1} X_{T_1} \stackrel{1} R  \stackrel{2} X_P   \stackrel{2}L \stackrel{2}X_T
\widetilde  r_{12}=q \cdots \stackrel{1}X_{T_2} \stackrel{1} L \stackrel{2} X_P   \stackrel{2}L \stackrel{2}X_T Q^{-1} \stackrel{1} X_{T_1} \stackrel{1} R
 \widetilde r_{12}\nonumber\\
 &{}&=q^{3/2} \cdots \stackrel{1}X_{T_2} \stackrel{1} L \stackrel{2} X_P   \stackrel{2}L \stackrel{2}X_T  \stackrel{2} R \stackrel{2} X_{T_1}
 \widetilde r_{12}=q^{3/2} \cdots \stackrel{2} X_P   \stackrel{2}L \stackrel{2}X_T  \stackrel{2} R \stackrel{2} X_{T_1}  Q \stackrel{1}X_{T_2} \stackrel{1} L
\widetilde  r_{12}\nonumber\\
 &{}&=q \cdots \stackrel{2} X_P   \stackrel{2}L \stackrel{2}X_T  \stackrel{2} R \stackrel{2} X_{T_1}  \stackrel{2} R  \stackrel{2}X_{T_2}
\widetilde  r_{12},\ \hbox{etc.}\nonumber
\eea
As a result, we obtain two new arcs two halves of which are ``reflected'' from the insertion of $\widetilde r_{12}$. We also see that the above reflections
respect the following mnemonic law: if we multiply all $R$ by $q^{-1/4}$ and all $L$ by $q^{1/4}$, the $q$-factors will be absorbed into the definitions of
$R$ and $L$.

\subsection{Riedemeister moves for quantum geodesic functions}\label{ss:Rieremeister}

For the case of CGL, we have three Riedemester moves for quantum geodesic function algebras.

{\bf(i)} {Given three quantum geodesic functions or quantum geodesic arcs $G_i^\hbar$, $i=1,2,3$  (we do not distinguish between geodesic functions and arcs in this relation), their product
$G^\hbar_1 G^\hbar_2 G^\hbar_3$ can be represented in two ways:}
$$
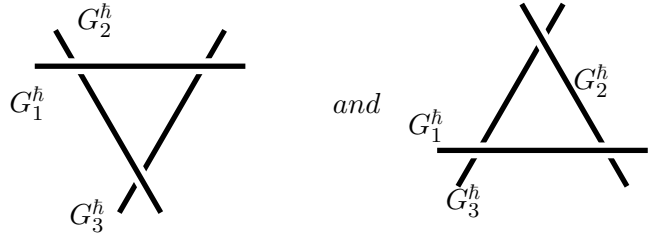
\begin{pspicture}(-2,-1.5)(2,1.5)
{\psset{unit=0.7}
\newcommand{\DISCMIX}{%
{\psset{unit=1}
\rput{240}(0,0){\psline[linewidth=6pt,linecolor=white](-2,0.7)(2,0.7)}
\rput{240}(0,0){\psline[linewidth=2pt,linecolor=black](-2,0.7)(2,0.7)}
\rput{120}(0,0){\psline[linewidth=6pt,linecolor=white](-2,0.7)(2,0.7)}
\rput{120}(0,0){\psline[linewidth=2pt,linecolor=black](-2,0.7)(2,0.7)}
\rput{0}(0,0){\psline[linewidth=6pt,linecolor=white](-2,0.7)(2,0.7)}
\rput{0}(0,0){\psline[linewidth=2pt,linecolor=black](-2,0.7)(2,0.7)}
}
}
\rput(0,0){\DISCMIX}
}
\rput(-1.8,0){\makebox(0,0)[lc]{$G^\hbar_1$}}
\rput(-0.9,1.1){\makebox(0,0)[lc]{$G^\hbar_2$}}
\rput(-1,-1.5){\makebox(0,0)[lc]{$G^\hbar_3$}}
\end{pspicture}
\begin{pspicture}(-3,-1.5)(3,1.5){
{\psset{unit=0.7}
\newcommand{\DISCMIX}{%
{\psset{unit=1}
\rput{240}(0,0){\psline[linewidth=6pt,linecolor=white](-2,-0.7)(2,-0.7)}
\rput{240}(0,0){\psline[linewidth=2pt,linecolor=black](-2,-0.7)(2,-0.7)}
\rput{120}(0,0){\psline[linewidth=6pt,linecolor=white](-2,-0.7)(2,-0.7)}
\rput{120}(0,0){\psline[linewidth=2pt,linecolor=black](-2,-0.7)(2,-0.7)}
\rput{0}(0,0){\psline[linewidth=6pt,linecolor=white](-2,-0.7)(2,-0.7)}
\rput{0}(0,0){\psline[linewidth=2pt,linecolor=black](-2,-0.7)(2,-0.7)}
}
}
\rput(0.5,-.2){\DISCMIX}
}
\rput(-2.5,0){\makebox(0,0)[lc]{$and  $}}
\rput(-1.5,-0.3){\makebox(0,0)[lc]{$G^\hbar_1$}}
\rput(0.7,.3){\makebox(0,0)[lc]{$G^\hbar_2$}}
\rput(-1,-1.3){\makebox(0,0)[lc]{$G^\hbar_3$}}
}
\end{pspicture}
$$
where the upper/lower crossing indicates the order of the corresponding terms in the quantum product. Using the quantum skein relation we prove that indeed the left hand side of this picture is equal to the right hand side. Infact, by resolving all crossings we obtain have eight diagrams on each side: in the left-hand side we have:
$$
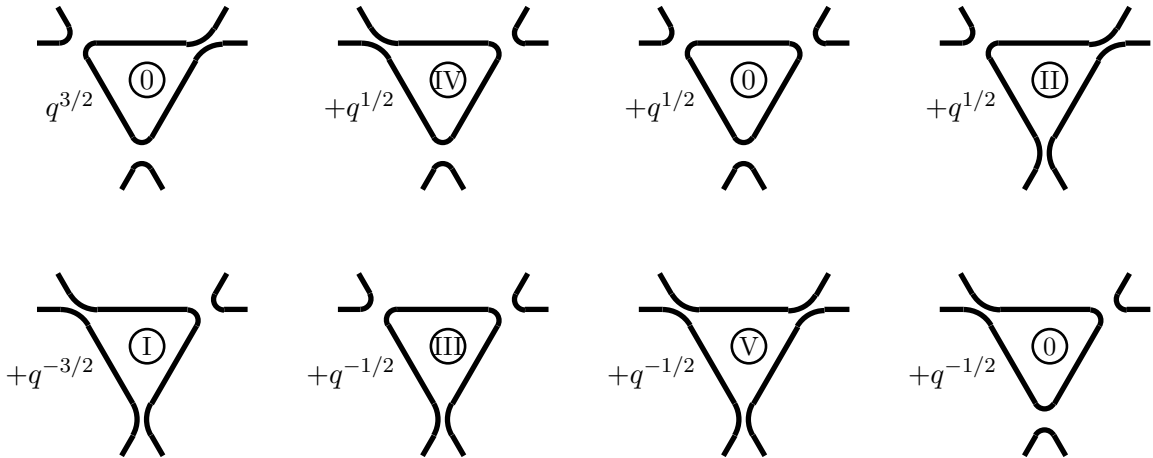
\begin{pspicture}(-2,-1.5)(2,1.5)
{\psset{unit=0.7}
\newcommand{\DISCMIX}{%
{\psset{unit=1}
\rput{240}(0,0){\psline[linewidth=6pt,linecolor=white](-2,0.7)(2,0.7)}
\rput{240}(0,0){\psline[linewidth=2pt,linecolor=black](-2,0.7)(2,0.7)}
\rput{120}(0,0){\psline[linewidth=6pt,linecolor=white](-2,0.7)(2,0.7)}
\rput{120}(0,0){\psline[linewidth=2pt,linecolor=black](-2,0.7)(2,0.7)}
\rput(0,0){\psline[linewidth=6pt,linecolor=white](-2,0.7)(2,0.7)}
\rput(0,0){\psline[linewidth=2pt,linecolor=black](-2,0.7)(2,0.7)}
}
}
\newcommand{\CROSSG}{%
{\psset{unit=1}
\rput(0,0){\pscircle[linewidth=1pt, linecolor=white, fillstyle=solid, fillcolor=white](-1.4,0){0.35}}
\rput(0,0){\psarc[linewidth=2pt](-1.4,-0.69){0.6}{60}{120}}
\rput(0,0){\psarc[linewidth=2pt](-1.4,0.69){0.6}{240}{300}}
}
}
\newcommand{\CROSSL}{%
{\psset{unit=1}
\rput(0,-0.05){\pscircle[linewidth=1pt, linecolor=white, fillstyle=solid, fillcolor=white](-1.4,0){0.35}}
\rput(0,-0.05){\psarc[linewidth=2pt](-1.8,0){0.2}{-60}{60}}
\rput(0,-0.05){\psarc[linewidth=2pt](-1,0){0.2}{120}{240}}
}
}
\rput(0,0){\DISCMIX}
\rput{210}(-0.2,-0.07){\CROSSG}
\rput{-270}(-0.15,0.1){\CROSSL}
\rput{-30}(0,0){\CROSSL}
\rput(-1,-0.5){\makebox(0,0)[rc]{$q^{3/2}$}}
\rput(0,0){\pscircle[linewidth=1pt](0,0){0.35}}
\rput(0,0){\makebox(0,0)[cc]{$0$}}
}
\end{pspicture}
\begin{pspicture}(-2,-1.5)(2,1.5)
{\psset{unit=0.7}
\newcommand{\DISCMIX}{%
{\psset{unit=1}
\rput{240}(0,0){\psline[linewidth=6pt,linecolor=white](-2,0.7)(2,0.7)}
\rput{240}(0,0){\psline[linewidth=2pt,linecolor=black](-2,0.7)(2,0.7)}
\rput{120}(0,0){\psline[linewidth=6pt,linecolor=white](-2,0.7)(2,0.7)}
\rput{120}(0,0){\psline[linewidth=2pt,linecolor=black](-2,0.7)(2,0.7)}
\rput(0,0){\psline[linewidth=6pt,linecolor=white](-2,0.7)(2,0.7)}
\rput(0,0){\psline[linewidth=2pt,linecolor=black](-2,0.7)(2,0.7)}
}
}
\newcommand{\CROSSG}{%
{\psset{unit=1}
\rput(0,0){\pscircle[linewidth=1pt, linecolor=white, fillstyle=solid, fillcolor=white](-1.4,0){0.35}}
\rput(0,0){\psarc[linewidth=2pt](-1.4,-0.69){0.6}{60}{120}}
\rput(0,0){\psarc[linewidth=2pt](-1.4,0.69){0.6}{240}{300}}
}
}
\newcommand{\CROSSL}{%
{\psset{unit=1}
\rput(0,-0.05){\pscircle[linewidth=1pt, linecolor=white, fillstyle=solid, fillcolor=white](-1.4,0){0.35}}
\rput(0,-0.05){\psarc[linewidth=2pt](-1.8,0){0.2}{-60}{60}}
\rput(0,-0.05){\psarc[linewidth=2pt](-1,0){0.2}{120}{240}}
}
}
\rput(0,0){\DISCMIX}
\rput{210}(-0.15,-0.1){\CROSSL}
\rput{-270}(-0.15,0.1){\CROSSL}
\rput{-30}(0,-0.05){\CROSSG}
\rput(-1,-0.5){\makebox(0,0)[rc]{$+q^{1/2}$}}
\rput(0,0){\pscircle[linewidth=1pt](0,0){0.35}}
\rput(0,0){\makebox(0,0)[cc]{\small IV}}
}
\end{pspicture}
\begin{pspicture}(-2,-1.5)(2,1.5)
{\psset{unit=0.7}
\newcommand{\DISCMIX}{%
{\psset{unit=1}
\rput{240}(0,0){\psline[linewidth=6pt,linecolor=white](-2,0.7)(2,0.7)}
\rput{240}(0,0){\psline[linewidth=2pt,linecolor=black](-2,0.7)(2,0.7)}
\rput{120}(0,0){\psline[linewidth=6pt,linecolor=white](-2,0.7)(2,0.7)}
\rput{120}(0,0){\psline[linewidth=2pt,linecolor=black](-2,0.7)(2,0.7)}
\rput(0,0){\psline[linewidth=6pt,linecolor=white](-2,0.7)(2,0.7)}
\rput(0,0){\psline[linewidth=2pt,linecolor=black](-2,0.7)(2,0.7)}
}
}
\newcommand{\CROSSG}{%
{\psset{unit=1}
\rput(0,0){\pscircle[linewidth=1pt, linecolor=white, fillstyle=solid, fillcolor=white](-1.4,0){0.35}}
\rput(0,0){\psarc[linewidth=2pt](-1.4,-0.69){0.6}{60}{120}}
\rput(0,0){\psarc[linewidth=2pt](-1.4,0.69){0.6}{240}{300}}
}
}
\newcommand{\CROSSL}{%
{\psset{unit=1}
\rput(0,-0.05){\pscircle[linewidth=1pt, linecolor=white, fillstyle=solid, fillcolor=white](-1.4,0){0.35}}
\rput(0,-0.05){\psarc[linewidth=2pt](-1.8,0){0.2}{-60}{60}}
\rput(0,-0.05){\psarc[linewidth=2pt](-1,0){0.2}{120}{240}}
}
}
\rput(0,0){\DISCMIX}
\rput{210}(-0.15,-0.1){\CROSSL}
\rput{-270}(-0.15,0.1){\CROSSL}
\rput{-30}(0,0){\CROSSL}
\rput(-1,-0.5){\makebox(0,0)[rc]{$+q^{1/2}$}}
\rput(0,0){\pscircle[linewidth=1pt](0,0){0.35}}
\rput(0,0){\makebox(0,0)[cc]{$0$}}
}
\end{pspicture}
\begin{pspicture}(-2,-1.5)(2,1.5)
{\psset{unit=0.7}
\newcommand{\DISCMIX}{%
{\psset{unit=1}
\rput{240}(0,0){\psline[linewidth=6pt,linecolor=white](-2,0.7)(2,0.7)}
\rput{240}(0,0){\psline[linewidth=2pt,linecolor=black](-2,0.7)(2,0.7)}
\rput{120}(0,0){\psline[linewidth=6pt,linecolor=white](-2,0.7)(2,0.7)}
\rput{120}(0,0){\psline[linewidth=2pt,linecolor=black](-2,0.7)(2,0.7)}
\rput(0,0){\psline[linewidth=6pt,linecolor=white](-2,0.7)(2,0.7)}
\rput(0,0){\psline[linewidth=2pt,linecolor=black](-2,0.7)(2,0.7)}
}
}
\newcommand{\CROSSG}{%
{\psset{unit=1}
\rput(0,0){\pscircle[linewidth=1pt, linecolor=white, fillstyle=solid, fillcolor=white](-1.4,0){0.35}}
\rput(0,0){\psarc[linewidth=2pt](-1.4,-0.69){0.6}{60}{120}}
\rput(0,0){\psarc[linewidth=2pt](-1.4,0.69){0.6}{240}{300}}
}
}
\newcommand{\CROSSL}{%
{\psset{unit=1}
\rput(0,-0.05){\pscircle[linewidth=1pt, linecolor=white, fillstyle=solid, fillcolor=white](-1.4,0){0.35}}
\rput(0,-0.05){\psarc[linewidth=2pt](-1.8,0){0.2}{-60}{60}}
\rput(0,-0.05){\psarc[linewidth=2pt](-1,0){0.2}{120}{240}}
}
}
\rput(0,0){\DISCMIX}
\rput{210}(-0.2,-0.07){\CROSSG}
\rput{-270}(-0.1,0.1){\CROSSG}
\rput{-30}(0,0){\CROSSL}
\rput(-1,-0.5){\makebox(0,0)[rc]{$+q^{1/2}$}}
\rput(0,0){\pscircle[linewidth=1pt](0,0){0.35}}
\rput(0,0){\makebox(0,0)[cc]{\small II}}
}
\end{pspicture}
$$
$$
\begin{pspicture}(-2,-1.5)(2,1.5)
{\psset{unit=0.7}
\newcommand{\DISCMIX}{%
{\psset{unit=1}
\rput{240}(0,0){\psline[linewidth=6pt,linecolor=white](-2,0.7)(2,0.7)}
\rput{240}(0,0){\psline[linewidth=2pt,linecolor=black](-2,0.7)(2,0.7)}
\rput{120}(0,0){\psline[linewidth=6pt,linecolor=white](-2,0.7)(2,0.7)}
\rput{120}(0,0){\psline[linewidth=2pt,linecolor=black](-2,0.7)(2,0.7)}
\rput(0,0){\psline[linewidth=6pt,linecolor=white](-2,0.7)(2,0.7)}
\rput(0,0){\psline[linewidth=2pt,linecolor=black](-2,0.7)(2,0.7)}
}
}
\newcommand{\CROSSG}{%
{\psset{unit=1}
\rput(0,0){\pscircle[linewidth=1pt, linecolor=white, fillstyle=solid, fillcolor=white](-1.4,0){0.35}}
\rput(0,0){\psarc[linewidth=2pt](-1.4,-0.69){0.6}{60}{120}}
\rput(0,0){\psarc[linewidth=2pt](-1.4,0.69){0.6}{240}{300}}
}
}
\newcommand{\CROSSL}{%
{\psset{unit=1}
\rput(0,-0.05){\pscircle[linewidth=1pt, linecolor=white, fillstyle=solid, fillcolor=white](-1.4,0){0.35}}
\rput(0,-0.05){\psarc[linewidth=2pt](-1.8,0){0.2}{-60}{60}}
\rput(0,-0.05){\psarc[linewidth=2pt](-1,0){0.2}{120}{240}}
}
}
\rput(0,0){\DISCMIX}
\rput{210}(-0.15,-0.1){\CROSSL}
\rput{-270}(-0.1,0.1){\CROSSG}
\rput{-30}(0,-0.05){\CROSSG}
\rput(-1,-0.5){\makebox(0,0)[rc]{$+q^{-3/2}$}}
\rput(0,0){\pscircle[linewidth=1pt](0,0){0.35}}
\rput(0,0){\makebox(0,0)[cc]{\small I}}
}
\end{pspicture}
\begin{pspicture}(-2,-1.5)(2,1.5)
{\psset{unit=0.7}
\newcommand{\DISCMIX}{%
{\psset{unit=1}
\rput{240}(0,0){\psline[linewidth=6pt,linecolor=white](-2,0.7)(2,0.7)}
\rput{240}(0,0){\psline[linewidth=2pt,linecolor=black](-2,0.7)(2,0.7)}
\rput{120}(0,0){\psline[linewidth=6pt,linecolor=white](-2,0.7)(2,0.7)}
\rput{120}(0,0){\psline[linewidth=2pt,linecolor=black](-2,0.7)(2,0.7)}
\rput(0,0){\psline[linewidth=6pt,linecolor=white](-2,0.7)(2,0.7)}
\rput(0,0){\psline[linewidth=2pt,linecolor=black](-2,0.7)(2,0.7)}
}
}
\newcommand{\CROSSG}{%
{\psset{unit=1}
\rput(0,0){\pscircle[linewidth=1pt, linecolor=white, fillstyle=solid, fillcolor=white](-1.4,0){0.35}}
\rput(0,0){\psarc[linewidth=2pt](-1.4,-0.69){0.6}{60}{120}}
\rput(0,0){\psarc[linewidth=2pt](-1.4,0.69){0.6}{240}{300}}
}
}
\newcommand{\CROSSL}{%
{\psset{unit=1}
\rput(0,-0.05){\pscircle[linewidth=1pt, linecolor=white, fillstyle=solid, fillcolor=white](-1.4,0){0.35}}
\rput(0,-0.05){\psarc[linewidth=2pt](-1.8,0){0.2}{-60}{60}}
\rput(0,-0.05){\psarc[linewidth=2pt](-1,0){0.2}{120}{240}}
}
}
\rput(0,0){\DISCMIX}
\rput{210}(-0.15,-0.1){\CROSSL}
\rput{-270}(-0.1,0.1){\CROSSG}
\rput{-30}(0,0){\CROSSL}
\rput(-1,-0.5){\makebox(0,0)[rc]{$+q^{-1/2}$}}
\rput(0,0){\pscircle[linewidth=1pt](0,0){0.35}}
\rput(0,0){\makebox(0,0)[cc]{\small III}}
}
\end{pspicture}
\begin{pspicture}(-2,-1.5)(2,1.5)
{\psset{unit=0.7}
\newcommand{\DISCMIX}{%
{\psset{unit=1}
\rput{240}(0,0){\psline[linewidth=6pt,linecolor=white](-2,0.7)(2,0.7)}
\rput{240}(0,0){\psline[linewidth=2pt,linecolor=black](-2,0.7)(2,0.7)}
\rput{120}(0,0){\psline[linewidth=6pt,linecolor=white](-2,0.7)(2,0.7)}
\rput{120}(0,0){\psline[linewidth=2pt,linecolor=black](-2,0.7)(2,0.7)}
\rput(0,0){\psline[linewidth=6pt,linecolor=white](-2,0.7)(2,0.7)}
\rput(0,0){\psline[linewidth=2pt,linecolor=black](-2,0.7)(2,0.7)}
}
}
\newcommand{\CROSSG}{%
{\psset{unit=1}
\rput(0,0){\pscircle[linewidth=1pt, linecolor=white, fillstyle=solid, fillcolor=white](-1.4,0){0.35}}
\rput(0,0){\psarc[linewidth=2pt](-1.4,-0.69){0.6}{60}{120}}
\rput(0,0){\psarc[linewidth=2pt](-1.4,0.69){0.6}{240}{300}}
}
}
\newcommand{\CROSSL}{%
{\psset{unit=1}
\rput(0,-0.05){\pscircle[linewidth=1pt, linecolor=white, fillstyle=solid, fillcolor=white](-1.4,0){0.35}}
\rput(0,-0.05){\psarc[linewidth=2pt](-1.8,0){0.2}{-60}{60}}
\rput(0,-0.05){\psarc[linewidth=2pt](-1,0){0.2}{120}{240}}
}
}
\rput(0,0){\DISCMIX}
\rput{210}(-0.2,-0.07){\CROSSG}
\rput{-270}(-0.1,0.1){\CROSSG}
\rput{-30}(0,-0.05){\CROSSG}
\rput(-1,-0.5){\makebox(0,0)[rc]{$+q^{-1/2}$}}
\rput(0,0){\pscircle[linewidth=1pt](0,0){0.35}}
\rput(0,0){\makebox(0,0)[cc]{\small V}}
}
\end{pspicture}
\begin{pspicture}(-2,-1.5)(2,1.5)
{\psset{unit=0.7}
\newcommand{\DISCMIX}{%
{\psset{unit=1}
\rput{240}(0,0){\psline[linewidth=6pt,linecolor=white](-2,0.7)(2,0.7)}
\rput{240}(0,0){\psline[linewidth=2pt,linecolor=black](-2,0.7)(2,0.7)}
\rput{120}(0,0){\psline[linewidth=6pt,linecolor=white](-2,0.7)(2,0.7)}
\rput{120}(0,0){\psline[linewidth=2pt,linecolor=black](-2,0.7)(2,0.7)}
\rput(0,0){\psline[linewidth=6pt,linecolor=white](-2,0.7)(2,0.7)}
\rput(0,0){\psline[linewidth=2pt,linecolor=black](-2,0.7)(2,0.7)}
}
}
\newcommand{\CROSSG}{%
{\psset{unit=1}
\rput(0,0){\pscircle[linewidth=1pt, linecolor=white, fillstyle=solid, fillcolor=white](-1.4,0){0.35}}
\rput(0,0){\psarc[linewidth=2pt](-1.4,-0.69){0.6}{60}{120}}
\rput(0,0){\psarc[linewidth=2pt](-1.4,0.69){0.6}{240}{300}}
}
}
\newcommand{\CROSSL}{%
{\psset{unit=1}
\rput(0,-0.05){\pscircle[linewidth=1pt, linecolor=white, fillstyle=solid, fillcolor=white](-1.4,0){0.35}}
\rput(0,-0.05){\psarc[linewidth=2pt](-1.8,0){0.2}{-60}{60}}
\rput(0,-0.05){\psarc[linewidth=2pt](-1,0){0.2}{120}{240}}
}
}
\rput(0,0){\DISCMIX}
\rput{210}(-0.15,-0.1){\CROSSL}
\rput{-270}(-0.15,0.1){\CROSSL}
\rput{-30}(0,-0.05){\CROSSG}
\rput(-1,-0.5){\makebox(0,0)[rc]{$+q^{-1/2}$}}
\rput(0,0){\pscircle[linewidth=1pt](0,0){0.35}}
\rput(0,0){\makebox(0,0)[cc]{\small $0$}}
}
\end{pspicture}
$$
and on the right-had side we have:
$$
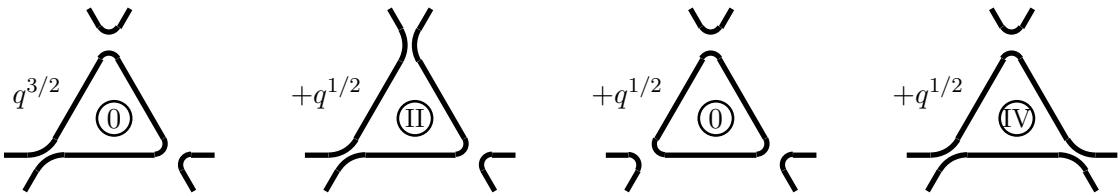
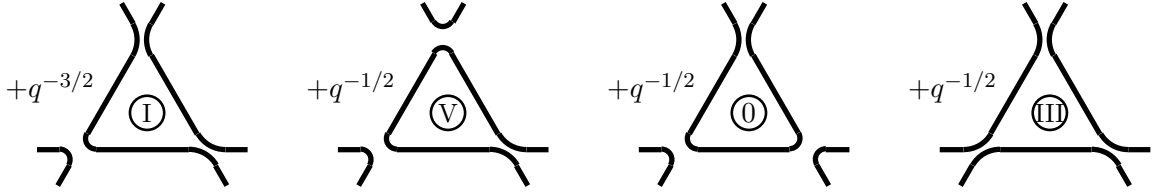
\begin{pspicture}(-2,-1.5)(2,1.5)
{\psset{unit=0.7}
\newcommand{\DISCMIX}{%
{\psset{unit=1}
\rput{240}(0,0){\psline[linewidth=6pt,linecolor=white](-2,-0.7)(2,-0.7)}
\rput{240}(0,0){\psline[linewidth=2pt,linecolor=black](-2,-0.7)(2,-0.7)}
\rput{120}(0,0){\psline[linewidth=6pt,linecolor=white](-2,-0.7)(2,-0.7)}
\rput{120}(0,0){\psline[linewidth=2pt,linecolor=black](-2,-0.7)(2,-0.7)}
\rput{0}(0,0){\psline[linewidth=6pt,linecolor=white](-2,-0.7)(2,-0.7)}
\rput{0}(0,0){\psline[linewidth=2pt,linecolor=black](-2,-0.7)(2,-0.7)}
}
}
\newcommand{\CROSSG}{%
{\psset{unit=1}
\rput(0,0){\pscircle[linewidth=1pt, linecolor=white, fillstyle=solid, fillcolor=white](-1.4,0){0.35}}
\rput(0,0){\psarc[linewidth=2pt](-1.4,-0.69){0.6}{60}{120}}
\rput(0,0){\psarc[linewidth=2pt](-1.4,0.69){0.6}{240}{300}}
}
}
\newcommand{\CROSSL}{%
{\psset{unit=1}
\rput(0,-0.05){\pscircle[linewidth=1pt, linecolor=white, fillstyle=solid, fillcolor=white](-1.4,0){0.35}}
\rput(0,-0.05){\psarc[linewidth=2pt](-1.8,0){0.2}{-60}{60}}
\rput(0,-0.05){\psarc[linewidth=2pt](-1,0){0.2}{120}{240}}
}
}
\rput(0,0){\DISCMIX}
\rput{-210}(-0.2,0.02){\CROSSL}
\rput{270}(-0.05,-0.05){\CROSSL}
\rput{30}(0,0.05){\CROSSG}
\rput(-1,0.5){\makebox(0,0)[rc]{$q^{3/2}$}}
\rput(0,0){\pscircle[linewidth=1pt](0,0){0.35}}
\rput(0,0){\makebox(0,0)[cc]{$0$}}
}
\end{pspicture}
\begin{pspicture}(-2,-1.5)(2,1.5)
{\psset{unit=0.7}
\newcommand{\DISCMIX}{%
{\psset{unit=1}
\rput{240}(0,0){\psline[linewidth=6pt,linecolor=white](-2,-0.7)(2,-0.7)}
\rput{240}(0,0){\psline[linewidth=2pt,linecolor=black](-2,-0.7)(2,-0.7)}
\rput{120}(0,0){\psline[linewidth=6pt,linecolor=white](-2,-0.7)(2,-0.7)}
\rput{120}(0,0){\psline[linewidth=2pt,linecolor=black](-2,-0.7)(2,-0.7)}
\rput{0}(0,0){\psline[linewidth=6pt,linecolor=white](-2,-0.7)(2,-0.7)}
\rput{0}(0,0){\psline[linewidth=2pt,linecolor=black](-2,-0.7)(2,-0.7)}
}
}
\newcommand{\CROSSG}{%
{\psset{unit=1}
\rput(0,0){\pscircle[linewidth=1pt, linecolor=white, fillstyle=solid, fillcolor=white](-1.4,0){0.35}}
\rput(0,0){\psarc[linewidth=2pt](-1.4,-0.69){0.6}{60}{120}}
\rput(0,0){\psarc[linewidth=2pt](-1.4,0.69){0.6}{240}{300}}
}
}
\newcommand{\CROSSL}{%
{\psset{unit=1}
\rput(0,-0.05){\pscircle[linewidth=1pt, linecolor=white, fillstyle=solid, fillcolor=white](-1.4,0){0.35}}
\rput(0,-0.05){\psarc[linewidth=2pt](-1.8,0){0.2}{-60}{60}}
\rput(0,-0.05){\psarc[linewidth=2pt](-1,0){0.2}{120}{240}}
}
}
\rput(0,0){\DISCMIX}
\rput{-210}(-0.2,0.02){\CROSSL}
\rput{270}(-0.1,-0.1){\CROSSG}
\rput{30}(0,0.05){\CROSSG}
\rput(-1,0.5){\makebox(0,0)[rc]{$+q^{1/2}$}}
\rput(0,0){\pscircle[linewidth=1pt](0,0){0.35}}
\rput(0,0){\makebox(0,0)[cc]{\small II}}
}
\end{pspicture}
\begin{pspicture}(-2,-1.5)(2,1.5)
{\psset{unit=0.7}
\newcommand{\DISCMIX}{%
{\psset{unit=1}
\rput{240}(0,0){\psline[linewidth=6pt,linecolor=white](-2,-0.7)(2,-0.7)}
\rput{240}(0,0){\psline[linewidth=2pt,linecolor=black](-2,-0.7)(2,-0.7)}
\rput{120}(0,0){\psline[linewidth=6pt,linecolor=white](-2,-0.7)(2,-0.7)}
\rput{120}(0,0){\psline[linewidth=2pt,linecolor=black](-2,-0.7)(2,-0.7)}
\rput{0}(0,0){\psline[linewidth=6pt,linecolor=white](-2,-0.7)(2,-0.7)}
\rput{0}(0,0){\psline[linewidth=2pt,linecolor=black](-2,-0.7)(2,-0.7)}
}
}
\newcommand{\CROSSG}{%
{\psset{unit=1}
\rput(0,0){\pscircle[linewidth=1pt, linecolor=white, fillstyle=solid, fillcolor=white](-1.4,0){0.35}}
\rput(0,0){\psarc[linewidth=2pt](-1.4,-0.69){0.6}{60}{120}}
\rput(0,0){\psarc[linewidth=2pt](-1.4,0.69){0.6}{240}{300}}
}
}
\newcommand{\CROSSL}{%
{\psset{unit=1}
\rput(0,-0.05){\pscircle[linewidth=1pt, linecolor=white, fillstyle=solid, fillcolor=white](-1.4,0){0.35}}
\rput(0,-0.05){\psarc[linewidth=2pt](-1.8,0){0.2}{-60}{60}}
\rput(0,-0.05){\psarc[linewidth=2pt](-1,0){0.2}{120}{240}}
}
}
\rput(0,0){\DISCMIX}
\rput{-210}(-0.2,0.02){\CROSSL}
\rput{270}(-0.05,-0.05){\CROSSL}
\rput{30}(-0.05,0.1){\CROSSL}
\rput(-1,0.5){\makebox(0,0)[rc]{$+q^{1/2}$}}
\rput(0,0){\pscircle[linewidth=1pt](0,0){0.35}}
\rput(0,0){\makebox(0,0)[cc]{$0$}}
}
\end{pspicture}
\begin{pspicture}(-2,-1.5)(2,1.5)
{\psset{unit=0.7}
\newcommand{\DISCMIX}{%
{\psset{unit=1}
\rput{240}(0,0){\psline[linewidth=6pt,linecolor=white](-2,-0.7)(2,-0.7)}
\rput{240}(0,0){\psline[linewidth=2pt,linecolor=black](-2,-0.7)(2,-0.7)}
\rput{120}(0,0){\psline[linewidth=6pt,linecolor=white](-2,-0.7)(2,-0.7)}
\rput{120}(0,0){\psline[linewidth=2pt,linecolor=black](-2,-0.7)(2,-0.7)}
\rput{0}(0,0){\psline[linewidth=6pt,linecolor=white](-2,-0.7)(2,-0.7)}
\rput{0}(0,0){\psline[linewidth=2pt,linecolor=black](-2,-0.7)(2,-0.7)}
}
}
\newcommand{\CROSSG}{%
{\psset{unit=1}
\rput(0,0){\pscircle[linewidth=1pt, linecolor=white, fillstyle=solid, fillcolor=white](-1.4,0){0.35}}
\rput(0,0){\psarc[linewidth=2pt](-1.4,-0.69){0.6}{60}{120}}
\rput(0,0){\psarc[linewidth=2pt](-1.4,0.69){0.6}{240}{300}}
}
}
\newcommand{\CROSSL}{%
{\psset{unit=1}
\rput(0,-0.05){\pscircle[linewidth=1pt, linecolor=white, fillstyle=solid, fillcolor=white](-1.4,0){0.35}}
\rput(0,-0.05){\psarc[linewidth=2pt](-1.8,0){0.2}{-60}{60}}
\rput(0,-0.05){\psarc[linewidth=2pt](-1,0){0.2}{120}{240}}
}
}
\rput(0,0){\DISCMIX}
\rput{-210}(-0.15,0.05){\CROSSG}
\rput{270}(-0.05,-0.05){\CROSSL}
\rput{30}(0,0.05){\CROSSG}
\rput(-1,0.5){\makebox(0,0)[rc]{$+q^{1/2}$}}
\rput(0,0){\pscircle[linewidth=1pt](0,0){0.35}}
\rput(0,0){\makebox(0,0)[cc]{\small IV}}
}
\end{pspicture}
$$
$$
\begin{pspicture}(-2,-1.5)(2,1.5)
{\psset{unit=0.7}
\newcommand{\DISCMIX}{%
{\psset{unit=1}
\rput{240}(0,0){\psline[linewidth=6pt,linecolor=white](-2,-0.7)(2,-0.7)}
\rput{240}(0,0){\psline[linewidth=2pt,linecolor=black](-2,-0.7)(2,-0.7)}
\rput{120}(0,0){\psline[linewidth=6pt,linecolor=white](-2,-0.7)(2,-0.7)}
\rput{120}(0,0){\psline[linewidth=2pt,linecolor=black](-2,-0.7)(2,-0.7)}
\rput{0}(0,0){\psline[linewidth=6pt,linecolor=white](-2,-0.7)(2,-0.7)}
\rput{0}(0,0){\psline[linewidth=2pt,linecolor=black](-2,-0.7)(2,-0.7)}
}
}
\newcommand{\CROSSG}{%
{\psset{unit=1}
\rput(0,0){\pscircle[linewidth=1pt, linecolor=white, fillstyle=solid, fillcolor=white](-1.4,0){0.35}}
\rput(0,0){\psarc[linewidth=2pt](-1.4,-0.69){0.6}{60}{120}}
\rput(0,0){\psarc[linewidth=2pt](-1.4,0.69){0.6}{240}{300}}
}
}
\newcommand{\CROSSL}{%
{\psset{unit=1}
\rput(0,-0.05){\pscircle[linewidth=1pt, linecolor=white, fillstyle=solid, fillcolor=white](-1.4,0){0.35}}
\rput(0,-0.05){\psarc[linewidth=2pt](-1.8,0){0.2}{-60}{60}}
\rput(0,-0.05){\psarc[linewidth=2pt](-1,0){0.2}{120}{240}}
}
}
\rput(0,0){\DISCMIX}
\rput{-210}(-0.15,0.05){\CROSSG}
\rput{270}(-0.1,-0.1){\CROSSG}
\rput{30}(-0.05,0.1){\CROSSL}
\rput(-1,0.5){\makebox(0,0)[rc]{$+q^{-3/2}$}}
\rput(0,0){\pscircle[linewidth=1pt](0,0){0.35}}
\rput(0,0){\makebox(0,0)[cc]{\small I}}
}
\end{pspicture}
\begin{pspicture}(-2,-1.5)(2,1.5)
{\psset{unit=0.7}
\newcommand{\DISCMIX}{%
{\psset{unit=1}
\rput{240}(0,0){\psline[linewidth=6pt,linecolor=white](-2,-0.7)(2,-0.7)}
\rput{240}(0,0){\psline[linewidth=2pt,linecolor=black](-2,-0.7)(2,-0.7)}
\rput{120}(0,0){\psline[linewidth=6pt,linecolor=white](-2,-0.7)(2,-0.7)}
\rput{120}(0,0){\psline[linewidth=2pt,linecolor=black](-2,-0.7)(2,-0.7)}
\rput{0}(0,0){\psline[linewidth=6pt,linecolor=white](-2,-0.7)(2,-0.7)}
\rput{0}(0,0){\psline[linewidth=2pt,linecolor=black](-2,-0.7)(2,-0.7)}
}
}
\newcommand{\CROSSG}{%
{\psset{unit=1}
\rput(0,0){\pscircle[linewidth=1pt, linecolor=white, fillstyle=solid, fillcolor=white](-1.4,0){0.35}}
\rput(0,0){\psarc[linewidth=2pt](-1.4,-0.69){0.6}{60}{120}}
\rput(0,0){\psarc[linewidth=2pt](-1.4,0.69){0.6}{240}{300}}
}
}
\newcommand{\CROSSL}{%
{\psset{unit=1}
\rput(0,-0.05){\pscircle[linewidth=1pt, linecolor=white, fillstyle=solid, fillcolor=white](-1.4,0){0.35}}
\rput(0,-0.05){\psarc[linewidth=2pt](-1.8,0){0.2}{-60}{60}}
\rput(0,-0.05){\psarc[linewidth=2pt](-1,0){0.2}{120}{240}}
}
}
\rput(0,0){\DISCMIX}
\rput{-210}(-0.15,0.05){\CROSSG}
\rput{270}(-0.05,-0.05){\CROSSL}
\rput{30}(-0.05,0.1){\CROSSL}
\rput(-1,0.5){\makebox(0,0)[rc]{$+q^{-1/2}$}}
\rput(0,0){\pscircle[linewidth=1pt](0,0){0.35}}
\rput(0,0){\makebox(0,0)[cc]{\small V}}
}
\end{pspicture}
\begin{pspicture}(-2,-1.5)(2,1.5)
{\psset{unit=0.7}
\newcommand{\DISCMIX}{%
{\psset{unit=1}
\rput{240}(0,0){\psline[linewidth=6pt,linecolor=white](-2,-0.7)(2,-0.7)}
\rput{240}(0,0){\psline[linewidth=2pt,linecolor=black](-2,-0.7)(2,-0.7)}
\rput{120}(0,0){\psline[linewidth=6pt,linecolor=white](-2,-0.7)(2,-0.7)}
\rput{120}(0,0){\psline[linewidth=2pt,linecolor=black](-2,-0.7)(2,-0.7)}
\rput{0}(0,0){\psline[linewidth=6pt,linecolor=white](-2,-0.7)(2,-0.7)}
\rput{0}(0,0){\psline[linewidth=2pt,linecolor=black](-2,-0.7)(2,-0.7)}
}
}
\newcommand{\CROSSG}{%
{\psset{unit=1}
\rput(0,0){\pscircle[linewidth=1pt, linecolor=white, fillstyle=solid, fillcolor=white](-1.4,0){0.35}}
\rput(0,0){\psarc[linewidth=2pt](-1.4,-0.69){0.6}{60}{120}}
\rput(0,0){\psarc[linewidth=2pt](-1.4,0.69){0.6}{240}{300}}
}
}
\newcommand{\CROSSL}{%
{\psset{unit=1}
\rput(0,-0.05){\pscircle[linewidth=1pt, linecolor=white, fillstyle=solid, fillcolor=white](-1.4,0){0.35}}
\rput(0,-0.05){\psarc[linewidth=2pt](-1.8,0){0.2}{-60}{60}}
\rput(0,-0.05){\psarc[linewidth=2pt](-1,0){0.2}{120}{240}}
}
}
\rput(0,0){\DISCMIX}
\rput{-210}(-0.2,0.02){\CROSSL}
\rput{270}(-0.1,-0.1){\CROSSG}
\rput{30}(-0.05,0.1){\CROSSL}
\rput(-1,0.5){\makebox(0,0)[rc]{$+q^{-1/2}$}}
\rput(0,0){\pscircle[linewidth=1pt](0,0){0.35}}
\rput(0,0){\makebox(0,0)[cc]{$0$}}
}
\end{pspicture}
\begin{pspicture}(-2,-1.5)(2,1.5)
{\psset{unit=0.7}
\newcommand{\DISCMIX}{%
{\psset{unit=1}
\rput{240}(0,0){\psline[linewidth=6pt,linecolor=white](-2,-0.7)(2,-0.7)}
\rput{240}(0,0){\psline[linewidth=2pt,linecolor=black](-2,-0.7)(2,-0.7)}
\rput{120}(0,0){\psline[linewidth=6pt,linecolor=white](-2,-0.7)(2,-0.7)}
\rput{120}(0,0){\psline[linewidth=2pt,linecolor=black](-2,-0.7)(2,-0.7)}
\rput{0}(0,0){\psline[linewidth=6pt,linecolor=white](-2,-0.7)(2,-0.7)}
\rput{0}(0,0){\psline[linewidth=2pt,linecolor=black](-2,-0.7)(2,-0.7)}
}
}
\newcommand{\CROSSG}{%
{\psset{unit=1}
\rput(0,0){\pscircle[linewidth=1pt, linecolor=white, fillstyle=solid, fillcolor=white](-1.4,0){0.35}}
\rput(0,0){\psarc[linewidth=2pt](-1.4,-0.69){0.6}{60}{120}}
\rput(0,0){\psarc[linewidth=2pt](-1.4,0.69){0.6}{240}{300}}
}
}
\newcommand{\CROSSL}{%
{\psset{unit=1}
\rput(0,-0.05){\pscircle[linewidth=1pt, linecolor=white, fillstyle=solid, fillcolor=white](-1.4,0){0.35}}
\rput(0,-0.05){\psarc[linewidth=2pt](-1.8,0){0.2}{-60}{60}}
\rput(0,-0.05){\psarc[linewidth=2pt](-1,0){0.2}{120}{240}}
}
}
\rput(0,0){\DISCMIX}
\rput{-210}(-0.15,0.05){\CROSSG}
\rput{270}(-0.1,-0.1){\CROSSG}
\rput{30}(0,0.05){\CROSSG}
\rput(-1,0.5){\makebox(0,0)[rc]{$+q^{-1/2}$}}
\rput(0,0){\pscircle[linewidth=1pt](0,0){0.35}}
\rput(0,0){\makebox(0,0)[cc]{\small III}}
}
\end{pspicture}
$$
Here, for convenience, we have indicated by Roman numerals homotopic terms in both sides of the equality. The terms indicated by ``$0$'' labels cancel separately on  both sides of the equality provided we set the empty loop equal to $-q-q^{-1}$.

{\bf (ii)} The second Riedemeister move reads
$$
\begin{pspicture}(-1.3,-1.5)(1.33,1.5)
{\psset{unit=1}
\newcommand{\DISCMIX}{%
{\psset{unit=1}
\psline[linewidth=2pt,linecolor=black](0.5,-0.7)(0.5,0.2)
\psline[linewidth=2pt,linecolor=black](-0.5,-0.7)(-0.5,0.2)
\psarc[linewidth=2pt](0,0.2){0.5}{0}{180}
\psline[linewidth=6pt,linecolor=white](-1,0)(1,0)
\psline[linewidth=2pt,linecolor=black](-1,0)(1,0)
}
}
\newcommand{\CROSSG}{%
{\psset{unit=1}
\rput(-0.5,0){\pscircle[linewidth=1pt, linecolor=white, fillstyle=solid, fillcolor=white](0,0){0.2}}
\rput(-0.7,0.2){\psarc[linewidth=2pt](0,0){0.2}{-90}{0}}
\rput(-0.3,-0.2){\psarc[linewidth=2pt](0,0){0.2}{90}{180}}
}
}
\newcommand{\CROSSL}{%
{\psset{unit=1}
\rput(-0.5,0){\pscircle[linewidth=1pt, linecolor=white, fillstyle=solid, fillcolor=white](0,0){0.2}}
\rput(-0.7,-0.2){\psarc[linewidth=2pt](0,0){0.2}{0}{90}}
\rput(-0.3,0.2){\psarc[linewidth=2pt](0,0){0.2}{180}{270}}
}
}
\rput(0,0){\DISCMIX}
}
\rput(-.7,0.3){\makebox(0,0)[rc]{$G^\hbar_1$}}
\rput(-0.5,-.6){\makebox(0,0)[lc]{$G^\hbar_2$}}
\end{pspicture}
\begin{pspicture}(-1.3,-1.5)(1.3,1.5)
{\psset{unit=1}
\newcommand{\DISCMIX}{%
{\psset{unit=1}
\psline[linewidth=2pt,linecolor=black](0.5,-0.7)(0.5,0.2)
\psline[linewidth=2pt,linecolor=black](-0.5,-0.7)(-0.5,0.2)
\psarc[linewidth=2pt](0,0.2){0.5}{0}{180}
\psline[linewidth=6pt,linecolor=white](-1,0)(1,0)
\psline[linewidth=2pt,linecolor=black](-1,0)(1,0)
}
}
\newcommand{\CROSSG}{%
{\psset{unit=1}
\rput(-0.5,0){\pscircle[linewidth=1pt, linecolor=white, fillstyle=solid, fillcolor=white](0,0){0.2}}
\rput(-0.7,0.2){\psarc[linewidth=2pt](0,0){0.2}{-90}{0}}
\rput(-0.3,-0.2){\psarc[linewidth=2pt](0,0){0.2}{90}{180}}
}
}
\newcommand{\CROSSL}{%
{\psset{unit=1}
\rput(-0.5,0){\pscircle[linewidth=1pt, linecolor=white, fillstyle=solid, fillcolor=white](0,0){0.2}}
\rput(-0.7,-0.2){\psarc[linewidth=2pt](0,0){0.2}{0}{90}}
\rput(-0.3,0.2){\psarc[linewidth=2pt](0,0){0.2}{180}{270}}
}
}
\rput(0,0){\DISCMIX}
\rput(0,0.){\CROSSG}
\rput{180}(-0.12,0){\CROSSG}
\rput(-1.2,0){\makebox(0,0)[rc]{$=$}}
\rput(-0.7,0.5){\makebox(0,0)[rc]{$q$}}
\rput(0,0){\pscircle[linewidth=1pt](0,0.3){0.2}}
\rput(0,0.3){\makebox(0,0)[cc]{\tiny $0$}}
}
\end{pspicture}
\begin{pspicture}(-1.3,-1.5)(1.3,1.5)
{\psset{unit=1}
\newcommand{\DISCMIX}{%
{\psset{unit=1}
\psline[linewidth=2pt,linecolor=black](0.5,-0.7)(0.5,0.2)
\psline[linewidth=2pt,linecolor=black](-0.5,-0.7)(-0.5,0.2)
\psarc[linewidth=2pt](0,0.2){0.5}{0}{180}
\psline[linewidth=6pt,linecolor=white](-1,0)(1,0)
\psline[linewidth=2pt,linecolor=black](-1,0)(1,0)
}
}
\newcommand{\CROSSG}{%
{\psset{unit=1}
\rput(-0.5,0){\pscircle[linewidth=1pt, linecolor=white, fillstyle=solid, fillcolor=white](0,0){0.2}}
\rput(-0.7,0.2){\psarc[linewidth=2pt](0,0){0.2}{-90}{0}}
\rput(-0.3,-0.2){\psarc[linewidth=2pt](0,0){0.2}{90}{180}}
}
}
\newcommand{\CROSSL}{%
{\psset{unit=1}
\rput(-0.5,0){\pscircle[linewidth=1pt, linecolor=white, fillstyle=solid, fillcolor=white](0,0){0.2}}
\rput(-0.7,-0.2){\psarc[linewidth=2pt](0,0){0.2}{0}{90}}
\rput(-0.3,0.2){\psarc[linewidth=2pt](0,0){0.2}{180}{270}}
}
}
\rput(0,0){\DISCMIX}
\rput(0,0.){\CROSSG}
\rput{180}(-0.12,0){\CROSSL}
\rput(-1.2,0){\makebox(0,0)[rc]{$+$}}
\rput(-0.7,0.5){\makebox(0,0)[rc]{$1$}}
}
\end{pspicture}
\begin{pspicture}(-1.3,-1.5)(1.3,1.5)
{\psset{unit=1}
\newcommand{\DISCMIX}{%
{\psset{unit=1}
\psline[linewidth=2pt,linecolor=black](0.5,-0.7)(0.5,0.2)
\psline[linewidth=2pt,linecolor=black](-0.5,-0.7)(-0.5,0.2)
\psarc[linewidth=2pt](0,0.2){0.5}{0}{180}
\psline[linewidth=6pt,linecolor=white](-1,0)(1,0)
\psline[linewidth=2pt,linecolor=black](-1,0)(1,0)
}
}
\newcommand{\CROSSG}{%
{\psset{unit=1}
\rput(-0.5,0){\pscircle[linewidth=1pt, linecolor=white, fillstyle=solid, fillcolor=white](0,0){0.2}}
\rput(-0.7,0.2){\psarc[linewidth=2pt](0,0){0.2}{-90}{0}}
\rput(-0.3,-0.2){\psarc[linewidth=2pt](0,0){0.2}{90}{180}}
}
}
\newcommand{\CROSSL}{%
{\psset{unit=1}
\rput(-0.5,0){\pscircle[linewidth=1pt, linecolor=white, fillstyle=solid, fillcolor=white](0,0){0.2}}
\rput(-0.7,-0.2){\psarc[linewidth=2pt](0,0){0.2}{0}{90}}
\rput(-0.3,0.2){\psarc[linewidth=2pt](0,0){0.2}{180}{270}}
}
}
\rput(0,0){\DISCMIX}
\rput(0,0.){\CROSSL}
\rput{180}(-0.12,0){\CROSSG}
\rput(-1.2,0){\makebox(0,0)[rc]{$+$}}
\rput(-0.7,0.5){\makebox(0,0)[rc]{$1$}}
\rput(0,0){\pscircle[linewidth=1pt](0,0.3){0.2}}
\rput(0,0.3){\makebox(0,0)[cc]{\tiny $0$}}
}
\end{pspicture}
\begin{pspicture}(-1.3,-1.5)(1.3,1.5)
{\psset{unit=1}
\newcommand{\DISCMIX}{%
{\psset{unit=1}
\psline[linewidth=2pt,linecolor=black](0.5,-0.7)(0.5,0.2)
\psline[linewidth=2pt,linecolor=black](-0.5,-0.7)(-0.5,0.2)
\psarc[linewidth=2pt](0,0.2){0.5}{0}{180}
\psline[linewidth=6pt,linecolor=white](-1,0)(1,0)
\psline[linewidth=2pt,linecolor=black](-1,0)(1,0)
}
}
\newcommand{\CROSSG}{%
{\psset{unit=1}
\rput(-0.5,0){\pscircle[linewidth=1pt, linecolor=white, fillstyle=solid, fillcolor=white](0,0){0.2}}
\rput(-0.7,0.2){\psarc[linewidth=2pt](0,0){0.2}{-90}{0}}
\rput(-0.3,-0.2){\psarc[linewidth=2pt](0,0){0.2}{90}{180}}
}
}
\newcommand{\CROSSL}{%
{\psset{unit=1}
\rput(-0.5,0){\pscircle[linewidth=1pt, linecolor=white, fillstyle=solid, fillcolor=white](0,0){0.2}}
\rput(-0.7,-0.2){\psarc[linewidth=2pt](0,0){0.2}{0}{90}}
\rput(-0.3,0.2){\psarc[linewidth=2pt](0,0){0.2}{180}{270}}
}
}
\rput(0,0){\DISCMIX}
\rput(0,0.){\CROSSL}
\rput{180}(-0.12,0){\CROSSL}
\rput(-1.2,0){\makebox(0,0)[rc]{$+$}}
\rput(-0.7,0.5){\makebox(0,0)[rc]{$q^{-1}$}}
\rput(0,0){\pscircle[linewidth=1pt](0,0.3){0.2}}
\rput(0,0.3){\makebox(0,0)[cc]{\tiny $0$}}
}
\end{pspicture}
\begin{pspicture}(-1.3,-1.5)(1.33,1.5)
{\psset{unit=1}
\newcommand{\DISCMIX}{%
{\psset{unit=1}
\psarc[linewidth=2pt](0,-0.7){0.5}{0}{180}
\psline[linewidth=6pt,linecolor=white](-1,0)(1,0)
\psline[linewidth=2pt,linecolor=black](-1,0)(1,0)
}
}
\newcommand{\CROSSG}{%
{\psset{unit=1}
\rput(-0.5,0){\pscircle[linewidth=1pt, linecolor=white, fillstyle=solid, fillcolor=white](0,0){0.2}}
\rput(-0.7,0.2){\psarc[linewidth=2pt](0,0){0.2}{-90}{0}}
\rput(-0.3,-0.2){\psarc[linewidth=2pt](0,0){0.2}{90}{180}}
}
}
\newcommand{\CROSSL}{%
{\psset{unit=1}
\rput(-0.5,0){\pscircle[linewidth=1pt, linecolor=white, fillstyle=solid, fillcolor=white](0,0){0.2}}
\rput(-0.7,-0.2){\psarc[linewidth=2pt](0,0){0.2}{0}{90}}
\rput(-0.3,0.2){\psarc[linewidth=2pt](0,0){0.2}{180}{270}}
}
}
\rput(0,0){\DISCMIX}
}
\rput(-1.2,0){\makebox(0,0)[rc]{$=$}}
\rput(-.5,0.3){\makebox(0,0)[rc]{$G^\hbar_1$}}
\rput(-0.5,-.8){\makebox(0,0)[lc]{$G^\hbar_2$}}
\end{pspicture}
$$
Here, again, all unwanted terms labeled ``$0$'' are mutually canceled provided the empty loop is equal to $-q-q^{-1}$.

{\bf (iii)} {The third Riedemeister move of two arcs terminating at the same cusp reads:}
$$
\begin{pspicture}(-2,-1.5)(2,1.5)
{\psset{unit=1.5}
\newcommand{\DISCMIX}{%
{\psset{unit=1}
\psclip{\pscircle[linewidth=1.5pt, linestyle=dashed](0,0){1}}
\rput(-0.1,0.86){\psarc[linewidth=1.5pt,linecolor=blue](0,0){1}{-180}{90}}
\rput(-0.1,-0.86){\psarc[linewidth=4pt,linecolor=white](0,0){1}{-90}{180}}
\rput(-0.1,-0.86){\psarc[linewidth=1.5pt,linecolor=blue](0,0){1}{-90}{180}}
\rput(-0.7,0){\pscircle[linewidth=0pt, linecolor=white,fillstyle=solid, fillcolor=white](0,0){0.3}}
\rput(-1,1){\pscircle[linewidth=1pt,linecolor=green, linestyle=dashed, fillstyle=solid, fillcolor=lightgray](0,0){1}}
\rput(-1,-1){\pscircle[linewidth=1pt,linecolor=green, linestyle=dashed, fillstyle=solid, fillcolor=lightgray](0,0){1}}
\psclip{\pscircle[linestyle=dashed,linewidth=1.5pt](-0.7,0){0.3}}
\rput(-1,2){\psarc[linewidth=1pt,linecolor=blue, linestyle=dashed](0,0){2}{270}{320}}
\rput(-1,-2){\psarc[linewidth=1pt,linecolor=blue, linestyle=dashed](0,0){2}{60}{90}}
\endpsclip
\endpsclip
}
}
\rput(0,0){\DISCMIX}
\rput(0.8,0.6){\makebox(0,0)[lb]{$\mathfrak a^\hbar_2$}}
\rput(0.8,-0.6){\makebox(0,0)[lt]{$\mathfrak a^\hbar_1$}}
}
\end{pspicture}
\begin{pspicture}(-2.5,-1.5)(2.5,1.5)
{\psset{unit=1.5}
\newcommand{\DISCMIX}{%
{\psset{unit=1}
\psclip{\pscircle[linewidth=1.5pt, linestyle=dashed](0,0){1}}
\rput(-0.1,0.86){\psarc[linewidth=1.5pt,linecolor=blue](0,0){1}{-180}{90}}
\rput(-0.1,-0.86){\psarc[linewidth=4pt,linecolor=white](0,0){1}{-90}{180}}
\rput(-0.1,-0.86){\psarc[linewidth=1.5pt,linecolor=blue](0,0){1}{-90}{180}}
\rput(0.45,0){\pscircle[linewidth=0pt, linecolor=white,fillstyle=solid, fillcolor=white](0,0){0.25}}
\psarc[linewidth=1.5pt,linecolor=blue](0.35,0.465){0.4}{250}{320}
\psarc[linewidth=1.5pt,linecolor=blue](0.35,-0.465){0.4}{40}{110}
\rput(-0.7,0){\pscircle[linewidth=0pt, linecolor=white,fillstyle=solid, fillcolor=white](0,0){0.3}}
\rput(-1,1){\pscircle[linewidth=1pt,linecolor=green, linestyle=dashed, fillstyle=solid, fillcolor=lightgray](0,0){1}}
\rput(-1,-1){\pscircle[linewidth=1pt,linecolor=green, linestyle=dashed, fillstyle=solid, fillcolor=lightgray](0,0){1}}
\psclip{\pscircle[linestyle=dashed,linewidth=1.5pt](-0.7,0){0.3}}
\rput(-1,2){\psarc[linewidth=1pt,linecolor=blue, linestyle=dashed](0,0){2}{270}{320}}
\rput(-1,-2){\psarc[linewidth=1pt,linecolor=blue, linestyle=dashed](0,0){2}{60}{90}}
\endpsclip
\endpsclip
}
}
\rput(0,0){\DISCMIX}
\rput(-1.1,0){\makebox(0,0)[rc]{$=q^{1/2}$}}
}
\end{pspicture}
\begin{pspicture}(-2.5,-1.5)(2.5,1.5)
{\psset{unit=1.5}
\newcommand{\DISCMIX}{%
{\psset{unit=1}
\psclip{\pscircle[linewidth=1.5pt, linestyle=dashed](0,0){1}}
\rput(-0.1,0.86){\psarc[linewidth=1.5pt,linecolor=blue](0,0){1}{-180}{90}}
\rput(-0.1,-0.86){\psarc[linewidth=4pt,linecolor=white](0,0){1}{-90}{180}}
\rput(-0.1,-0.86){\psarc[linewidth=1.5pt,linecolor=blue](0,0){1}{-90}{180}}
\rput(0.36,0){\pscircle[linewidth=0pt, linecolor=white,fillstyle=solid, fillcolor=white](0,0){0.19}}
\psarc[linewidth=1.5pt,linecolor=blue](0.165,0){0.1}{-70}{70}
\psarc[linewidth=1.5pt,linecolor=blue](0.585,0){0.1}{120}{250}
\rput(-0.7,0){\pscircle[linewidth=0pt, linecolor=white,fillstyle=solid, fillcolor=white](0,0){0.3}}
\rput(-1,1){\pscircle[linewidth=1pt,linecolor=green, linestyle=dashed, fillstyle=solid, fillcolor=lightgray](0,0){1}}
\rput(-1,-1){\pscircle[linewidth=1pt,linecolor=green, linestyle=dashed, fillstyle=solid, fillcolor=lightgray](0,0){1}}
\rput(1.41,-1.41){\psarc[linewidth=4pt,linecolor=white](0,0){2}{125}{145}}
\rput(1.41,1.41){\psarc[linewidth=4pt,linecolor=white](0,0){2}{215}{235}}
\rput(1.41,-1.41){\psarc[linewidth=2pt,linecolor=red](0,0){2}{127}{143}}
\rput(1.41,1.41){\psarc[linewidth=2pt,linecolor=red](0,0){2}{217}{233}}
\psclip{\pscircle[linestyle=dashed,linewidth=1.5pt](-0.7,0){0.3}}
\rput(-1,2){\psarc[linewidth=1pt,linecolor=blue, linestyle=dashed](0,0){2}{270}{320}}
\rput(-1,-2){\psarc[linewidth=1pt,linecolor=blue, linestyle=dashed](0,0){2}{60}{90}}
\endpsclip
\endpsclip
}
}
\rput(0,0){\DISCMIX}
\rput(-1.1,0){\makebox(0,0)[rc]{$+q^{-1/2}$}}
}
\end{pspicture}
$$
where the second diagram does not contribute if we set the empty loop starting and terminating at the same bordered cusp to be zero.

\subsection{Quantum cluster algebras of geometric type}\label{ss:q-geom}

When quantizing the shear coordinates we associate to $Z_\alpha$ and $\pi_j$ Hermitian operators $Z_\alpha^\hbar$, $\pi_j^\hbar$
with constant commutation relations (\ref{qqq}).

We now fix a spine $\widehat{\mathcal G}_{g,s,n}\in\widehat\Gamma_{g,s,n}$. For any arc ${\mathfrak a}$ (not necessarily belonging to
CGL$_{\mathfrak a}^{\text{max}}$ dual to $\widehat{\mathcal G}_{g,s,n}$) we define the quantum $\lambda$-length (geodesic function) to be
\be
\lambda^\hbar_{\mathfrak a}:=G_{\mathfrak a}:=\tr \bigl[ X_{\pi_{j_1}} \tilde L X_{Z_{\alpha_1}} \tilde R \dots X_{Z_j} F_{\omega_j} X_{Z_j} \dots
X_{Z_{\alpha_n}} \tilde R X_{\pi_{j_2}}K\bigr],
\label{ORDER}
\ee
where $\tilde L=q^{1/4}L$, $\tilde R=q^{-1/4} R$, the matrices $F_\omega$ and $K$ are the same as in the classical case, and
the quantum ordering of operators coincides with the natural ordering of the matrix product.

\begin{theorem}[Laurent positivity]\label{th:q-cluster}
{Let
\be
\label{eq:ring}
\mathcal Z:={\mathbb Z}_{\ge0}\bigl[(\lambda^\hbar_\alpha)^{\pm1},q^{\pm1/4},\omega_j\bigr]\slash_\sim
\ee
be the ring of polynomials with nonnegative integer
coefficients where $\lambda^\hbar_\alpha$ are quantum $\lambda$-lengths comprising a CGL$_{\mathfrak a}^{\text{max}}$,  up to the equivalence relation $\sim$  defined by
 the quantum commutation relations:
$$
q^{I({\mathfrak a}_1,{\mathfrak a}_2)/4}
\lambda_{{\mathfrak a}_1}^\hbar\lambda_{{\mathfrak a}_2}^\hbar
=q^{-I({\mathfrak a}_1,{\mathfrak a}_2)/4}
\lambda_{{\mathfrak a}_2}^\hbar\lambda_{{\mathfrak a}_1}^\hbar,
$$
where $q$ and $\omega_j$ commute with all variables.}

{Then the quantum $\lambda$-length of any other arc in the given cusped Riemann surface belongs to
$\mathcal Z$ and  the shear coordinates determined by the spine $\widehat{\mathcal G}_{g,s,n}\in\widehat\Gamma_{g,s,n}$
dual to the above CGL$_{\mathfrak a}^{\text{max}}$ are monomials in quantum $\lambda$-variables}; explicitly,
\be
e^{Z_e^\hbar/2}=q^{S/16}(\lambda_b^\hbar)^{1/2}(\lambda_d^\hbar)^{1/2}(\lambda_a^\hbar)^{-1/2}(\lambda_c^\hbar)^{-1/2},
\ \hbox{(cf. Fig.~\ref{fi:cross})},
\label{LL1}
\ee
for internal edges that are not incident to loops; here
$$
S=I({\mathfrak a_b},{\mathfrak a_d})-I({\mathfrak a_b},{\mathfrak a_a})-I({\mathfrak a_b},{\mathfrak a_c})
-I({\mathfrak a_d},{\mathfrak a_a})-I({\mathfrak a_d},{\mathfrak a_c})+I({\mathfrak a_a},{\mathfrak a_b});
$$
\be
e^{Z^\hbar_j}=q^{-I({\mathfrak a_b},{\mathfrak a_a})/4}\lambda_b^\hbar (\lambda_a^\hbar)^{-1}\ \hbox{(cf. Fig.~\ref{fi:clustergeneral})}
\label{LL2}
\ee
for internal edges incident to loops, and
\be
e^{\pi^\hbar_j/2}=q^{R/16}(\lambda_c^\hbar)^{1/2}(\lambda_b^\hbar)^{1/2}(\lambda_a^\hbar)^{-1/2}\ \hbox{(cf. Fig.~\ref{fi:cusp})}
\label{LL3}
\ee
for external edges, where $R=I({\mathfrak a_c},{\mathfrak a_b})-I({\mathfrak a_c},{\mathfrak a_a})-I({\mathfrak a_b},{\mathfrak a_a})$.
\end{theorem}

\proof
{To prove that the shear coordinates determined by the spine $\widehat{\mathcal G}_{g,s,n}\in\widehat\Gamma_{g,s,n}$
dual to the above CGL$_{\mathfrak a}^{\text{max}}$ are monomials in quantum $\lambda$-variables, we use Theorem 4.4: since the  $\lambda$-variables  are monomials in the exponentiated shear coordinates, we can invert all formulae and express the exponentiated shear coordinates in terms of  $\lambda$-variables.
In particular formulae (\ref{LL1},\ref{LL2},\ref{LL3}) can be derived in this way, and the powers of $q$ follow from the
Hermiticity property of quantum shear coordinates. }

The
quantum $\lambda$-length $\lambda_{\mathfrak a}$ corresponding to \emph{any} arc ${\mathfrak a}$
(entering \emph{some} CGL$_{\mathfrak a}^{\text{max}}$) is expressed by Lemma~\ref{lm:ordering} as an ordered quantum polynomial
in $e^{\pm Z^\hbar_\alpha/2}$, $e^{\pm Z^\hbar_j}$, $e^{\pi^\hbar_j/2}$, and $\omega_j$. All $\lambda$-lengths enter these expressions
in integer, not half-integer, powers. To see this, let us consider the product of matrices in Lemma~\ref{lm:ordering}: disregarding left and right turns and $\omega_j$, we have an (ordered) string of shear coordinates $\pi^\hbar_{j_1},Z^\hbar_{\alpha_1},\dots,
Z^\hbar_{\alpha_k}, Z^\hbar_j, Z^\hbar_j, Z^\hbar_{\alpha_{k+1}},\dots, Z^\hbar_{\alpha_n}, \pi^\hbar_{j_2}$
(cf. expression (\ref{ORDER})). Expression  (\ref{ORDER}) is a polynomial in $e^{\pm Z^\hbar_{\alpha_k}/2}$,
$e^{\pm Z^\hbar_{j}}$, $\omega_j$ and is clearly proportional to $e^{\pi^\hbar_{j_1}/2}e^{\pi^\hbar_{j_2}/2}$ with coefficients that are in turn
Laurent polynomials of $q^{1/4}$ with positive integer coefficients.
It is easy to see that expressing the terms of this operatorial expansion
in terms of $\lambda$-lengths from the special CGL$_{\mathfrak a}^{\text{max}}$ using formulas (\ref{LL1})--(\ref{LL3})
we have that every $\lambda$-length from this set enters every term of
this expansion even number of times (every time in power $+1/2$ or $-1/2$)
so the total power of any $\lambda$-length is necessarily integer in every term of expansion of (\ref{ORDER}), which completes the
proof of the theorem.\endproof

\begin{remark}\label{rm:closed}
Because the quantum geodesic function $G_\gamma$
of every \emph{closed} geodesic $\gamma$ in $\Sigma_{g,s,n}$ is also a quantum polynomial
in $e^{\pm Z_\alpha/2}$, $e^{\pm Z_j}$, $\omega_j$, and $q^{\pm 1/2}$
with positive integer coefficients, this geodesic function can be again expressed as a Laurent polynomial in $\lambda$-lengths
from a given CGL$_{\mathfrak a}^{\text{max}}$. $\lambda$-lengths thus indeed provide an alternative parameterization of the complete
set of geodesic functions for $\Sigma_{g,s,n}$ with $n>0$.
\end{remark}

\subsubsection{Quantum mutations of quantum cluster variables}
For quantum arcs we have the following mutation rules:
\begin{itemize}
\item Mutating a general inner arc $\lambda_e$ (neither a boundary arc nor an arc bounding a monogon)
for the resulting quantum arc $\lambda^\hbar_f$ we obtain
\be
\lambda^\hbar_f
=\lambda^\hbar_a \bigl(\lambda^\hbar_e\bigr)^{-1}\lambda^\hbar_c+
\lambda^\hbar_b \bigl(\lambda^\hbar_e\bigr)^{-1}\lambda^\hbar_d
\quad \hbox{(cf. Fig~\ref{fi:cross})}.
\label{q-mutation1}
\ee
Here all combinations of four bordered cusps can be identified (for instance, for $\Sigma_{g,s,1}$ all these cusps coincide and
all arcs start and terminate at this single cusp), but for all these combinations we have that
$$
\Bigl(\lambda^\hbar_a \bigl(\lambda^\hbar_e\bigr)^{-1}\lambda^\hbar_c\Bigr)^*=
\lambda^\hbar_c \bigl(\lambda^\hbar_e\bigr)^{-1}\lambda^\hbar_a=\lambda^\hbar_a \bigl(\lambda^\hbar_e\bigr)^{-1}\lambda^\hbar_c
$$
and
$$
\Bigl(\lambda^\hbar_b \bigl(\lambda^\hbar_e\bigr)^{-1}\lambda^\hbar_d\Bigr)^*=
\lambda^\hbar_d \bigl(\lambda^\hbar_e\bigr)^{-1}\lambda^\hbar_b=\lambda^\hbar_b \bigl(\lambda^\hbar_e\bigr)^{-1}\lambda^\hbar_d,
$$
so formula (\ref{q-mutation1}) holds.
\item For a quantum arc $\lambda_c^\hbar$ that bounds a monogon, we have
\be
\lambda_d^\hbar=\lambda_a^\hbar\bigl(\lambda_c^\hbar\bigr)^{-1}\lambda_a^\hbar
+\lambda_b^\hbar\bigl(\lambda_c^\hbar\bigr)^{-1}\lambda_b^\hbar
+\omega_j q^{-I({\mathfrak a}_a,{\mathfrak a}_c)/4+I({\mathfrak a}_b,{\mathfrak a}_c)/4}
\lambda_a^\hbar\bigl(\lambda_c^\hbar\bigr)^{-1}\lambda_b^\hbar
\quad \hbox{(cf. Fig.~\ref{fi:clustergeneral})},
\label{q-mutation2}
\ee
and whereas the first two summands in the right-hand side are obviously self adjoint,
it is the only case of quantum mutations (for $\omega_j\ne 0$) where an explicit $q$-factor appears. Note that $\lambda_a^\hbar$
always commutes with $\lambda_b^\hbar$ and either $I({\mathfrak a}_a,{\mathfrak a}_c)=I({\mathfrak a}_b,{\mathfrak a}_c)=0$
or one of these intersection indices vanishes and the other is equal to $\pm 4$, so possible powers of $q$ in (\ref{q-mutation2})
are $-1,0,1$.
\item No mutation of bordering arcs are allowed.
\end{itemize}

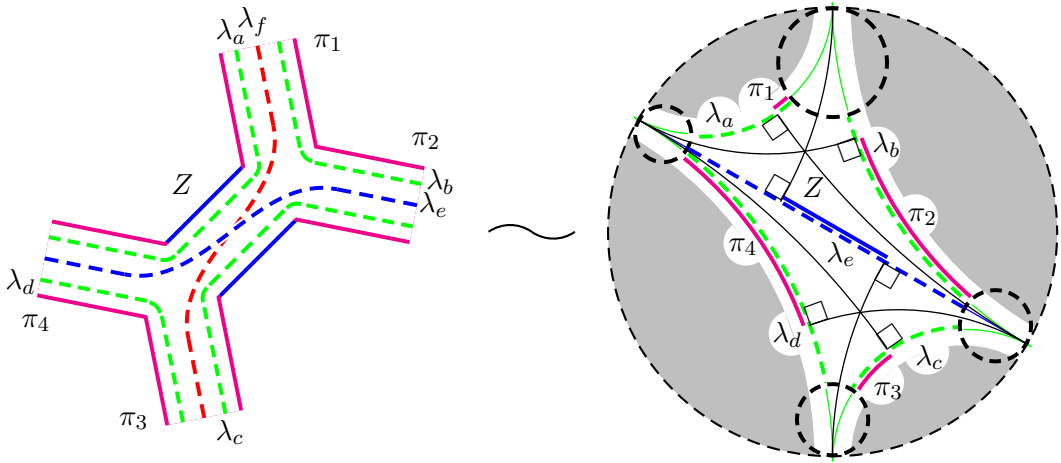
\begin{figure}[tb]
\begin{pspicture}(-3,-3)(4,3){
\newcommand{\DISC}{%
{\psset{unit=1}
\rput{30}(0,0){\psarc[linewidth=1.5pt,linecolor=green,linestyle=dashed](6,0){5.2}{150}{210}}
\rput{210}(0,0){\psarc[linewidth=1.5pt,linecolor=green,linestyle=dashed](6,0){5.2}{150}{210}}
\rput{120}(0,0){\psarc[linewidth=1.5pt,linecolor=green,linestyle=dashed](3.46,0){1.73}{120}{240}}
\rput{300}(0,0){\psarc[linewidth=1.5pt,linecolor=green,linestyle=dashed](3.46,0){1.73}{120}{240}}
\rput{90}(0,0){\pscircle[linecolor=white,fillstyle=solid,fillcolor=white](2.25,0){0.75}}
\rput{-30}(0,0){\pscircle[linecolor=white,fillstyle=solid,fillcolor=white](2.5,0){0.5}}
\rput{-90}(0,0){\pscircle[linecolor=white,fillstyle=solid,fillcolor=white](2.5,0){0.5}}
\rput{150}(0,0){\pscircle[linecolor=white,fillstyle=solid,fillcolor=white](2.6,0){0.4}}
\rput{90}(0,0){
\psclip{\pscircle[linecolor=white](2.25,0){0.75}}
\rput{-60}(0,0){\psarc[linewidth=.5pt,linecolor=green](6,0){5.2}{150}{180}}
\rput{30}(0,0){\psarc[linewidth=.5pt,linecolor=green](3.46,0){1.73}{180}{240}}
\endpsclip
}
\rput{-30}(0,0){
\psclip{\pscircle[linecolor=white](2.5,0){0.5}}
\rput{60}(0,0){\psarc[linewidth=.5pt,linecolor=green](6,0){5.2}{180}{210}}
\rput{-30}(0,0){\psarc[linewidth=.5pt,linecolor=green](3.46,0){1.73}{120}{180}}
\endpsclip
}
\rput{-90}(0,0){
\psclip{\pscircle[linecolor=white](2.5,0){0.5}}
\rput{-60}(0,0){\psarc[linewidth=.5pt,linecolor=green](6,0){5.2}{150}{180}}
\rput{30}(0,0){\psarc[linewidth=.5pt,linecolor=green](3.46,0){1.73}{180}{240}}
\endpsclip
}
\rput{150}(0,0){
\psclip{\pscircle[linecolor=white](2.6,0){0.4}}
\rput{60}(0,0){\psarc[linewidth=.5pt,linecolor=green](6,0){5.2}{180}{210}}
\rput{-30}(0,0){\psarc[linewidth=.5pt,linecolor=green](3.46,0){1.73}{120}{180}}
\endpsclip
}
\psclip{\pscircle[linecolor=white](0,0){3}}
\rput{30}(0,0){\psarc[linewidth=60pt,linecolor=lightgray](6,0){3.9}{150}{210}}
\rput{210}(0,0){\psarc[linewidth=60pt,linecolor=lightgray](6,0){3.9}{150}{210}}
\rput{120}(0,0){\psarc[linewidth=60pt,linecolor=lightgray](3.46,0){.43}{120}{240}}
\rput{300}(0,0){\psarc[linewidth=60pt,linecolor=lightgray](3.46,0){.43}{120}{240}}
\endpsclip
\rput{90}(0,0){\pscircle[linecolor=black,linewidth=1.5pt,linestyle=dashed](2.25,0){0.75}}
\rput{-30}(0,0){\pscircle[linecolor=black,linewidth=1.5pt,linestyle=dashed](2.5,0){0.5}}
\rput{-90}(0,0){\pscircle[linecolor=black,linewidth=1.5pt,linestyle=dashed](2.5,0){0.5}}
\rput{150}(0,0){\pscircle[linecolor=black,linewidth=1.5pt,linestyle=dashed](2.6,0){0.4}}
\rput(0,0){\pscircle[linecolor=black,linestyle=dashed]{3}}
}
}
\rput{45}(-4,0){
\rput(0,0){\psline[linewidth=30pt,linecolor=blue](-1,0)(1,0)}
\rput(0,0){\psline[linewidth=30pt,linecolor=magenta](1,0)(2,1.5)}
\rput(0,0){\psline[linewidth=30pt,linecolor=magenta](1,0)(2,-1.5)}
\rput(0,0){\psline[linewidth=30pt,linecolor=magenta](-1,0)(-2,1.5)}
\rput(0,0){\psline[linewidth=30pt,linecolor=magenta](-1,0)(-2,-1.5)}
\rput(0,0){\psline[linewidth=27pt,linecolor=white](-1,0)(1,0)}
\rput(0,0){\psline[linewidth=27pt,linecolor=white](1,0)(2,1.5)}
\rput(0,0){\psline[linewidth=27pt,linecolor=white](1,0)(2,-1.5)}
\rput(0,0){\psline[linewidth=27pt,linecolor=white](-1,0)(-2,1.5)}
\rput(0,0){\psline[linewidth=27pt,linecolor=white](-1,0)(-2,-1.5)}
\rput(0,0){\psline[linewidth=1.5pt,linecolor=green,linestyle=dashed](-0.7,0.3)(0.7,0.3)}
\rput(0,0){\psline[linewidth=1.5pt,linecolor=green,linestyle=dashed](-0.7,-0.3)(0.7,-0.3)}
\rput(0,0){\psline[linewidth=1.5pt,linecolor=green,linestyle=dashed](0.96,0.46)(1.76,1.66)}
\rput(0,0){\psline[linewidth=1.5pt,linecolor=green,linestyle=dashed](1.44,0.14)(2.24,1.34)}
\rput(0,0){\psline[linewidth=1.5pt,linecolor=green,linestyle=dashed](0.96,-0.46)(1.76,-1.66)}
\rput(0,0){\psline[linewidth=1.5pt,linecolor=green,linestyle=dashed](1.44,-0.14)(2.24,-1.34)}
\rput(0,0){\psline[linewidth=1.5pt,linecolor=green,linestyle=dashed](-0.96,0.46)(-1.76,1.66)}
\rput(0,0){\psline[linewidth=1.5pt,linecolor=green,linestyle=dashed](-1.44,0.14)(-2.24,1.34)}
\rput(0,0){\psline[linewidth=1.5pt,linecolor=green,linestyle=dashed](-0.96,-0.46)(-1.76,-1.66)}
\rput(0,0){\psline[linewidth=1.5pt,linecolor=green,linestyle=dashed](-1.44,-0.14)(-2.24,-1.34)}
\rput(1.6,0){\psarc[linewidth=1.5pt,linecolor=green,linestyle=dashed](0,0){.2}{150}{210}}
\rput{180}(-1.6,0){\psarc[linewidth=1.5pt,linecolor=green,linestyle=dashed](0,0){.2}{150}{210}}
\rput{120}(.75,0.5){\psarc[linewidth=1.5pt,linecolor=green,linestyle=dashed](0,0){.2}{150}{210}}
\rput{240}(.75,-0.5){\psarc[linewidth=1.5pt,linecolor=green,linestyle=dashed](0,0){.2}{150}{210}}
\rput{300}(-.75,-0.5){\psarc[linewidth=1.5pt,linecolor=green,linestyle=dashed](0,0){.2}{150}{210}}
\rput{60}(-.75,0.5){\psarc[linewidth=1.5pt,linecolor=green,linestyle=dashed](0,0){.2}{150}{210}}
\rput(0,0){\psline[linewidth=1.5pt,linecolor=red, linestyle=dashed](1.4,0.6)(2,1.5)}
\rput(0,0){\psline[linewidth=1.5pt,linecolor=blue, linestyle=dashed](1.4,-0.6)(2,-1.5)}
\rput(0,0){\psline[linewidth=1.5pt,linecolor=blue, linestyle=dashed](-1.4,0.6)(-2,1.5)}
\rput(0,0){\psline[linewidth=1.5pt,linecolor=red, linestyle=dashed](-1.4,-0.6)(-2,-1.5)}
\psbezier[linewidth=1.5pt,linecolor=red,linestyle=dashed](-1.4,-0.6)(-0.8,0.3)(.8,-0.3)(1.4,0.6)
\psbezier[linewidth=3pt,linecolor=white](-1.4,0.6)(-0.8,-0.3)(.8,0.3)(1.4,-0.6)
\psbezier[linewidth=1.5pt,linecolor=blue,linestyle=dashed](-1.4,0.6)(-0.8,-0.3)(.8,0.3)(1.4,-0.6)
}
\rput(-4,0){
\rput(-0.5,0.5){\makebox(0,0)[rb]{$Z$}}
\rput(0.2,2.5){\makebox(0,0)[rb]{$\lambda_a$}}
\rput(0.5,2.6){\makebox(0,0)[rb]{$\lambda_f$}}
\rput(2.6,0.5){\makebox(0,0)[lb]{$\lambda_b$}}
\rput(2.5,0.2){\makebox(0,0)[lb]{$\lambda_e$}}
\rput(-0.2,-2.5){\makebox(0,0)[lt]{$\lambda_c$}}
\rput(-2.6,-0.5){\makebox(0,0)[rt]{$\lambda_d$}}
\rput(1.1,2.4){\makebox(0,0)[lb]{$\pi_1$}}
\rput(2.4,1.1){\makebox(0,0)[lb]{$\pi_2$}}
\rput(-1.1,-2.4){\makebox(0,0)[rt]{$\pi_3$}}
\rput(-2.4,-1.1){\makebox(0,0)[rt]{$\pi_4$}}

}
\rput{55}(0,0){
\psarc(0.5,0){0.5}{180}{250}
\psarc(-0.5,0){0.5}{0}{70}
}
\rput(4,0){\DISC
\rput{-30}(0,0){\psline[linewidth=1.5pt,linecolor=blue,linestyle=dashed](-2.2,0)(2,0)}
\rput{-30}(0,0){\psline[linewidth=.5pt,linecolor=blue](-3,0)(-2.2,0)}
\rput{-30}(0,0){\psline[linewidth=.5pt,linecolor=blue](2,0)(3,0)}
}
\rput(4,0){
\rput{150}(0,0){\psarc[linewidth=.5pt](6,0){5.2}{180}{210}}
\rput{-30}(0,0){\psarc[linewidth=.5pt](6,0){5.2}{180}{210}}
\rput{-30}(0,0){\psarc[linewidth=.5pt](3,10.4){10.4}{245}{270}}
\rput{150}(0,0){\psarc[linewidth=.5pt](3,10.4){10.4}{245}{270}}
\rput{-30}(0,0){\psarc[linewidth=.5pt](-3,3.46){3.46}{270}{320}}
\rput{150}(0,0){\psarc[linewidth=.5pt](-3,3.46){3.46}{270}{320}}
\rput{-30}(0,0){\psline[linewidth=1.5pt,linecolor=blue](-0.8,0.07)(.8,0.07)}
\rput{30}(0,0){\psarc[linewidth=1.5pt,linecolor=magenta](6,0){5.12}{170}{199}}
\rput{210}(0,0){\psarc[linewidth=1.5pt,linecolor=magenta](6,0){5.12}{170}{201}}
\rput{120}(0,0){\psarc[linewidth=1.5pt,linecolor=magenta](3.46,0){1.65}{185}{193}}
\rput{300}(0,0){\psarc[linewidth=1.5pt,linecolor=magenta](3.46,0){1.65}{185}{207}}
\rput(-1.3,1.4){\rput{10}(-0.2,0.1){\pswedge[linecolor=white,fillstyle=solid,fillcolor=white](0,0){0.3}{0}{180}}
\makebox(0,0)[rb]{$\lambda_a$}}
\rput(0.9,1){\rput{-70}(-0.3,0.1){\pswedge[linecolor=white,fillstyle=solid,fillcolor=white](0,0){0.3}{0}{180}}
\makebox(0,0)[rb]{$\lambda_b$}}
\rput(0.3,-0.3){\makebox(0,0)[rc]{$\lambda_e$}}
\rput(1.1,-1.5){\rput{-155}(0.15,-0.1){\pswedge[linecolor=white,fillstyle=solid,fillcolor=white](0,0){0.3}{0}{180}}
\makebox(0,0)[lt]{$\lambda_c$}}
\rput(-0.8,-1.2){\rput{110}(0.25,-0.15){\pswedge[linecolor=white,fillstyle=solid,fillcolor=white](0,0){0.3}{0}{180}}
\makebox(0,0)[lt]{$\lambda_d$}}
\rput{-30}(0,0){\rput(0.78,-0.15){\makebox(0,0)[lc]{$\square$}}}
\rput{-30}(0,0){\rput(-0.78,0.15){\makebox(0,0)[rc]{$\square$}}}
\rput{-55}(0,0){\rput(-1.72,0.3){\makebox(0,0)[lt]{$\square$}}}
\rput{125}(0,0){\rput(-1.72,0.3){\makebox(0,0)[lt]{$\square$}}}
\rput{20}(0,0){\rput(0.43,1.07){\makebox(0,0)[lt]{$\square$}}}
\rput{200}(0,0){\rput(0.43,1.07){\makebox(0,0)[lt]{$\square$}}}
\rput(-0.4,0.6){\makebox(0,0)[lc]{$Z$}}
\rput(-0.75,1.75){\rput{40}(-0.2,0){\pswedge[linecolor=white,fillstyle=solid,fillcolor=white](0,0){0.3}{0}{180}}
\makebox(0,0)[rb]{$\pi_1$}}
\rput(1,0.2){\rput{-60}(0.1,0){\pswedge[linecolor=white,fillstyle=solid,fillcolor=white](0,0){0.3}{0}{180}}
\makebox(0,0)[lc]{$\pi_2$}}
\rput(0.55,-2){\rput{-135}(0.1,0){\pswedge[linecolor=white,fillstyle=solid,fillcolor=white](0,0){0.3}{0}{180}}
\makebox(0,0)[lt]{$\pi_3$}}
\rput(-1,-0.2){\rput{130}(-0.15,0.05){\pswedge[linecolor=white,fillstyle=solid,fillcolor=white](0,0){0.3}{0}{180}}
\makebox(0,0)[rc]{$\pi_4$}}
}
}
\end{pspicture}
\caption{\small $\Sigma_{0,1,4}$---a decorated ideal quadrangle. We indicate all five shear coordinates; four of them ($\pi_i$)
correspond to external sides that
constitute the boundary of a hole containing four bordered cusps decorated with horocycles, the fifth coordinate $Z$
corresponds to the inner edge. In this example, the signed distance $Z$ is negative.
Dashed lines correspond to the $\lambda$-lengths, solid lines correspond to the shear
coordinates.}
\label{fi:quadrangle}
\end{figure}

\begin{example}\label{ex:quadrangle}
We now consider in details the example of $\Sigma_{0,1,4}$ represented by an ideal quadrangle in Fig.~\ref{fi:quadrangle}
In this case, we have five shear coordinates: $\pi_i$, $i=1,\dots,4$ and $Z$ and six arcs indicated by dashed lines in the left-hand side
of the figure. The lambda lengths of all six possible arcs are
\beq
\begin{array}{lll}
\lambda_a=e^{\pi_1/2+\pi_4/2+Z/2}, & \lambda_b=e^{\pi_1/2+\pi_2/2}, & \lambda_c=e^{\pi_2/2+\pi_3/2+Z/2},\cr
\lambda_d=e^{\pi_3/2+\pi_4/2}, & \lambda_e=e^{\pi_2/2+\pi_4/2+Z/2}, & \lambda_f=e^{\pi_1/2+\pi_3/2+Z/2}+e^{\pi_1/2+\pi_3/2-Z/2},
\end{array}
\eeq
the  nontrivial commutation relations are
\beq
[\pi_1,\pi_2]=[\pi_2,Z]=[Z,\pi_1]=[\pi_3,\pi_4]=[\pi_4,Z]=[Z,\pi_3]=2\pi i\hbar,
\eeq
and the only nonhomogeneous commutation relation is between $\lambda_e$ and $\lambda_f$:
\beq
\lambda_e\lambda_f=q^{1/2}\lambda_a\lambda_c+q^{-1/2}\lambda_b\lambda_d;\qquad
\lambda_f\lambda_e=q^{-1/2}\lambda_a\lambda_c+q^{1/2}\lambda_b\lambda_d.
\eeq
\end{example}

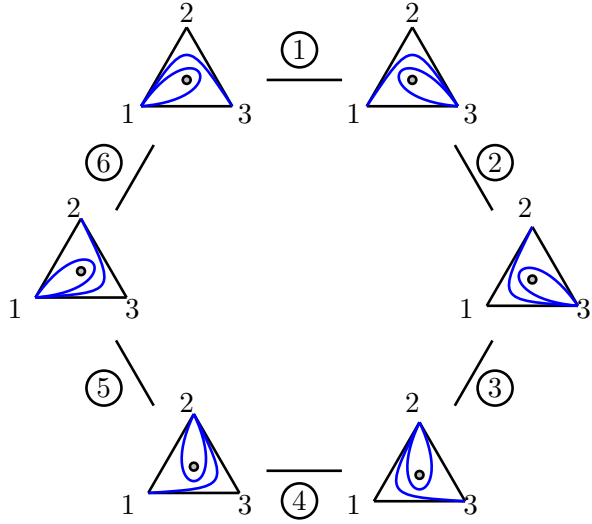
\begin{figure}[tb]
\begin{pspicture}(-4,-3.5)(4,3.5){
{\psset{unit=1}
\newcommand{\TRILEFT}{%
{\psset{unit=0.7}
\rput(0,0){\psline[linewidth=1pt,linecolor=black](-0.866,-0.5)(0.866,-0.5)}
\rput{120}(0,0){\psline[linewidth=1pt,linecolor=black](-0.866,-0.5)(0.866,-0.5)}
\rput{240}(0,0){\psline[linewidth=1pt,linecolor=black](-0.866,-0.5)(0.866,-0.5)}
\rput(0,0){\psbezier[linewidth=1pt,linecolor=blue](-0.866,-0.5)(0,0.8)(0,.8)(0.866,-0.5)}
\rput{120}(0,0){\psbezier[linewidth=1pt,linecolor=blue](0,1)(-0.8,-0.7)(0.8,-0.7)(0,1)}
\rput(0,0){\pscircle[linecolor=black,linewidth=1pt,fillstyle=solid,fillcolor=lightgray](0,0){0.1}}
}
}
\newcommand{\TRIRIGHT}{%
{\psset{unit=0.7}
\rput(0,0){\psline[linewidth=1pt,linecolor=black](-0.866,-0.5)(0.866,-0.5)}
\rput{120}(0,0){\psline[linewidth=1pt,linecolor=black](-0.866,-0.5)(0.866,-0.5)}
\rput{240}(0,0){\psline[linewidth=1pt,linecolor=black](-0.866,-0.5)(0.866,-0.5)}
\rput(0,0){\psbezier[linewidth=1pt,linecolor=blue](-0.866,-0.5)(0,0.8)(0,.8)(0.866,-0.5)}
\rput{240}(0,0){\psbezier[linewidth=1pt,linecolor=blue](0,1)(-0.8,-0.7)(0.8,-0.7)(0,1)}
\rput(0,0){\pscircle[linecolor=black,linewidth=1pt,fillstyle=solid,fillcolor=lightgray](0,0){0.1}}
}
}
\newcommand{\NUM}{%
{\psset{unit=0.9}
\rput(-0.866,-0.5){\makebox(0,0)[cc]{$1$}}
\rput(0,1){\makebox(0,0)[cc]{$2$}}
\rput(0.866,-0.5){\makebox(0,0)[cc]{$3$}}
}
}
\newcommand{\CIR}[1]{%
{\psset{unit=1}
\rput(0,0){\pscircle[linecolor=black,linewidth=1pt](0,0){0.25}}
\rput(0,0){\makebox(0,0)[cc]{${#1}$}}
}
}

\rput(0,0){\psline[linewidth=1pt,linecolor=black](-1.5,2.6)(1.5,2.6)}
\rput{60}(0,0){\psline[linewidth=1pt,linecolor=black](-1.5,2.6)(1.5,2.6)}
\rput{120}(0,0){\psline[linewidth=1pt,linecolor=black](-1.5,2.6)(1.5,2.6)}
\rput{180}(0,0){\psline[linewidth=1pt,linecolor=black](-1.5,2.6)(1.5,2.6)}
\rput{240}(0,0){\psline[linewidth=1pt,linecolor=black](-1.5,2.6)(1.5,2.6)}
\rput{300}(0,0){\psline[linewidth=1pt,linecolor=black](-1.5,2.6)(1.5,2.6)}
\rput(0,0){\pscircle[linecolor=white,linewidth=1pt,fillstyle=solid,fillcolor=white](-1.5,2.6){1}}
\rput{60}(0,0){\pscircle[linecolor=white,linewidth=1pt,fillstyle=solid,fillcolor=white](-1.5,2.6){1}}
\rput{120}(0,0){\pscircle[linecolor=white,linewidth=1pt,fillstyle=solid,fillcolor=white](-1.5,2.6){1}}
\rput{180}(0,0){\pscircle[linecolor=white,linewidth=1pt,fillstyle=solid,fillcolor=white](-1.5,2.6){1}}
\rput{240}(0,0){\pscircle[linecolor=white,linewidth=1pt,fillstyle=solid,fillcolor=white](-1.5,2.6){1}}
\rput{300}(0,0){\pscircle[linecolor=white,linewidth=1pt,fillstyle=solid,fillcolor=white](-1.5,2.6){1}}
\rput(-1.5,2.6){\NUM}
\rput(1.5,2.6){\NUM}
\rput(3,0){\NUM}
\rput(-1.5,-2.6){\NUM}
\rput(1.5,-2.6){\NUM}
\rput(-3,0){\NUM}
\rput(-1.5,2.6){\TRILEFT}
\rput(1.5,2.6){\TRIRIGHT}
\rput(3,0){\rput{120}(0,0){\TRILEFT}}
\rput(1.5,-2.6){\rput{120}(0,0){\TRIRIGHT}}
\rput(-1.5,-2.6){\rput{240}(0,0){\TRILEFT}}
\rput(-3,0){\rput{240}(0,0){\TRIRIGHT}}
\rput(0,3){\CIR{1}}
\rput(2.6,1.5){\CIR{2}}
\rput(2.6,-1.5){\CIR{3}}
\rput(0,-3){\CIR{4}}
\rput(-2.6,-1.5){\CIR{5}}
\rput(-2.6,1.5){\CIR{6}}
}
}
\end{pspicture}
\caption{\small quantum cluster algebra structure for $\Sigma_{0,2,3}$---triangle with the hole inside. We have six seeds
related by six quantum mutations.}
\label{fi:triangle}
\end{figure}

\begin{example}
Quantum cluster algebras associated with polygons---Riemann surfaces $\Sigma_{0,1,n}$---are of finite type, as well as those associated
with the ``punctured'' polygons---Riemann surfaces $\Sigma_{0,2,n}$ in which all cusps are associated with the same boundary component. Let us consider the example of a triangle with one hole inside (cf. Fig.~7 in \cite{FST}); we let $\lambda_{i,j}$ denote the quantum $\lambda$-lengths of bordering arcs (frozen variables) joining vertices $i$ and $j$, we let $\widehat{\lambda}_{i,j}$ denote the quantum $\lambda$-lengths of the (unique) inner arcs joining the same vertices, and $\widehat{\lambda}_{i,i}$ the quantum $\lambda$-length of the loop starting and terminating at the $i$th vertex and going around the inner hole. We have six different seeds in total and they are related by six quantum mutations depicted in Fig.~\ref{fi:triangle}:
\bea
&{1:}&\widetilde{\lambda}_{33}=\widetilde{\lambda}_{13}(\widetilde{\lambda}_{11})^{-1}\widetilde{\lambda}_{13}+\omega \widetilde{\lambda}_{13}\lambda_{13}(\widetilde{\lambda}_{11})^{-1}+{\lambda}_{13}(\widetilde{\lambda}_{11})^{-1}\lambda_{13},
\nonumber\\
&{2:}&\widetilde{\lambda}_{23}=q^{1/4}\widetilde{\lambda}_{33}\lambda_{12}(\widetilde{\lambda}_{13})^{-1}+{\lambda}_{13}(\widetilde{\lambda}_{13})^{-1}\lambda_{23},
\nonumber\\
&{3:}&\widetilde{\lambda}_{22}=\widetilde{\lambda}_{23}(\widetilde{\lambda}_{33})^{-1}\widetilde{\lambda}_{23}+\omega \widetilde{\lambda}_{23}\lambda_{23}(\widetilde{\lambda}_{33})^{-1}+{\lambda}_{23}(\widetilde{\lambda}_{33})^{-1}\lambda_{23},
\nonumber\\
&{4:}&\widetilde{\lambda}_{12}=q^{1/4}\widetilde{\lambda}_{22}\lambda_{13}(\widetilde{\lambda}_{23})^{-1}+{\lambda}_{23}(\widetilde{\lambda}_{23})^{-1}\lambda_{12},
\nonumber\\
&{5:}&\widetilde{\lambda}_{11}=\widetilde{\lambda}_{12}(\widetilde{\lambda}_{22})^{-1}\widetilde{\lambda}_{12}+\omega \widetilde{\lambda}_{12}\lambda_{12}(\widetilde{\lambda}_{22})^{-1}+{\lambda}_{12}(\widetilde{\lambda}_{22})^{-1}\lambda_{12},
\nonumber\\
&{6:}&\widetilde{\lambda}_{12}=q^{1/4}\widetilde{\lambda}_{11}\lambda_{23}(\widetilde{\lambda}_{12})^{-1}+{\lambda}_{12}(\widetilde{\lambda}_{12})^{-1}\lambda_{13}.
\nonumber
\eea
\end{example}

\section{Conclusion}
In this paper we have developed a new surgery that allows passing from Riemann surfaces with holes to Riemann surfaces with bordered cusps by colliding holes of the original Riemann surface. We gave a quantitative description of the newly obtained Riemann surfaces with decorated bordered cusps in terms of the extended shear coordinates and derived explicit combinatorial formulas for geodesic functions of
closed geodesics and $\lambda$-lengths of arcs---geodesics stretched between decorated bordered cusps in terms of the extended shear coordinates.  We postulate the Poisson and quantum commutation relations on the set of extended shear coordinates that are MCG invariant and generate the Goldman brackets on the set of geodesic and arc functions. For generalized laminations (\cite{MW}, \cite{MSW2}) comprising both closed curves and arcs, we have found that maximum systems of arcs CGL$_{\mathfrak a}^{\text{max}}$ are quantum tori: their items (corresponding to compatible regular arcs in the terminology of \cite{FT}) have homogeneous commutation relations, transform in accordance with generalized mutation rules (see \cite{ChSh}) for quantum cluster algebras of Berenstein and Zelevinsky and can be therefore identified with seeds of these quantum cluster algebras. We have also found the explicit quantum ordering for quantum arcs proving that thus ordered expressions satisfy quantum skein relations.

In the forthcoming paper \cite{ChM-monodromy} we shall use the quantum ordering results of this paper for deriving explicit quantum algebras of monodromy matrices for the general $n$-point  SL$_2$ Schlesinger system~\cite{Dub},~\cite{DM}. It is also tempting to transfer our approach to
quantum cluster algebras to quiver algebras of geometric origin studied in \cite{Orlov}.

An interesting example of generalised cluster algebras has appeared recently in the paper by Gekhtman, Shapiro, and Vainshtein \cite{GSV15} where the authors constructed log-canonical (or Darboux) coordinates for $GL_n$ algebras and demonstrated that they transform under the generalised cluster mutations.
It is tempting to compare our approach with that of \cite{GSV15}.

Results of this paper were first reported by the first named author on the Nielsen Retreat of QGM, {\AA}rhus University, 26-29 October 2014. Simultaneously, the papers \cite{FG2} and \cite{All} had appeared dealing with similar issues. In particular, Allegretti had also introduced additional shear-type variables associated to external edges of an ideal triangle decomposition of a bordered cusped Riemann surfaces and observed (Lemma~6.3 in \cite{All}) the monoidal relation between exponentiated shear coordinates and $\lambda$-lengths. However, neither Poisson nor quantum algebras of arc functions were considered there.

\subsection*{Acknowledgments}
The authors wish to  thank Volodya Rubtsov for
several enlightening conversations on the main constructions of the paper.
The authors are also grateful to Misha Shapiro and Anton Zeitlin for the useful
discussion. The work of L. O. Chekhov (the results of Sections 2 and 5) was
supported by the Russian Science Foundation (project 14-50-00005) and was performed
in the Steklov Mathematical Institute of Russian Academy of Sciences.


\begin{thebibliography}{99}

\footnotesize\itemsep=0pt

\bibitem{All}
D. Allegretti, {\it Laminations from the symplectic double}, arXiv:1410.3035v1, 64 pp.

\bibitem{BerZel}
A. Berenstein and A. Zelevinsky, {\it Quantum cluster algebras} {\sl Advances Math.} {\bf195} (2005) 405--455; math/0404446.

\bibitem{Bon2}
F.~Bonahon, {\it Shearing hyperbolic surfaces, bending
pleated surfaces and Thurston's symplectic
form}, {\sl Ann. Fac. Sci. Toulouse Math 6} {\bf 5} (1996), 233-297.



\bibitem{BH}
Brezin E.,  Hikami S.,
Random Matrix, Singularities and Open/Close Intersection Numbers,
arXiv:1502.01416

\bibitem{B1}
Buryak A.,Open intersection numbers and the wave function of the KdV hierarchy, arXiv:1409.7957

\bibitem{Ch1a}
Chekhov L. O., {\it Riemann surfaces with orbifold point},
{\sl Proc. Steklov Math. Inst.}, {\bf266} (2009), pp.1-26


\bibitem{Ch2}
Chekhov L.~O., {\it Orbifold Riemann surfaces and geodesic algebras},
{\sl J. Phys. A: Math. Theor.} {\bf42} (2009) Paper 304007, 32 pp. (electronic)

\bibitem{ChF1}
 Chekhov L., Fock V., {\it A quantum Techm\"uller space}, {\sl Theor. Math.
Phys.} {\bf120} (1999), 1245--1259, math.QA/9908165.

\bibitem{ChF2}
Chekhov L., Fock V., {\it Quantum mapping class group,
pentagon relation, and geodesics},
{\sl Proc. Steklov Math. Inst.}
{\bf 226} (1999), 149--163.



\bibitem{ChM}
Chekhov L.O., Mazzocco M., {\it Isomonodromic deformations and twisted Yangians arising in
Teichm\"uller theory}, {\sl Advances Math.} {\bf 226(6)} (2011) 4731-4775, arXiv:0909.5350.

\bibitem{ChM-D4}
L.O. Chekhov, M. Mazzocco,
{\it Shear coordinate description of the quantized versal unfolding of a $D_4$ singularity}
{\sl J. Phys. A: Math. Theor.} {\bf 43} (2010) 442002 (9pp.)

\bibitem{ChM2}
Chekhov L.O. and Mazzocco M., {\it Quantum ordering for quantum geodesic functions of orbifold Riemann surfaces},
{\sl Amer. Math. Soc. Translations--Ser. 2}  {\bf 234} (2014) 93-116, arXiv:1309.3493.

\bibitem{ChM-monodromy}
Chekhov L.O., Mazzocco M. and Rubtsov V., {\it Quantum monodromy algebras and decorated character varieties,}
(in preparation).

\bibitem{CMR}
Chekhov L., Mazzocco M., Rubtsov V.,
Painlev\'e monodromy manifolds, decorated character varieties and cluster algebras, {\it IMRN,} arXiv:1511.03851.


\bibitem{ChP1}
L.O.Chekhov and R.C. Penner,
{\it Introduction to quantum Thurston theory},
{\sl Russ. Math. Surv.} {\bf 58(6)} (2003) 1141-1183.



\bibitem{ChSh}
L. Chekhov and M. Shapiro
{\it Teichm\"uller spaces of Riemann surfaces with orbifold points of arbitrary order and cluster variables},
{\sl Intl. Math. Res. Notices} 2013; doi: 10.1093/imrn/rnt016. (ArXiv:1111.3963, 20pp)

\bibitem{Dub}
Dubrovin B., Geometry of $2$D topological field theories, Integrable systems and quantum groups
(Montecatini Terme, 1993), {\it Lecture Notes in Math.,}\/ {\bf 1620}, Springer, Berlin, (1996) 120--348.


\bibitem{DM}
Dubrovin B.A., Mazzocco M., Monodromy of certain Painlev\'e-VI transcendents and ref\/lection group,
{\it Invent. Math.} {\bf 141} (2000), 55--147.




\bibitem{FK}
L. D. Faddeev and R. M. Kashaev, {\it Quantum dilogarithm}, {\sl Modern Phys. Lett.}
{\bf A9} (1994) 427--434, hep-th/9310070.

\bibitem{Faddeev}
L.~D.~Faddeev, {\it Discrete Heisenberg--Weyl group and modular group},
{\sl Lett. Math. Phys.}, {\bf34}, (1995), 249--254.

\bibitem{Fock1}
Fock V.V., {\it Combinatorial description of the moduli space of
projective structures}, hep-th/9312193.


\bibitem{FG3} V.~V.~Fock and A.~B.~Goncharov, {\it Cluster Poisson varieties at infinity}, {\sl Selecta Math. New Series}  (2016)
doi:10.1007/S00029-016-0282-6, pp.1-21.



\bibitem{FG1} V.~V.~Fock and A.~B.~Goncharov, {\it Moduli spaces of local systems and higher Teichm\"uller
theory}, {\sl Publ. Math. Inst. Hautes \'Etudes Sci.} 103 (2006), 1-211, math.AG/0311149 v4.

\bibitem{FG}
V.~V.~Fock and A.~B.~Goncharov, {\it Dual Teichm\"uller and lamination spaces},
Chapter 15 in: Handbook on Teichmuller Theory, Vol.1 (IRMA Lectures in Mathematics
and Physics, Vol.11), ed. A.Papadopoulos, IRMA Publ., Strasbourg, France. pp.647-684;
math.DG/0510312.

\bibitem{FG2} V.~Fock and A.~Goncharov, {\it Symplectic double for moduli spaces of $G$-local systems on surfaces} (2014) arXiv:1410.3526, 37pp.

\bibitem{Fock-Rosly}
V.~V.~Fock and A.~A.~Rosly,
{\it Moduli space of flat connections as a Poisson manifold},
{\it Internat. J. Modern Phys. B} {\bf 11} (1997), no. 26-27, 3195--3206.


\bibitem{FT} 
Fomin S. and Thurston D.,
{\it Cluster algebras and triangulated surfaces. Part II: Lambda lengths}, arXiv:1210.5569.


\bibitem{FST} 
Fomin S., Shapiro M., and Thurston D.,
{\it Cluster algebras and triangulated surfaces. Part I: Cluster complexes}, {\sl Acta Math.} {\bf 201} (2008), no. 1, 83--146.




\bibitem{FZ}
S.~Fomin and A.~Zelevinsky, {\it Cluster algebras I: Foundations}, {\sl J. Amer. Math. Soc.}
{\bf 15}(2) (2002) 497--529.

\bibitem{FZ1}
S.~Fomin and A.~Zelevinsky, {\it The Laurent phenomenon}, Adv. in Appl. Math. 28 (2002), no. 2, 119--144

\bibitem{FZ2}
S.~Fomin and A.~Zelevinsky, {\it Cluster algebra II: : Finite type classification} {\sl Invent. Math.}, {\bf 154} (2003), no. 1, 63--121.


\bibitem{GSV} M. Gekhtman, M. Shapiro, and A. Vainshtein, {\it Cluster algebra and Poisson geometry},
{\sl Moscow Math. J.} {\bf3}(3) (2003) 899-934.

\bibitem{GSV15} M. Gekhtman, M. Shapiro, and A. Vainshtein, {\it Generalized cluster structure on the Drinfeld double of $GL_n$}, arXiv:1507.00452.

\bibitem{Gold}
Goldman W.M., {\it Invariant functions on Lie groups and Hamiltonian
f\/lows of surface group representations}, {\sl Invent. Math.} {\bf85}
1986), 263--302.










\bibitem{Kashaev} R. M. Kashaev, {\it Quantization of Teichm\"uller spaces and
the quantum dilogarithm}, {\sl Lett. Math. Phys.} {\bf 43}(1998),105--115, q-alg/9706018.

\bibitem{Kashaev-Dehn}
R.~M.~Kashaev, {\it On the spectrum of Dehn
twists in quantum Teichm\"uller theory},
in: {\sl Physics and Combinatorics}, (Nagoya 2000). River Edge, NJ, World Sci. Publ.,
2001, 63--81; math.QA/0008148.

\bibitem{KP}
R.~M.~Kaufmann and R.~C.~Penner, {\it Closed/open string diagrammatics}, {\sl Nucl. Phys.}
{\bf B748} (2006) 335--379.




\bibitem{KulSk} P. P. Kulish and E. K. Sklyanin, {\it Quantum spectral transform method: Recent developments},
in {\sl Integrable Quantum Field Theories} (Lect. Notes in Physics,: Vol. 151), Berlin, Springer, 1982, pp. 61--119.

\bibitem{MM}
Mazzocco M.,
Confluences of the Painlev\'e equations, Cherednik algebras and q-Askey scheme, arXiv:1307.6140.

\bibitem{Molev}
Molev A., Yangians and classical Lie algebras.
{\it Mathematical Surveys and Monographs,}\/ {\bf 143}, American Mathematical Society, Providence, RI, (2007).


\bibitem{MR} A. Molev, E. Ragoucy, {\it Symmetries and invariants of twisted quantum algebras and associated Poisson algebras},
{\sl Rev. Math. Phys.}, {\bf 20}(2) (2008) 173--198.

\bibitem{MRS} A. Molev, E. Ragoucy, P. Sorba, {\it Coideal subalgebras in quantum affine algebras}, {\sl Rev. Math. Phys.}, {\bf 15} (2003) 789--822.

\bibitem{MSW1} G. Musiker, R. Schiffler, and L. Williams,
{\it Positivity for cluster algebras from surfaces}, {\sl Adv. Math.}, {\bf 227(6)} (2011) 2241-2308.

\bibitem{MSW2} G. Musiker, R. Schiffler, and L. Williams,
{\it Bases for cluster algebras from surfaces}, {\sl Compositio Math.}, {\bf 149(2)} (2013) 217-263; arXiv:1110.4364.

\bibitem{MW} G. Musiker and L. Williams, {\it Matrix formulae and skein relations for cluster algebras from surfaces}  {\sl Intl. Math. Res. Notices}
{\bf 2013(13)} (2013) 2891-2944.

\bibitem{Naz} M. L. Nazarov, {\it Quantum Berezinian and the classical Capelli identity}, {\sl Lett. Math. Phys.}
{\bf 21} (1991) 123--131.




\bibitem{Orlov}
D. Orlov, {\it Geomertic realisations of quiver algebras}, {\sl Proc. Steklov Math. Inst.}, {\bf 290:1} (2015) 70--83 arXiv:1503.03174.

\bibitem{PST}
Pandharipande R., Solomon J. and Walcher J., Intersection theory on moduli of disks, open KdV and Virasoro,
arXiv:1409.2191

\bibitem{Penn1}
Penner R.C., {\it The decorated Teichm\"uller space of Riemann surfaces},
{\sl Comm. Math. Phys.} {\bf113} (1988), 299--339.



\bibitem{ThSh}
W.~P.~Thurston, {\it Minimal stretch maps between
hyperbolic surfaces}, preprint (1984),
math.GT/9801039.




\end{thebibliography}
\end{document}